\documentclass[useAMS,usenatbib,letter]{mn2e}
\usepackage{amsmath}
\usepackage{graphicx}
\usepackage{amsfonts}
\usepackage{pifont}
\newcommand{\bt}{\mbox{\boldmath{$\theta$}}}

\newcommand{\f}{{\cal F}}
\newcommand{\g}{{\cal G}}
\newcommand{\bx}{{\mathbf x}}

\newcommand{\fmap}{F\!M_{\rm ap}}
\newcommand{\map}{M_{\rm ap}}

\begin{document}

\label{firstpage}
\title[Mass Substructure in Abell 1689]{New constraints on the complex mass substructure in Abell 1689 from gravitational flexion} \author[A. Leonard, L. J. King, \& D. M. Goldberg] {Adrienne
  Leonard$^{1,2}$\thanks{Email: adrienne.leonard@cea.fr}, Lindsay
  J. King$^2$, and David M. Goldberg$^3$\\$^1$ DSM/IRFU/SAp, CEA Centre de Saclay, Orme des Merisiers, F-91191 GIF-sur-YVETTE CEDEX, France\\$^2$Institute of Astronomy \& Kavli Institute for Cosmology, University of Cambridge, Madingley Rd, Cambridge CB3 0HA, UK \\$^3$ Department of Physics, Drexel University, 3141 Chestnut St, Philadelphia, PA 19104, USA}
\date{}
\maketitle

\pagerange{\pageref{firstpage}--\pageref{lastpage}} \pubyear{2010}

\begin{abstract}
In a recent publication, the flexion aperture mass statistic was found to provide a robust and effective method by which substructure in galaxy clusters might be mapped. Moreover, we suggested that the masses and mass profile of structures might be constrained using this method. In this paper, we apply the flexion aperture mass technique to HST ACS images of Abell 1689. We demonstrate that the flexion aperture mass statistic is sensitive to small-scale structures in the central region of the cluster. While the central potential is not constrained by our method, due largely to missing data in the central 0.5$^\prime$ of the cluster, we are able to place constraints on the masses and mass profiles of prominent substructures. We identify 4 separate mass peaks, and use the peak aperture mass signal and zero signal radius in each case to constrain the masses and mass profiles of these substructures. The three most massive peaks exhibit complex small-scale structure, and the masses indicated by the flexion aperture mass statistic suggest that these three peaks represent the dominant substructure component of the cluster ($\sim 7\times 10^{14}h^{-1}M_\odot$). Their complex structure indicates that the cluster -- far from being relaxed -- may have recently undergone a merger. The smaller, subsidiary peak is located coincident with a group of galaxies within the cluster, with mass $\sim 1\times10^{14}h^{-1}M_\odot$. These results are in excellent agreement with previous substructure studies of this cluster. 
\end{abstract}

\section{Introduction}

Accurately characterising the mass profiles of large structures in the universe yields important constraints on our cosmological model and theories of structure formation. While N-body simulations carried out under the assumption of the standard concordance cosmological model suggest that cold dark matter haloes are well fit by Navarro-Frenk-White mass density profiles (Navarro, Frenk \& White, 1997), gravitational lensing observations of field galaxies have found total density profiles that are consistently close to isothermal (Treu \& Koopmans, 2002; Rusin, Kochanek \& Keeton, 2003; Rusin \& Kochanek, 2005; Koopmans et al., 2006; Gavazzi et al., 2007; Parker et al., 2007; Czoske et al., 2008; Dye et al., 2008; Tu et al., 2009), reflecting the impact of baryonic physics. 
The NFW model has only recently become favoured over the isothermal model in studies of clusters of galaxies, away from their inner regions (e.g. Carlberg et al., 1997; van der Marel et al., 2000; Athreya et al., 2002; Lin et al., 2004; Hansen et al., 2005;  Katgert, Biviano \& Mazure, 2005; {\L}okas et al., 2006; Rines \& Diaferio, 2006; Wojtak et al., 2007; Okabe et al., 2009). 

Moreover, whilst N-body simulations suggest an anti-correlation between mass and NFW concentration parameter in massive haloes (e.g. Navarro et al., 1997; Shaw et al., 2006), and such a correlation has been seen in lensing studies of field galaxies (Mandelbaum et al., 2008), some clusters of galaxies appear to have much higher concentrations than would be expected. One example of such a cluster is Abell 1689, where previous strong and weak lensing studies have found best-fit mass profiles that differ from one another by up to a factor of $4$ in mass and $3$ in concentration parameter (see, for example, Peng et al., 2009 and references therein). 

Measurements of flexion offer an independent method by which the density profiles of clusters of galaxies and their associated substructures might be characterised (Goldberg \& Bacon, 2005; Bacon et al., 2006). Indeed, flexion measurements -- being a direct probe of the \textit{gradient} of the lensing convergence -- might well allow a clear discrimination between mass profiles. In a recent publication, Lasky and Fluke (2009) have shown that the first flexion signal from an SIS profile differs from that of an NFW profile substantially at moderate angular separation between the source image and the lens, whilst the second flexion signal shows strong variation when the NFW concentration parameter is varied. 

In Leonard, King \& Wilkins (2009), an aperture mass statistic for gravitational flexion was derived and shown to provide a robust method by which massive structures may be detected and mapped out. Moreover, it was shown to be more effective at detecting small-scale substructure than traditional weak lensing methods. Further exploration of this method (Leonard \& King, 2010) demonstrated, with simulated data, that multiple aperture mass reconstructions, performed under differing aperture parameters, might allow one to determine the mass and mass profile of a structure or substructure in an entirely nonparametric way, even with relatively low signal to noise in the reconstructions. 

In this paper, we demonstrate this method using flexion data from HST ACS images of Abell 1689. These data were previously used in a galaxy-galaxy flexion study of the cluster, as well as to generate a parametric model of the substructures in the galaxy cluster (Leonard et al., 2007). In \S~\ref{sec:formalism}, we briefly review the weak lensing and flexion formalism, the flexion aperture mass statistic $\fmap$, and describe a method for model discrimination using $\fmap$ reconstructions. In \S~\ref{sec:A1689}, we review previous lensing studies of Abell 1689, and outline the data and flexion measurement methods. In \S~\ref{sec:results}, we present the results of both our flexion aperture mass reconstructions, and place constraints on the locations, masses, and mass profiles of the substructures detected by the $\fmap$ method. We discuss our results in the context of previous studies of the cluster in \S~\ref{sec:discuss}, and conclude with a summary of our findings in \S~\ref{sec:conclusions}.

Throughout the text, we assume a standard $\Lambda$CDM cosmology with $\Omega_{\rm m}=0.27$, $\Omega_\Lambda=0.73$, and $H_0=100h$\,km\,s$^{-1}$\,Mpc$^{-1}$.

\section{Review of Formalism}
\label{sec:formalism}

In this section, we briefly review the formalism underlying traditional weak lensing and flexion studies, and present the aperture mass statistic for flexion. While this statistic does not directly provide a determination of the masses of structures detected, it does provide a robust, non-parametric method for detection of mass concentrations (see, e.g. Schneider, 1996; Schneider et al., 1998; Leonard, King \& Wilkins, 2009). 

\subsection{Shear and Flexion}

First, we begin by describing the origin of the shear and flexion terms in weak lensing. As described in Goldberg \& Bacon (2005), in the case of a smoothly varying lens field, the lens equation describing the mapping between coordinates in the source plane, $\beta_i$, and coordinates in the lens plane, $\theta_i$, can be expressed as
\begin{equation}
\beta_i\simeq {\cal A}_{ij}\theta_j+\frac{1}{2}D_{ijk}\theta_j\theta_k\ ,
\end{equation}
where
\begin{equation}
{\cal A}=\left(\begin{array}{cccc} 1-\kappa-\gamma_1 & -\gamma_2 \\ -\gamma_2 & 1-\kappa+\gamma_1\end{array}\right)
\end{equation}
is the magnification matrix, which is related to the dimensionless projected surface mass density $\kappa$ of the lens and the complex shear $\gamma$, and the $D$ operators $D_{ijk}=\partial_k{\cal A}_{ij}$ are given by (Bacon et al., 2006)
\begin{eqnarray}
D_{ij1}=-\frac{1}{2}\left(\begin{array}{cccc} 3\f_1+\g_1 & \f_2+\g_2\\ \f_2+\g_2 & \f_1-\g_1\end{array}\right)\ ,\nonumber\\
D_{ij2}=-\frac{1}{2}\left(\begin{array}{cccc}\f_2+\g_2 & \f_1-\g_1\\\f_1-\g_1 & 3\f_2-\g_2\end{array}\right)\ ,
\end{eqnarray}
where $\f$ and $\g$ are first and second flexion, respectively. 

While the convergence gives rise to an isotropic magnification of a background image, the shear and flexion operators yield anisotropic distortions in galaxy images. The complex shear $\gamma=\gamma_1+i\gamma_2=|\gamma|e^{2i\phi}$ induces a tangentially-aligned ellipticity in galaxy images, while the first flexion, $\f=\f_1+i\f_2=|\f|e^{i\phi}$, is a direct probe of the gradient of the convergence, and gives rise to a skewness and a shift in the centroid of the brightness distribution of the galaxy image. Second flexion, $\g=\g_1+i\g_2=|\g|e^{3i\phi}$, has $m=3$ rotational symmetry, and gives rise to a bending or arciness in galaxy images. As described in Goldberg \& Leonard (2007; see also Leonard et al., 2007; Okura et al., 2007, 2008), reliable measurements of this latter distortion have not, to date, been obtained; therefore for the remainder of this paper, we focus on measurements of first flexion only. 

Each of these distortions may be directly related back to the convergence through a convolution of the form
\begin{equation}
\kappa(\bt)=\int d^2\theta^\prime {\cal D}_M(\bt-\bt^\prime)\ M_E(\bt^\prime)\ , 
\end{equation}
where $M_E$ is the component of the shear (flexion) aligned tangentially (radially) with respect to the centre of the lens, and the convolution kernels are given by
\begin{eqnarray}
{\cal D}_\gamma&=&-\frac{1}{(\theta_1-i\theta_2)^2}\ , \\
{\cal D}_\f&=&\ \frac{1}{2\pi(\theta_1-i\theta_2)}\ , \\
{\cal D}_\g&=&-\frac{\theta_1+i\theta_2}{2\pi(\theta_1-i\theta_2)^2}\ .
\end{eqnarray}
Lasky \& Fluke (2009) have presented analytic expressions for the convergence, shear and flexion for a number of commonly-used mass profiles. Relevant results for the singular isothermal sphere and NFW models considered in this paper are given in Appendix \ref{app:profiles}.

\subsection{The Flexion Aperture Mass Statistic $\fmap$}

The flexion is a method by which the first flexion is related to the underlying lens convergence. In this method, apertures of radius $R$ are laid down at various points over the field. Within each aperture, the convergence is convolved with a mass filter function $w(r)$. This convolution is then related to the measured flexion convolved with an appropriate filter function $Q(r)$; i.e. for an aperture located at angular position $\bx_0$, the flexion aperture mass statistic is defined as
\begin{eqnarray}
\fmap(\bx_0; R)&=&\int_{|\bx|\le R} d^2\bx\ \kappa(\bx+\bx_0)w(|\bx|) \label{eq:1st}\\
&=& \int_{|\bx|\le R} d^2\bx\ \f_E(\bx;\bx_0) Q(|\bx|)\label{eq:2nd}\ ,
\end{eqnarray}
where $\bx$ is the co-ordinate with respect to the centre of the aperture and $\f_E$ is the compontent of the flexion oriented radially with respect to the aperture centre. Note that the definition of $\fmap$ given by equation \ref{eq:1st} is formally identical to the definition of the shear aperture mass statistic (see Schneider 1996). The two methods are distinct only in the measurable quantity considered and the choice of filter function $Q$, which is related to the mass filter function $w$ (see equation \ref{eq:qw} below). 

The mass filter function is subject to the condition
\begin{equation}
\int_0^R\ x\ w(x)dx=0\ ,
\end{equation}
in order to ensure that the aperture mass statistic is invariant under a transformation $\kappa(\bx) \rightarrow \kappa(\bx) + \kappa_0$, where $\kappa_0$ is an additive constant. Further, the mass filter function, $w$, must go to zero smoothly at the aperture boundary. The filter functions $w$ and $Q$ can be related to each other by considering the relationship between the convergence $\kappa$ and the flexion $\f$ (see Leonard, King \& Wilkins, 2009)
\begin{eqnarray}
Q(x)&=&-\frac{1}{x}\int_0^x\ x^\prime w(x^\prime)\ dx^\prime\ ,\nonumber\\
w(x)&=&-\frac{1}{x}Q(x)-\frac{dQ}{dx}\ .\label{eq:qw}
\end{eqnarray}

\begin{figure}
\center
\includegraphics[width=0.25\textwidth]{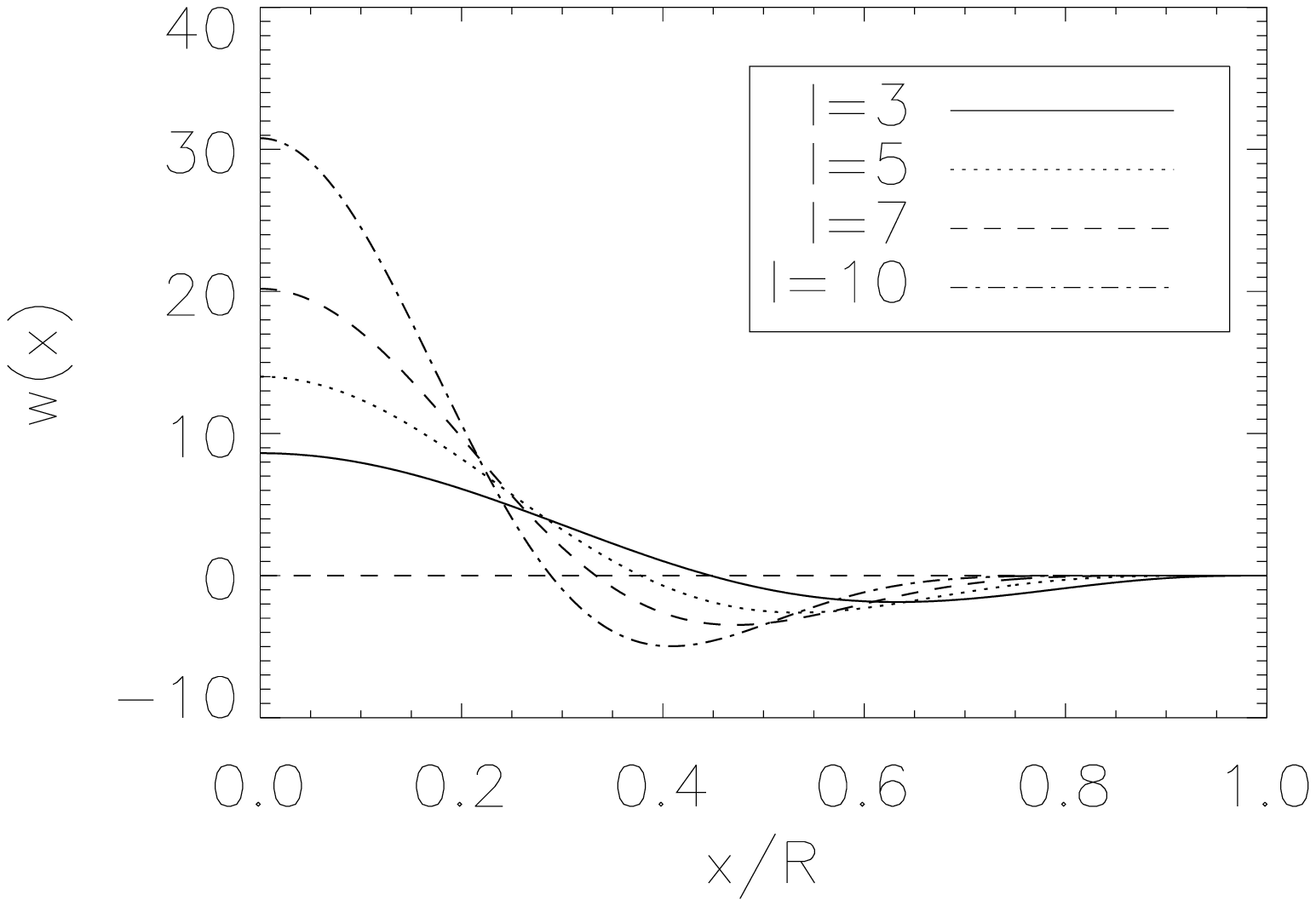}\includegraphics[width=0.25\textwidth]{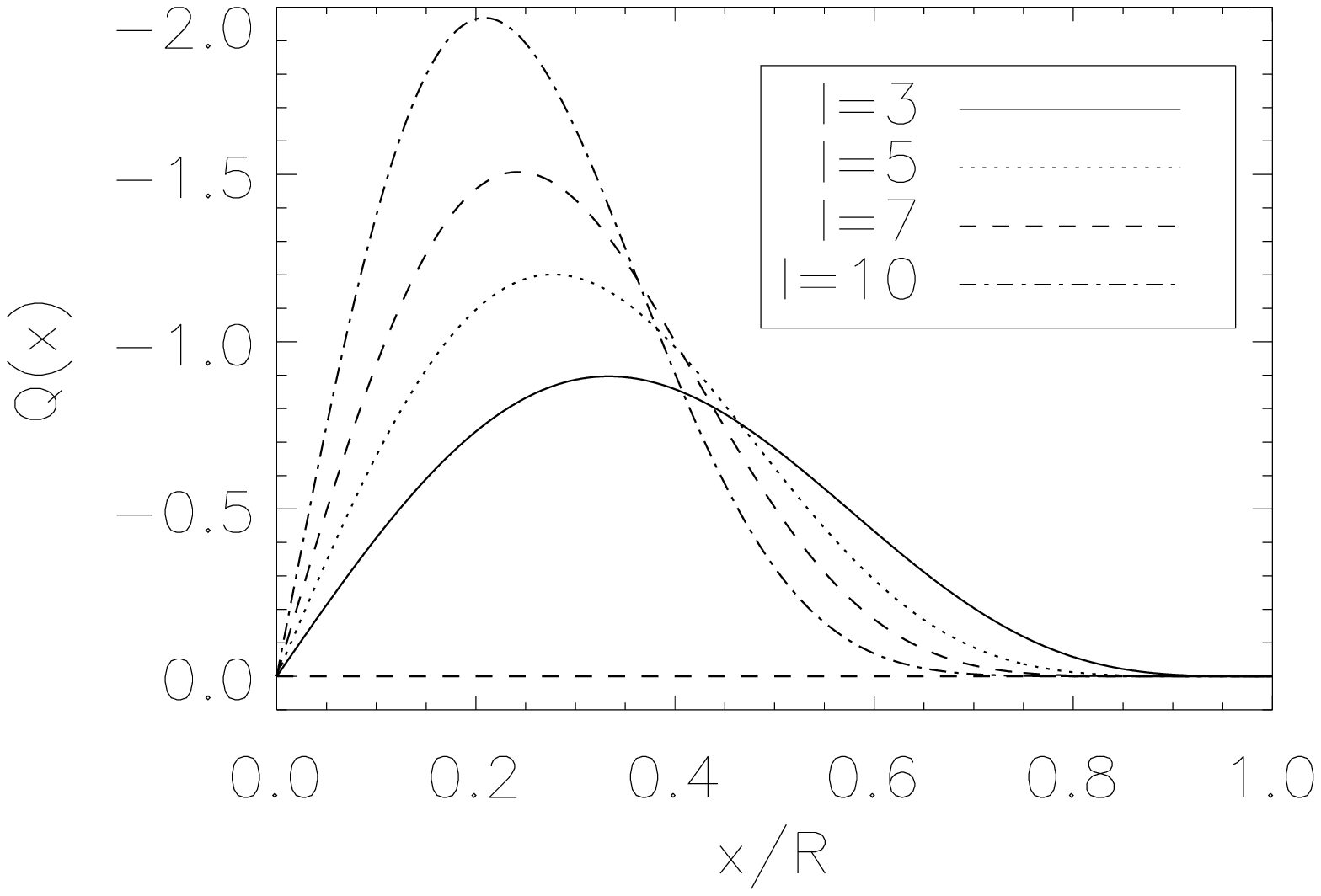}
\caption{Mass (\textit{left panel}) and flexion (\textit{right panel}) filter functions as described in the text for $\ell=$3, 5, 7, and 10.\label{fg:filters}}
\end{figure}

In this paper we choose our mass filter function $w$ to be given by a family of polynomial functions (Schneider et al., 1998)
\begin{equation}
w(x)=A\frac{(2+\ell)^2}{\pi}\left(1-\frac{x^2}{R^2}\right)^\ell\left(\frac{1}{2+\ell}-\frac{x^2}{R^2}\right)\ ,
\end{equation}
where
\begin{equation}
A=\frac{4}{\sqrt{\pi}}\frac{\Gamma\left(\frac{7}{2}+\ell\right)}{\Gamma\left(3+\ell\right)}
\end{equation}
is a normalisation constant, and $\ell$ is an integer defining the polynomial order of the filter function. This family of mass filter functions yields the flexion filter functions
\begin{equation}
Q(x)=-A\frac{2+\ell}{2\pi}\ x\left(1-\frac{x^2}{R^2}\right)^{1+\ell}\ .
\end{equation}
These filter functions are plotted in Figure\,\ref{fg:filters} for various values of $\ell$.

As discussed in Schneider et al. (1998) and Leonard, King \& Wilkins (2009), these filter functions are not designed to provide optimal signal to noise for any given mass profile; rather, the advantage of using such a family of filter functions is that they provide a robust way to detect real structures from lensing measurements. Moreover, higher-order polynomials provide more sharply peaked filter functions, which are sensitive to structures on smaller scales. Therefore, changing the aperture size and the polynomial order will provide information on a range of different physical scales. 

Filter functions that better match particular shear profiles have been suggested in the literature; for example Schirmer (2004) proposed a filter more appropriate to the NFW profile in the context of searching for galaxy clusters using weak lensing shear data. In addition, Bartelmann, King \& Schneider (2001) considered the impact of lens ellipticity on the measured shear aperture mass ($\map$) signal, concluding that for a moderate ellipticity there is only a few percent difference compared with a spherically symmetric lens. Similar corrections would be expeted for the $\fmap$ statistic, as it is formally identical to that for shear.

One of the advantages of using aperture mass techniques is that the noise in the reconstruction is fairly straightforward to estimate, and the noise properties of the aperture mass maps are generally well-understood. The noise may be estimated by repeatedly rotating the individual flexion vectors through a random angle, and repeating the measurement. The standard deviation of an ensemble of such randomisations gives a good estimate of the reconstruction noise. The mean signal from these randomisations also gives an indication of the level to which systematic errors affect the measurement. Zero signal is expected; deviation from this indicates a systematic effect. For a more complete discussion of the noise properties of the aperture mass statistic, the reader is referred to Schneider (1996).

Systematic errors may also be probed by considering the so-called B-mode signal, obtained by rotating the flexion measurements by $\pi/2$ radians and repeating the aperture filtering. Gravitational lensing naturally does not give rise to B-modes; therefore a B-mode signal in the resulting aperture mass maps is indicative of residual systematic errors in the measurement, e.g. arising from incomplete correction of galaxy ellipticities for the impact of telescope optics.

\subsection{Determining Masses and Profiles with $\fmap$}

In Leonard \& King (2010), we considered the radially-averaged flexion aperture mass signal from isothermal and NFW profile lenses with varying concentration parameter, and investigated the evolution of the signal under changes to the aperture parameters (the radius $R$ and polynomial order $\ell$). Specifically, measurements of the peak signal and zero-signal radius of the aperture mass signal for various aperture parameters allowed us to discriminate between mass models using simulated flexion data. In this paper, we apply this method to ACS images of Abell 1689. 

The method centres on comparing data to simulated template profiles, and cross-matching data and templates as the aperture parameters are varied. For each aperture mass reconstruction, mass peaks are identified in the signal to noise map. The aperture mass signal is then circularly-averaged around the centre of the signal-to-noise peak, and the peak signal and zero-signal radius are computed. The template models are then searched, and those models showing a peak signal and a zero-signal radius that match the data, within the error bars, are identified. As the aperture parameters are varied, the set of template models compatible with all the data points is reduced, as a result of the fact that the peak signal and zero signal radii from different mass models vary differently under changes to the aperture parameters. 

The template profiles span a range of masses between $10^{11}-2\times10^{15}h^{-1}M_\odot$, and  use both SIS and NFW models with concentration parameter sampled at integer values in the range $c=3-12$. The template models are generated assuming a constant background source redshift of $z_{\rm s}=1.0$ and a lens redshift of $z_{\rm d}=0.2$. 

These distances come into the normalisation of the flexion signals in different ways for the SIS and NFW density profiles. As described in Leonard \& King (2010), the SIS model induces a flexion related to the source and lens separations through
\begin{equation}
\f_{\rm SIS}\propto \frac{D_{\rm ds}}{D_{\rm s}}\ ,
\end{equation}
whilst the redshift dependence of the NFW flexion signal is given by
\begin{equation}
\f_{\rm NFW}\propto H^2(z_{\rm d})\frac{D_{\rm ds}D_{\rm d}^2}{D_{\rm s}}\ ,
\end{equation}
where $H(z_{\rm d})=H_0\sqrt{\Omega_{\rm M}(1+z_{\rm d})^3+\Omega_\Lambda}$. By appropriately scaling the template profiles, it is straightforward to compare measurements with different source and lens redshifts. 

One might imagine that in assuming a constant background source redshift, we might introduce a bias in our measurements, which may manifest as a tilt in the radial profile of the aperture mass statistic due to magnification effects in the neighbourhood of mass concentrations. However, we ignore this effect for the purposes of this paper. Given that the correction factors described above are fairly small under changes to the source and lens redshifts, and given that the $\fmap$ reconstructions themselves are rather noisy, it is unlikely that this bias will be a dominant source of error.

\section{Abell 1689}
\label{sec:A1689}
\subsection{Gravitational Lensing Studies}
Abell 1689 is a massive cluster of galaxies located at a redshift of $z=0.187$ (Frye et al., 2007), which exhibits the largest known Einstein radius for a gravitational lens observed to date: $\theta_E=45''$ for $z_{\rm s}=1$ (see, e.g., Tyson, Wenk \& Valdes., 1990; Miralda-Escud\'{e} \& Babul, 1995; Broadhurst et al., 2005a,b). This cluster has been the subject of numerous X-ray and gravitational lensing studies, and the mass estimates derived from these two methods have often been found to disagree by a factor of 2 or more (see Peng et al., 2009, and references therein). The X-ray morphology of the cluster appears to be fairly regular, implying hydrostatic equilibrium. However Andersson \& Madejski (2004) find an asymmetric temperature distribution as well as a high-redshift substructure within the cluster, which suggests that Abell 1689 is currently undergoing a merger. 

The existence of substructures, contributing $\sim 7\%$ of the total mass within $250h^{-1}$ kpc ($\sim 1.34^\prime$), has been confirmed using gravitational lensing (e.g. Broadhurst et al., 2005b; Diego et al., 2005; Halkola, Seitz, \& Pannella, 2006; Zekser et al., 2006; Limousin et al., 2007; Saha et al., 2006), though these substructures may be extended structures projected along the line of sight but external to the cluster, rather than merging clumps and groups within the cluster system. Indeed, {\L}okas et al. (2006) have studied the redshift distribution of galaxies in the cluster and its surroundings and concluded that there are structures along the line of sight that will affect lensing, but not dynamical, mass estimates. Such structures could help to explain the discrepancy in mass estimates between those derived through lensing measurements and those derived from X-ray studies, the latter of which tend to be systematically lower. 

Even amongst gravitational lensing studies themselves, however, discrepancies are seen. Strong lensing mass estimates appear to be higher than those obtained using weak lensing alone, with values ranging from $M_{200}=0.85 \times 10^{15}\,M_\odot$ (King, Clowe, \& Schneider, 2002) using weak lensing to $M_{200}=3.05 \times 10^{15}\,M_\odot$ (Halkola et al., 2006) using strong lensing.  Furthermore, gravitational lensing studies of Abell 1689 and other clusters often show higher concentration parameters than predicted by the mass-concentration relation seen in N-body simulations (e.g. Shaw et al., 2006), and show a wide spread in values across the lensing measurements. 

It has been argued that using a prolate halo model aligned along the line of sight, rather than a spherically-symmetric model, can help bring the measured concentration parameter more in line with that predicted by simulations (Oguri et al., 2005; Corless \& King, 2008; Corless, King, \& Clowe, 2009). Indeed, Shaw et al. (2006) find that higher mass haloes tend to be prolate, rather than spherical. Moreover, Morandi, Pedersen \& Limousin (2010) have found that taking proper account of the triaxiality of DM haloes can help resolve the discrepancy between lensing and X-ray mass estimates of clusters. However, such models have thus far not reduced the measured concentration parameter appreciably in studies of Abell 1689. 

In addition, there is little agreement regarding the importance of substructure in the mass models used to fit the lensing data. For example, Limousin et al. (2007) found it necessary to include in their strong lensing model a secondary mass peak, offset from the central core, in order to best fit the strong lensing features. On the other hand, others (e.g. Broadhurst et al. 2005a) have found a good fit with a single NFW halo. 

Most recently, Coe et al. (2010) have presented a combined strong and weak lensing analysis of the cluster, providing the highest resolution mass reconstruction of this cluster to date, and using the same ACS images considered here, supplemented with ground-based imaging. Their reconstruction shows complex substructure, and they find a best fit NFW profile with $M_{200}=1.8\times 10^{15}h^{-1}M_\odot$ and $c=9.2$ from combined strong and weak lensing constraints.

It is clear that Abell 1689 is a cluster with a complicated structure, and that there is still much work to be done to understand and accurately map the structure of the cluster. As flexion directly probes the gradient of the convergence, it is sensitive to structure on smaller scales than traditional weak lensing studies and can probe a greater distance from the centre of the cluster than strong lensing studies are able to. For this reason, flexion studies might be of great importance in the study of Abell 1689 and similar clusters. 

\subsection{HST ACS Data and Analysis Pipeline}
The data consist of 20 HST ACS images of Abell 1689, taken using the WFC. Each image has been reduced through the HST ACS pipeline and contamination from cosmic rays removed. These 2300-2400 second exposures cover a square field of view of $3.4'$ on a side, have an angular resolution of 0.05$''$\,pixel$^{-1}$, and are described in full detail in Broadhurst et al. (2005a). Of these images, 4 were taken using the F475W
filter (G-band), 4 using the F625W filter (R-band), 5 using the F775W
filter (I-band) and 7 using the F850LP filter (Z-band). The data analysis pipeline used in this paper is described in full in Leonard et al. (2007), and the key points are summarised below.

\subsection{Catalogue Generation}
For the purposes of source detection, stacking the images within each colour band using the SWarp software package (Bertin et al., 2002) was found to be beneficial. The software first estimates and subtracts off the background in each individual frame. Then the images are resampled and projected on to the output frame using a tangential (gnomonic) projection. We opted to use the median stacked image, rather than the mean, in order to avoid any biasing of the output image by spurious hot spots and bad pixels. A single co-added image was created for each of the G-, R- and I-bands. In the Z-band (as discussed in Leonard et al., 2007) it was found that very little benefit was gained in source detection from stacking 7 images as opposed to 3 or 4; therefore, two co-added images were created. 

Source extraction was carried out in two stages using SExtractor (Bertin \& Arnouts, 1996), using a modification of the ``hot" and ``cold" strategy of Rix et al. (2004). During the first pass, a catalogue of known foreground objects including stars and cluster members either listed in Duc et al. (2002) or listed as such in the NASA/IPAC Extragalactic Database (NED) was provided, and only those objects detected. These sources were then masked out by setting their pixel values equal to random noise, making use of the variance maps produced by SExtractor, and thus simulating an emptier field. SExtractor was then run on the masked images using a lower detection threshold, in order to detect the smaller, fainter background galaxies of interest. A final catalogue of objects was generated including all those objects that were detected in at least two co-added images, including either the I- or R-band image. 

The stacking process described above clearly will affect the PSF of the resulting median-stacked image in a complex way; each frame is offset and rotated slightly from the others within the stack, therefore the resulting PSF will be impossible to model simply. However, Leonard et al. found the effect of the telescope point spread function to be significant only for shear measurements, but not for flexion measurements, provided the source is well-resolved. This is because the induced flexion resulting from the PSF is related to the image size by (see Leonard et al 2007 for full details of this calculation):
\begin{equation}
\f_{\rm induced}\sim \f_{\rm PSF}\frac{a^4_{\rm PSF}}{a^4_{\rm source}+a^4_{\rm PSF}}\ ,
\end{equation}
where $\f_{\rm PSF}$ is the flexion measured in the PSF (typically of order $10^{-3}-10^{-4}/^{\prime\prime}$), $a_{\rm PSF}$ is the semi-major axis of the PSF ($\sim 0.1^{\prime\prime}$) and $a_{\rm source}$ is the semi-major axis of the source. Therefore flexion measurements were carried out on the stacked images, which offer improved signal to noise, and all sources with $a_{\rm source} < a_{\rm PSF}$ were rejected from the analysis. Flexion measurements were carried out using the HOLICs technique (Okura et al., 2007; Goldberg \& Leonard, 2007). 

In addition to PSF effects, images from the ACS WFC suffer from a geometric image distortion resulting from the off-axis location of the camera on the Hubble Space Telescope. This distortion has been modelled by Meurer et al. (2003), and is well fit by a $4^{th}$ order polynomial. This fit was used to evaluate the shear and flexion induced by the geometric distortion as a function of location on the chip. Leonard et al. (2007) found that here, again, the effect on the flexion was negligible ($\sim 10^{-4}/^{\prime\prime}$), and note also that the geometric distortion is greatly reduced in the stacked frames as a result of the reprojection of the images during the stacking process. 

Sources were rejected at various points in the analysis pipeline, in order to reduce contamination from erroneous noisy measurements and to ensure that only well-resolved sources were used. Full details of the selection criteria can be found in Leonard et al. (2007). The resulting background source count is $\sim 75\,$arcmin$^{-2}$. As in Leonard et al. (2007), we assume a median source redshift of $z_{\rm s}=0.9$, in accord with the measurements of Limousin et al. (2007)

\subsection{Aperture Mass Method}

The aperture mass statistic was computed using the polynomial filter described above, and using $1000\times1000$ evenly spaced apertures across the field. Due to the high resolution of the images and the resulting high background source count, including a large number of randomisations in the noise calculation proved rather time-consuming. Therefore, only 100 randomisations were carried out for each aperture mass reconstruction. This was found to be sufficient, however; increasing the number of randomisations for a given reconstruction was not found to significantly alter the error estimates. 

$\fmap$ reconstructions were carried out for aperture radii $R=[45^{\prime\prime},60^{\prime\prime},75^{\prime\prime},90^{\prime\prime}]$ and filter functions with polynomial order $\ell=[3,5,7,10]$. This set of aperture parameters will allow us to map the structure in the cluster on a range of scales, and provide a consistency check to help eliminate peaks arising purely out of noise. Moreover, having 16 different reconstructions should allow to discriminate between mass models for any structures detected; Leonard \& King (2010) demonstrated that, particularly when probing low-mass structures, a larger number of reconstructions is helpful in discriminating between competing mass profiles. For each reconstruction, a B-mode map was also generated in order to assess the level of systematic errors within the maps. 

In order to more quantitatively characterise the structures detected, an aggregate signal to noise map for $\fmap$, consisting of the average signal to noise of all 16 reconstructions in each case, is generated. Such a map is useful for identifying persistent structures across the 16 reconstructions, and is used solely for this purpose. As discussed in detail in Schirmer (2004) in the context of shear, significant $\map$ peaks that are in fact due to noise disappear at filter scales only slightly larger or smaller than the scale where they are detected. In contrast, real peaks are detected as significant over a much broader range of filter scales. A similar argument will apply for the $\fmap$ statistic.

Using the \textit{CLFIND} algorithm developed by Williams, de Geus \& Blitz (1994), we identify all significant peaks within the map down to a limiting signal to noise ratio. This software allows us to identify peaks even where they occur as a sub-peak in a larger structure. This is useful where structures have been broken up into smaller, blended substructures in the reconstructions. A location for each peak is then identified by computing the signal-to-noise weighted centroid. 

Similarly, for each individual reconstruction, peaks are identified and cross-matched with the aggregate map. A radial profile for each is computed by averaging the signal in radial annuli about the signal-to-noise weighted centroid. Note that the centroid used for a given peak is not fixed; i.e. the centroids for each peak are independently calculated in each $\fmap$ reconstruction, and the radial average is computed with respect to that centroid, rather than the centroid computed in the aggregate map. In addition, a B-mode radial profile is computed for each peak. Radial profiles exhibiting a B-mode signal not consistent with zero, within the error bars, are rejected from further analysis. 

\subsection{Model-selection and mass determination}

As described in Leonard \& King (2010), the radial profiles of the $\fmap$ signal can be used to discriminate between competing mass profiles. For each radial profile computed, a polynomial function is used to fit the signal, and the first real, positive root taken to be the zero-signal radius $R_0$. The radial profile of the associated noise map is also computed, and this yields error bars on each point in the radial profile for a given peak. 

\begin{figure}
\includegraphics[width=0.25\textwidth]{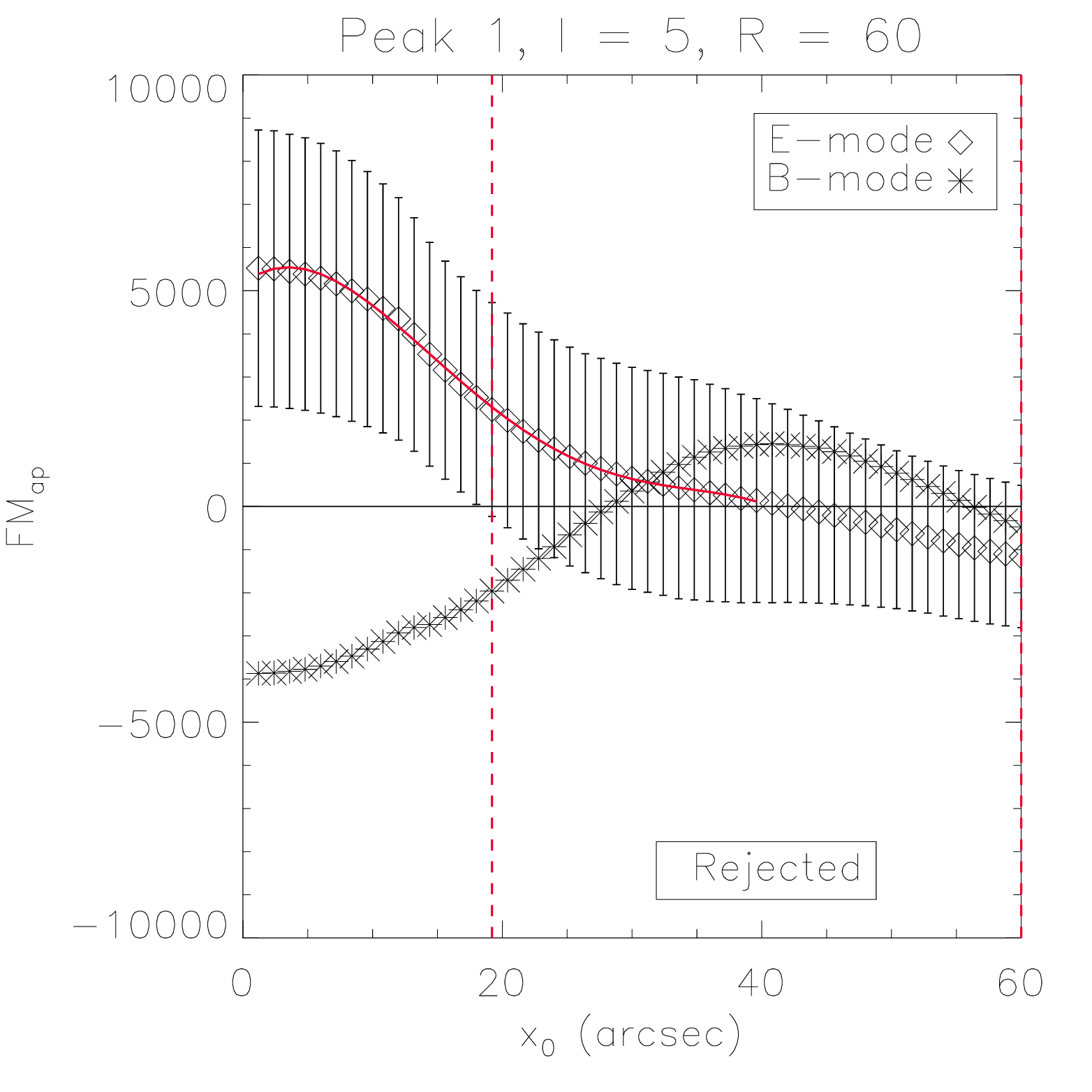}\includegraphics[width=0.25\textwidth]{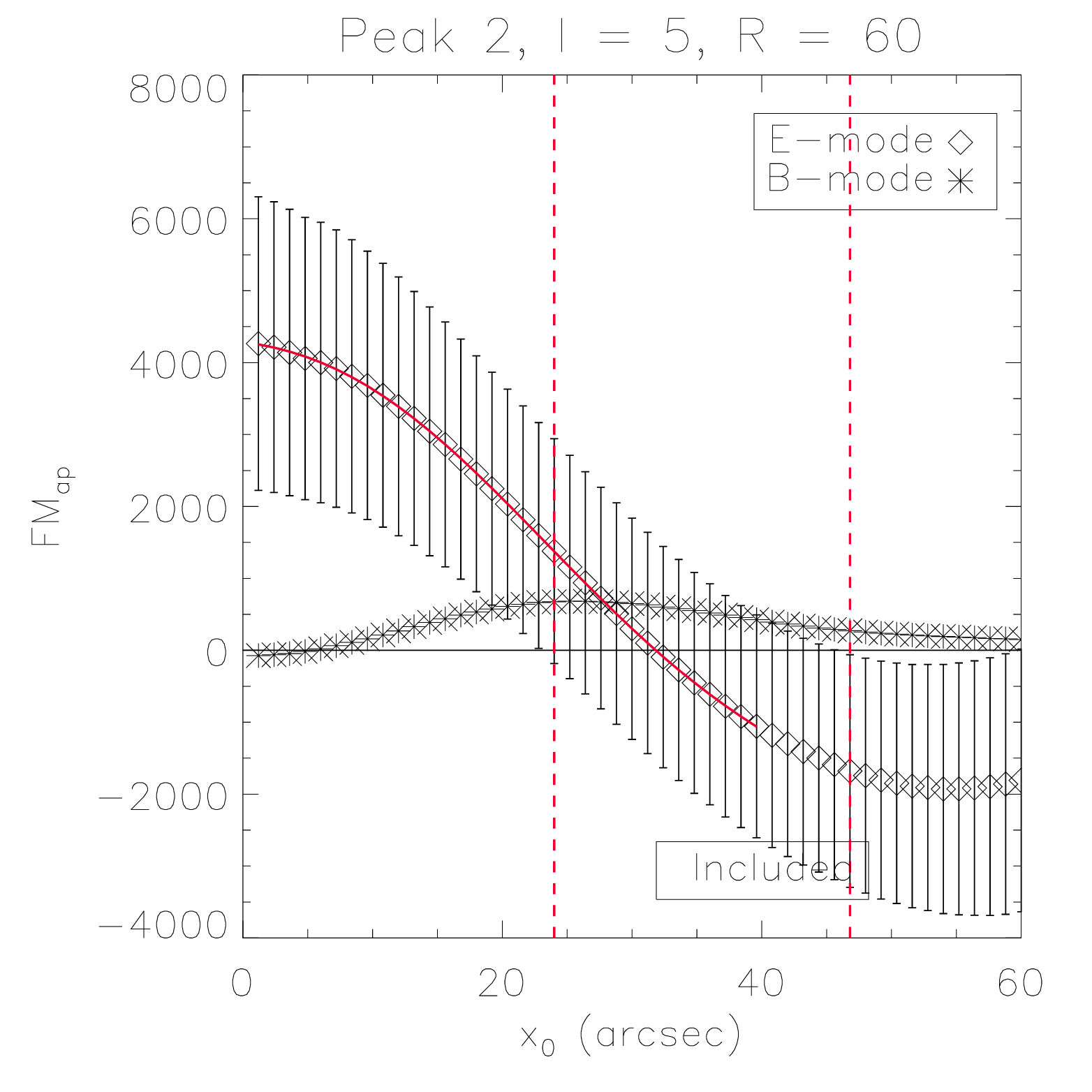}
\includegraphics[width=0.25\textwidth]{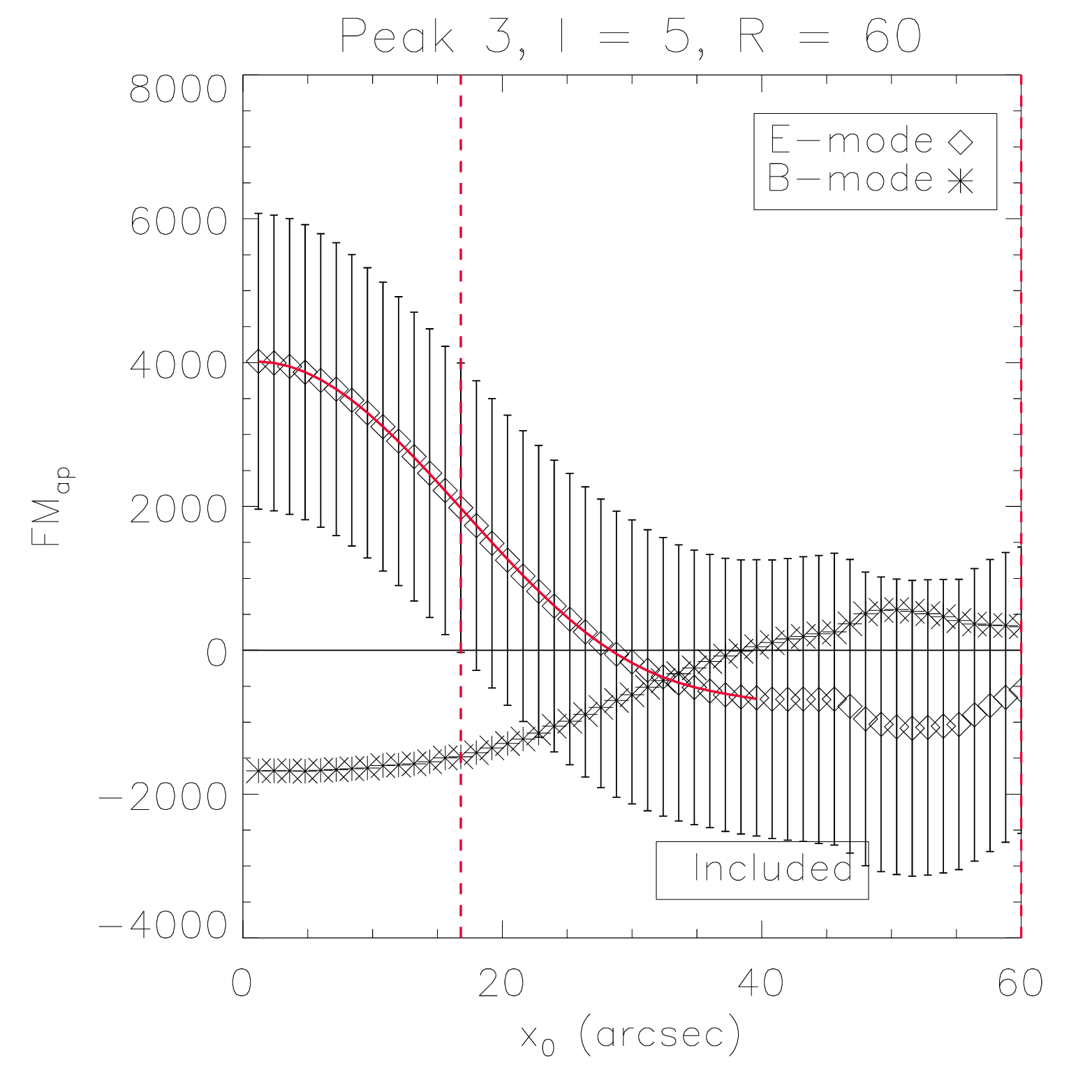}\includegraphics[width=0.25\textwidth]{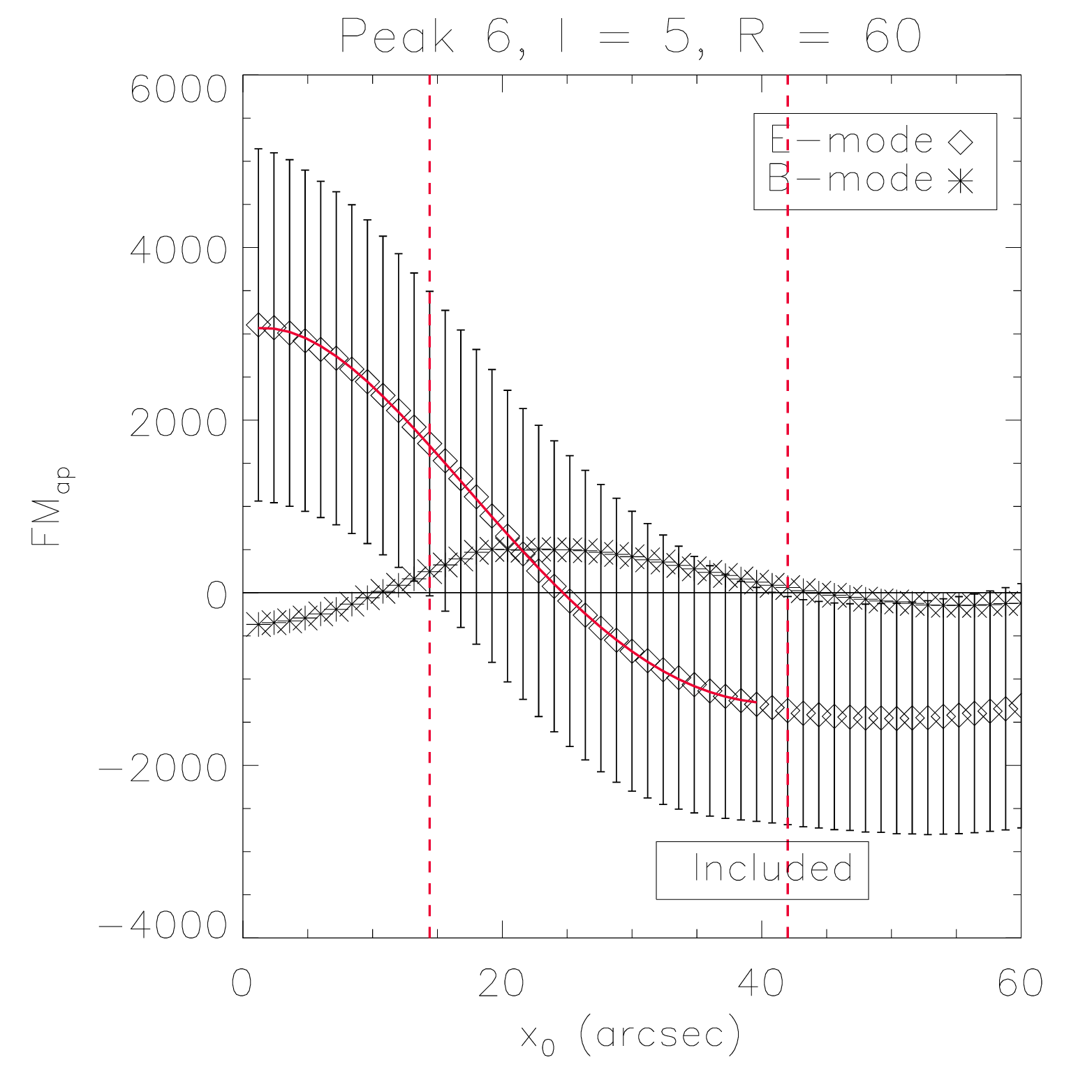}
\caption{Sample radial profiles for each of four peaks detected in the $\fmap$ reconstruction with $\ell = 5$ and $R=60^{\prime\prime}$ (see \S~\ref{sec:results} below) showing the $\fmap$ signal and associated error bars, the B-mode signal, the polynomial fit used to determine $R_0$ (solid curve), and the range of $R_0$ considered to be compatible with the data in the subsequent analysis (indicated by vertical dashed lines). \label{fg:profs}}
\end{figure}

The radial profiles can then be compared with the template models, and compatible regions of parameter space isolated for each peak. The catalogue is first searched for models whose peak signal falls within the error bar associated with the measured peak of the radial profile. Models in this subset are then excluded if their zero-signal radius falls outside the range allowed by the measurement given the error bars. The allowed range is taken to be all those values of $x_0$ for which the signal is compatible with zero given the $1-\sigma$ error bars (see Figure\,\ref{fg:profs}). In a number of cases, particularly those in which the peak has substantial associated substructure, this only allows us to place a lower limit on the zero-signal radius. 

This is a very conservative constraint, allowing a large range of $R_0$ values to be considered in each case. However, given the low signal to noise seen in our maps (see \S~\ref{sec:results} below), it seemed prudent to allow a larger range of $R_0$, so as not to present overly optimistic results.

Figure\,\ref{fg:profs} shows four example radial profiles to illustrate the method described above. In this figure, we plot the radially-averaged $\fmap$ signal for each of four peaks identified in the $\fmap$ reconstruction with $\ell = 5$ and $R=60^{\prime\prime}$ and the associated error bars. We overlay the polynomial fit used to determine the value of $R_0$ as a solid line. Dashed vertical lines indicate the range of $R_0$ considered to be compatible with the data; the range of $m_{\rm peak}$ considered is that covered by the $x_0=0$ error bar. We also plot the radial profile of the B-mode, to demonstrate the criteria for inclusion in the subsequent analysis.

The left panels show examples of peaks for which we were only able to place a lower limit on the range of $R_0$ compatible with the data, whilst the right panels highlight examples for which a smaller range of $R_0$ is considered. In addition, the figure indicates which peaks are included or rejected based on their B-modes. The peaks in the right column clearly do not exhibit appreciable B-modes and are included. The profile shown in the top left panel of the figure has B-modes that are clearly not consistent with zero within the error bars, and is therefore rejected. The peak in the bottom left panel again shows an appreciable B-mode signal; however the signal is compatible with zero within the error bars for all values of $x_0$, and therefore this peak is included in the subsequent analysis. Such borderline cases are relatively rare though, occurring for approximately 1 out of every 6-7 profiles considered.

For each model included for a given value of $\ell$ and $R$, we compute a fractional deviation from the measured signal, given by:
\begin{equation}
\Delta_{\ell,R}^2=\frac{\left(m_{\rm peak}-m_{\rm p, model}\right)^2}{m_{\rm peak}^2}+\frac{\left(R_0-R_{0,{\rm model}}\right)^2}{R_0^2}\ .
\end{equation}
After iterating over the available reconstructions, a weight ${\cal W}$ can be assigned to each model
\begin{equation}
{\cal W}=\frac{1}{\left\langle\Delta\right\rangle}\ ,
\end{equation}
where
\begin{equation}
\left\langle\Delta\right\rangle=\frac{1}{N_{\rm det}}\sum_{\rm det} \Delta_{\ell,R}
\end{equation}
is fractional deviation for the model averaged over all realisations for which it was identified. In effect, $\left\langle\Delta\right\rangle$ gives us a sort of mean $\chi^2$ statistic in parameter space for a given model as compared with the data. More accurately, $\left\langle\Delta\right\rangle$ gives the mean distance (in parameter space normalised by the measured $\fmap$ parameters) of a given model from the data. The statistic ensures that greater weight is given to models for which $N_{\rm det}$, the number of realisations for which the model was identified as compatible with the data, is higher and for which the mean deviation between the model and the data is smaller. 

Note that we treat the E-mode data for those profiles included in subsequent analysis entirely uncritically. In other words, no down-weighting is applied to those cases which have marginal B-modes as compared to those with none. This will introduce errors in the subsequent analysis; however we do not expect this to bias our results significantly, as such borderline cases are relatively rare.

This method allows us to identify models that are compatible not only with the peak signal and zero signal radius in a given realisation, but also with the evolution of these parameters as the aperture parameters are varied. So, while a simple $\chi^2$-fitting of a class of model (such as the NFW model) to the data will tell us how well a given model fits a given radial profile at fixed $\ell$ and $R$, the ${\cal W}$ statistic tells us how well a given model fits the \textit{ensemble} of $\fmap$ radial profiles. In this way, ${\cal W}$ acts as an unnormalised pseudo-likelihood distribution, yielding higher values for models which better fit the data, and are therefore more likely representative of the true matter density profile. 

Whilst this is a non-traditional statistic, and -- being unnormalised -- may be seen as somewhat difficult to interpret, there are several advantages to this approach. Firstly, a full Bayesian treatment of the flexion aperture mass statistic requires both a knowledge of the intrinsic distribution of flexion in unlensed galaxies, and the response of this distribution to aperture mass filtering on different scales. To date, the intrinsic flexion distribution in field galaxies not been measured or characterised, but a simple gaussian approximation is unlikely to suffice (Rowe et al. 2007), as is usually taken to be the case for shear.

In addition, while Bayesian methods yield straightforward error estimates on the best-fit parameters (which are not obtained using our method), they do not allow to compare between different \textit{families} of models. In other words, a maximum-likelihood estimator can compare NFW models with different masses and concentration parameters, but cannot directly compare that class of models to the class of SIS models simultaneously. Our method allows for such a comparison -- by defining the ${\cal W}$ statistic in exactly the same way for the SIS and NFW models, we are able to quantitatively compare the goodness-of-fit of a given SIS model with an NFW.

This is not to imply that standard statistical estimators are not important. In fact, in order to fully quantify the goodness-of-fit of a given model, parametric modelling will be essential. What we present here is simply a fast and effective method to identify substructure, to discriminate between competing families of models, and to eliminate areas of parameter space incompatible with the data efficiently.

\section{Results}
\label{sec:results}
\subsection{$\fmap$ reconstructions and peak identification}

\begin{figure*}
\center
\includegraphics[width=0.25\textwidth]{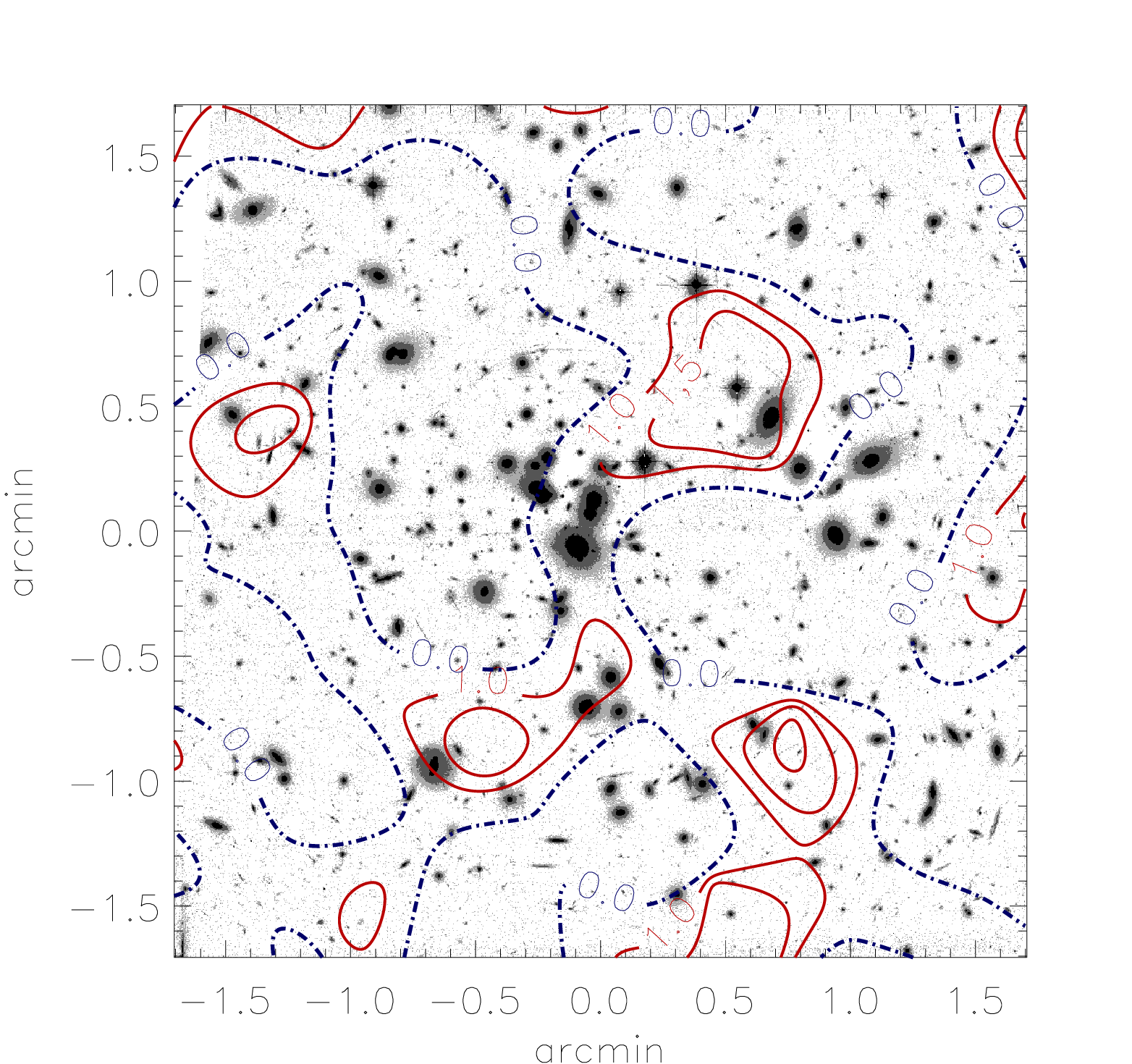}\includegraphics[width=0.25\textwidth]{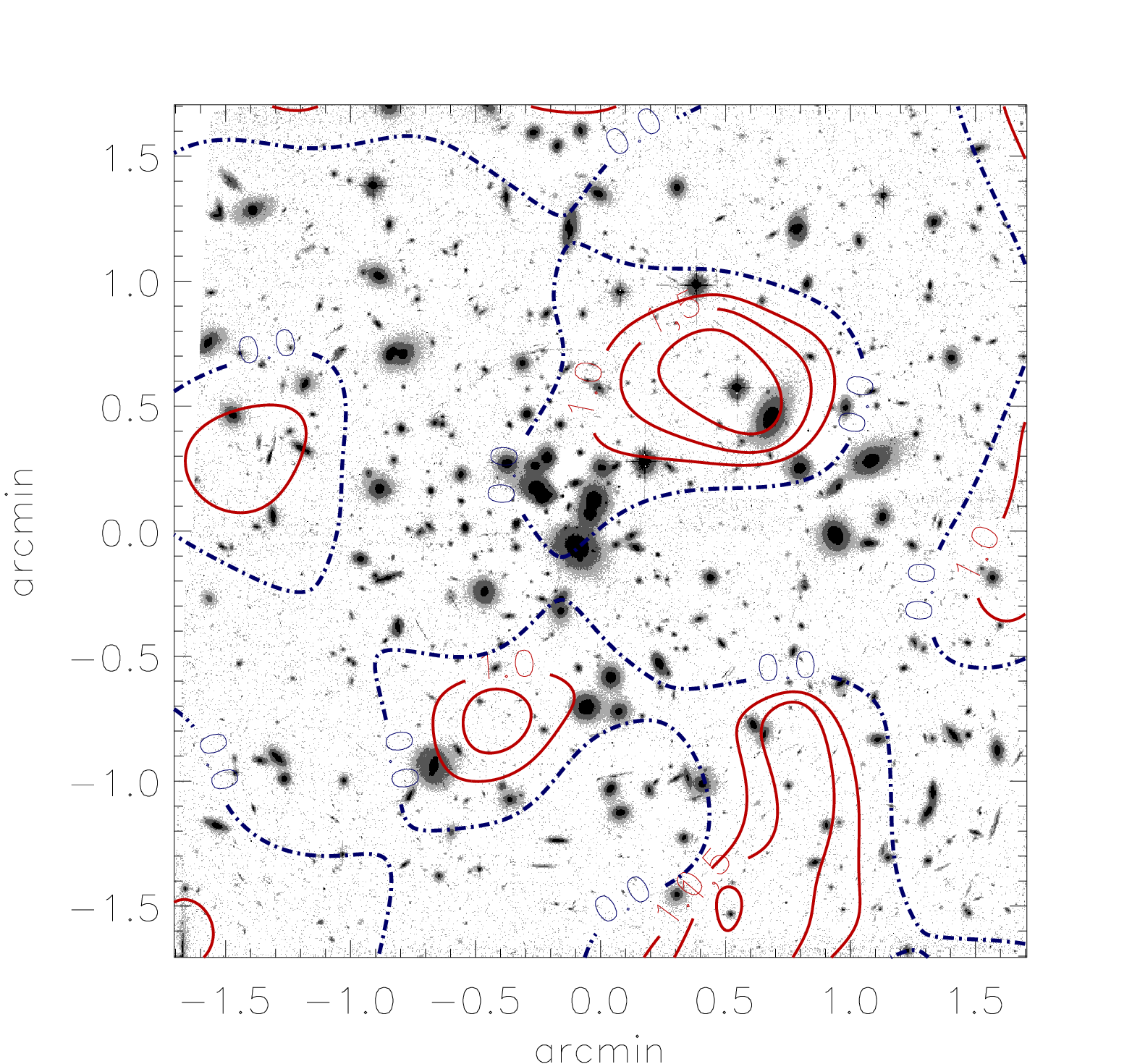}\includegraphics[width=0.25\textwidth]{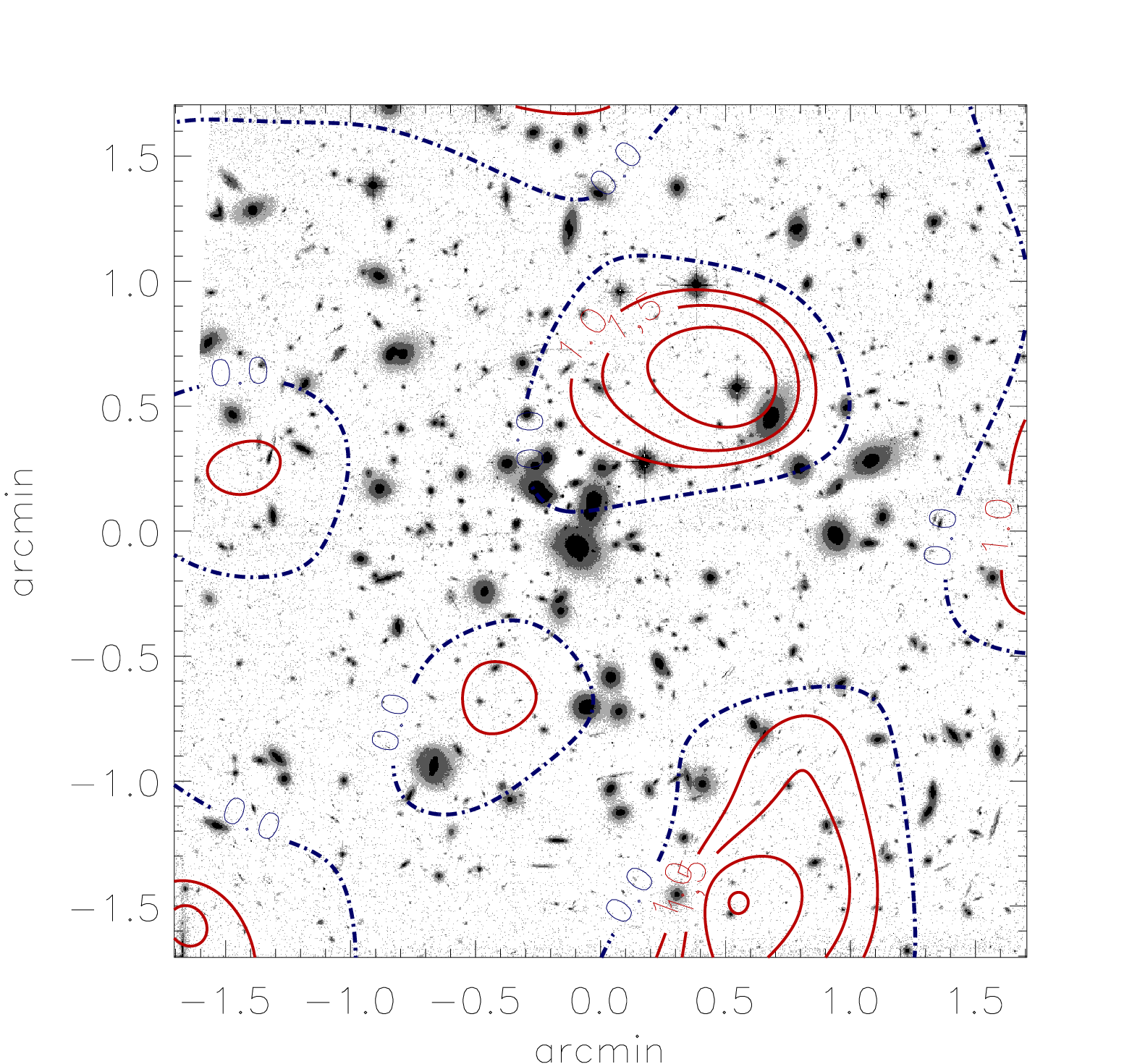}\includegraphics[width=0.25\textwidth]{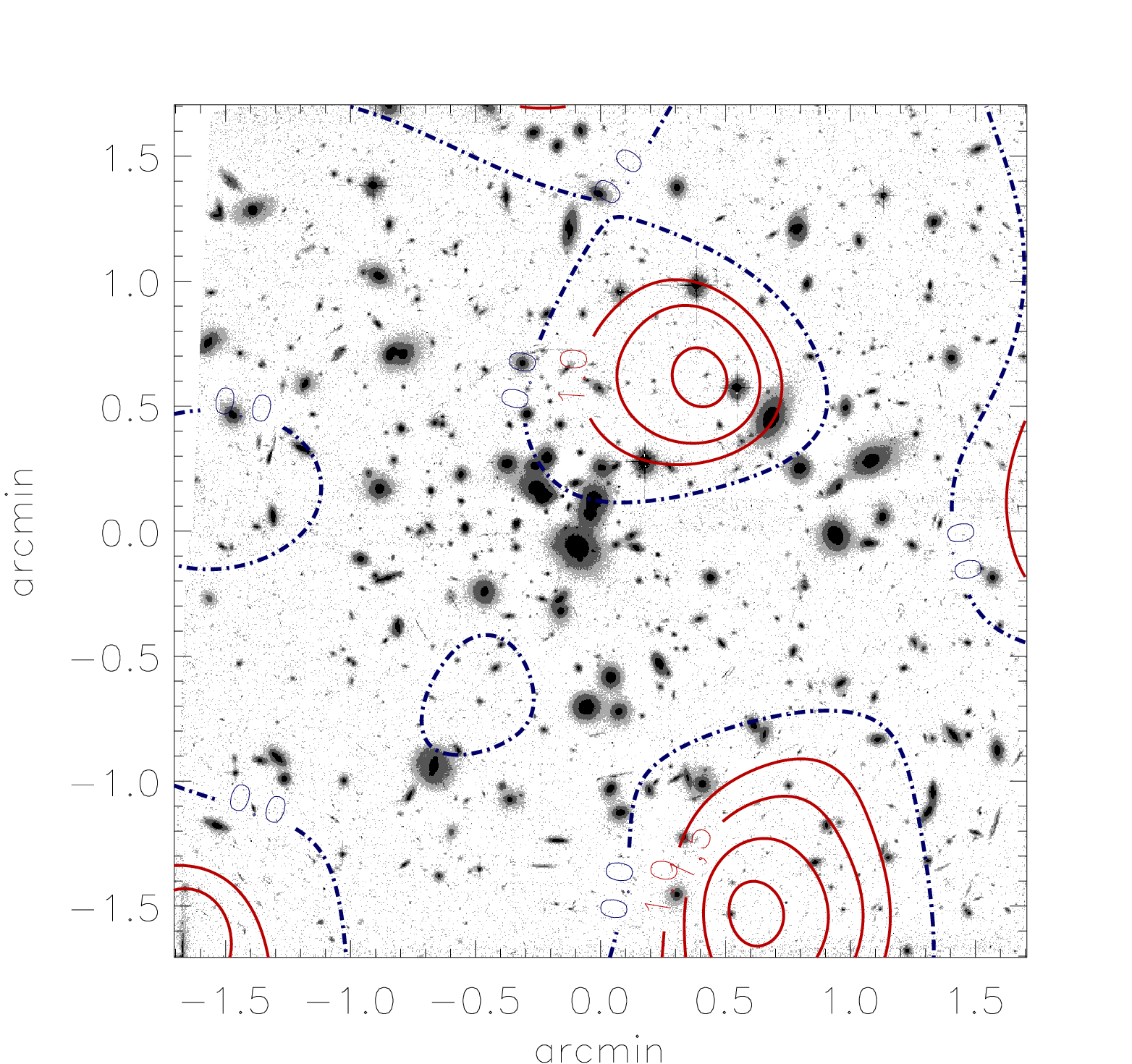}\vspace{15pt}
\includegraphics[width=0.25\textwidth]{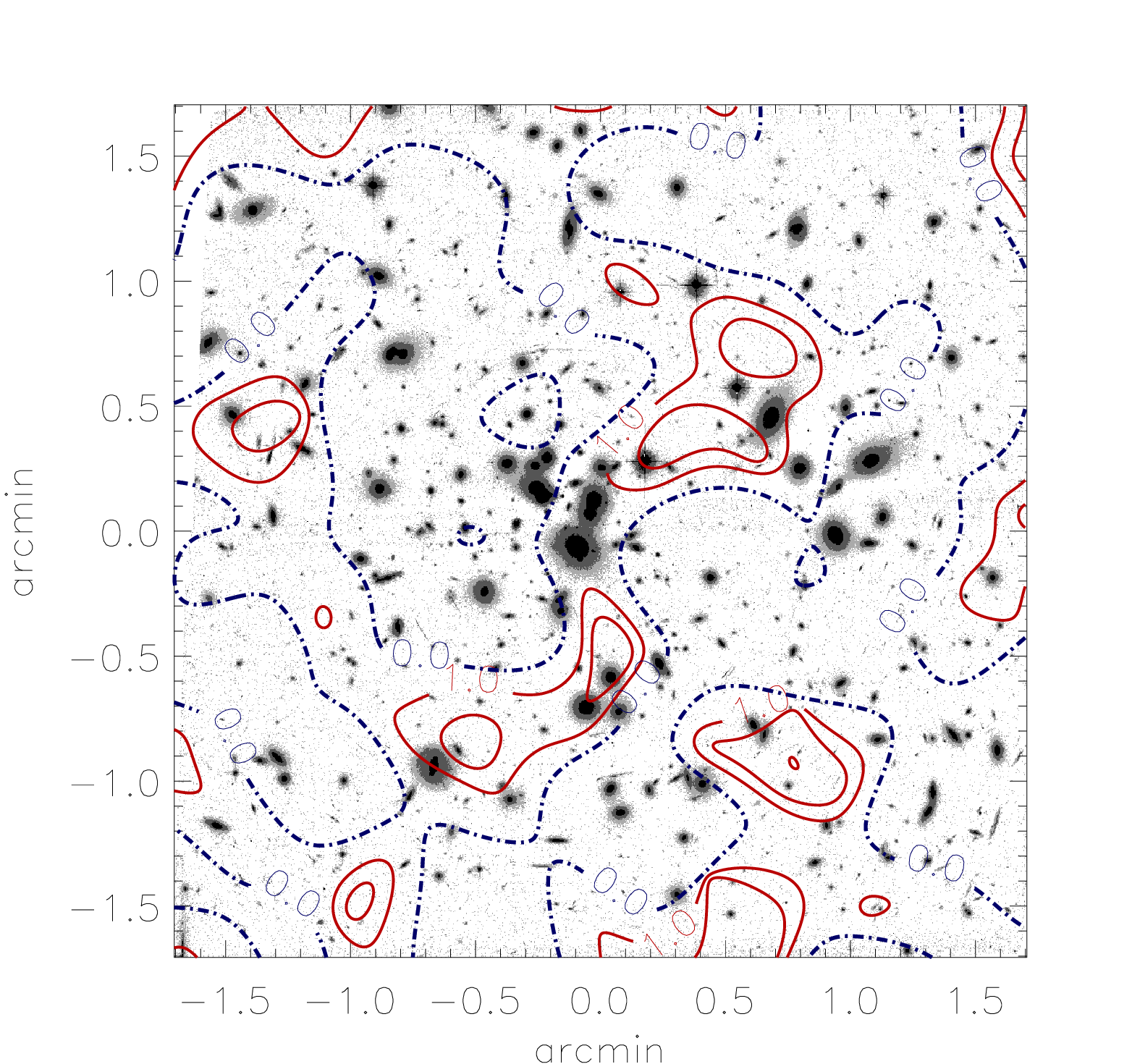}\includegraphics[width=0.25\textwidth]{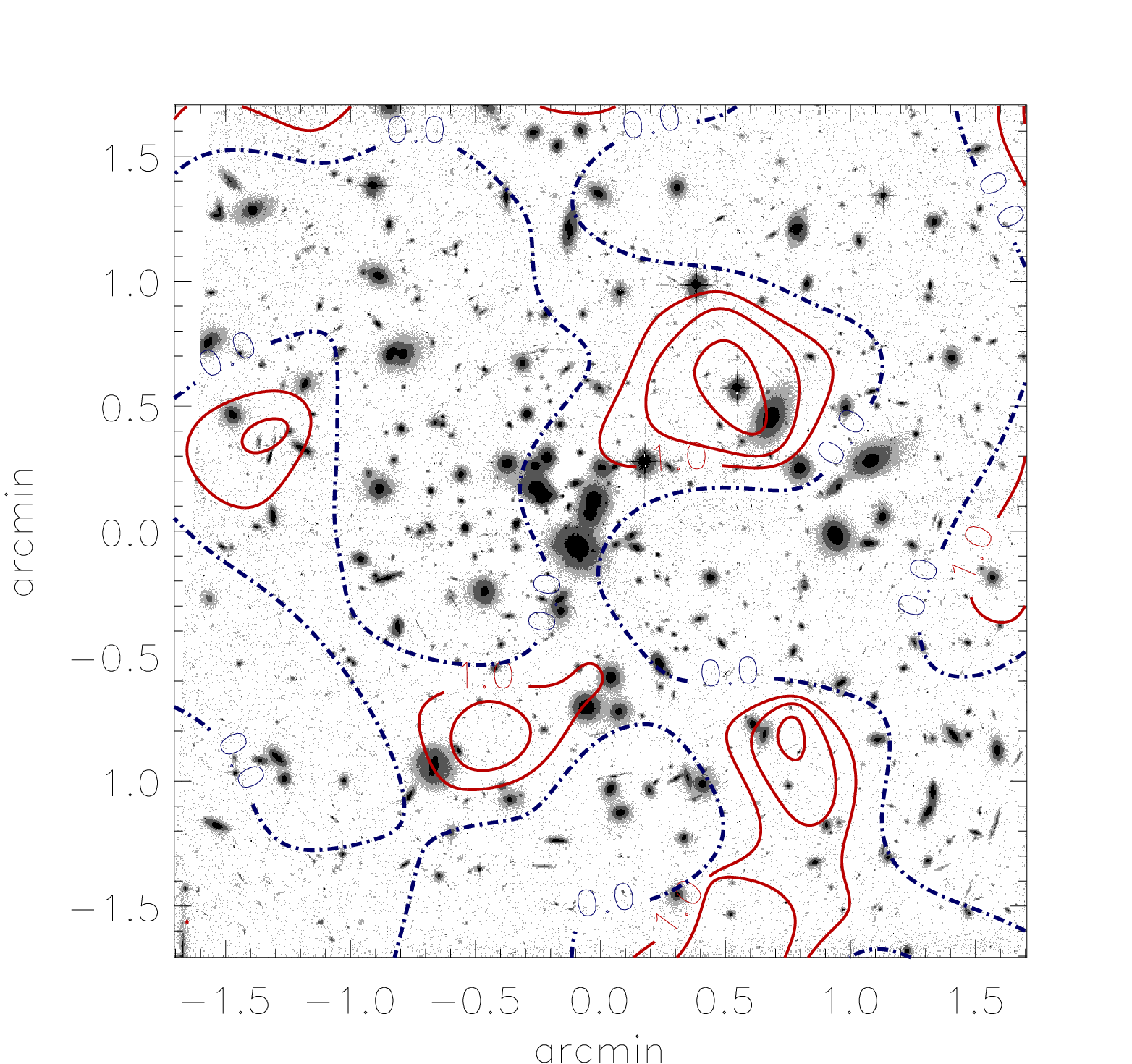}\includegraphics[width=0.25\textwidth]{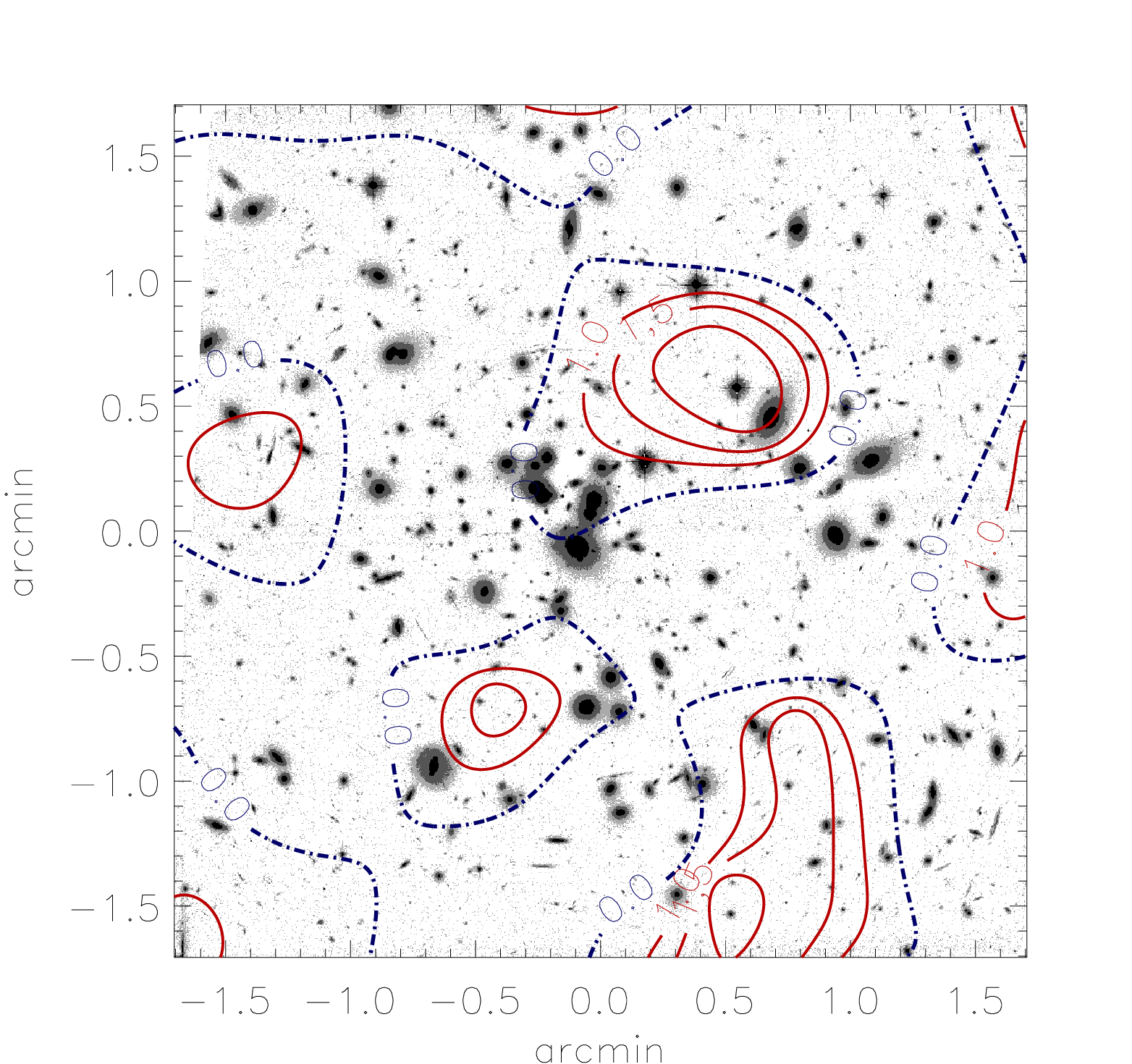}\includegraphics[width=0.25\textwidth]{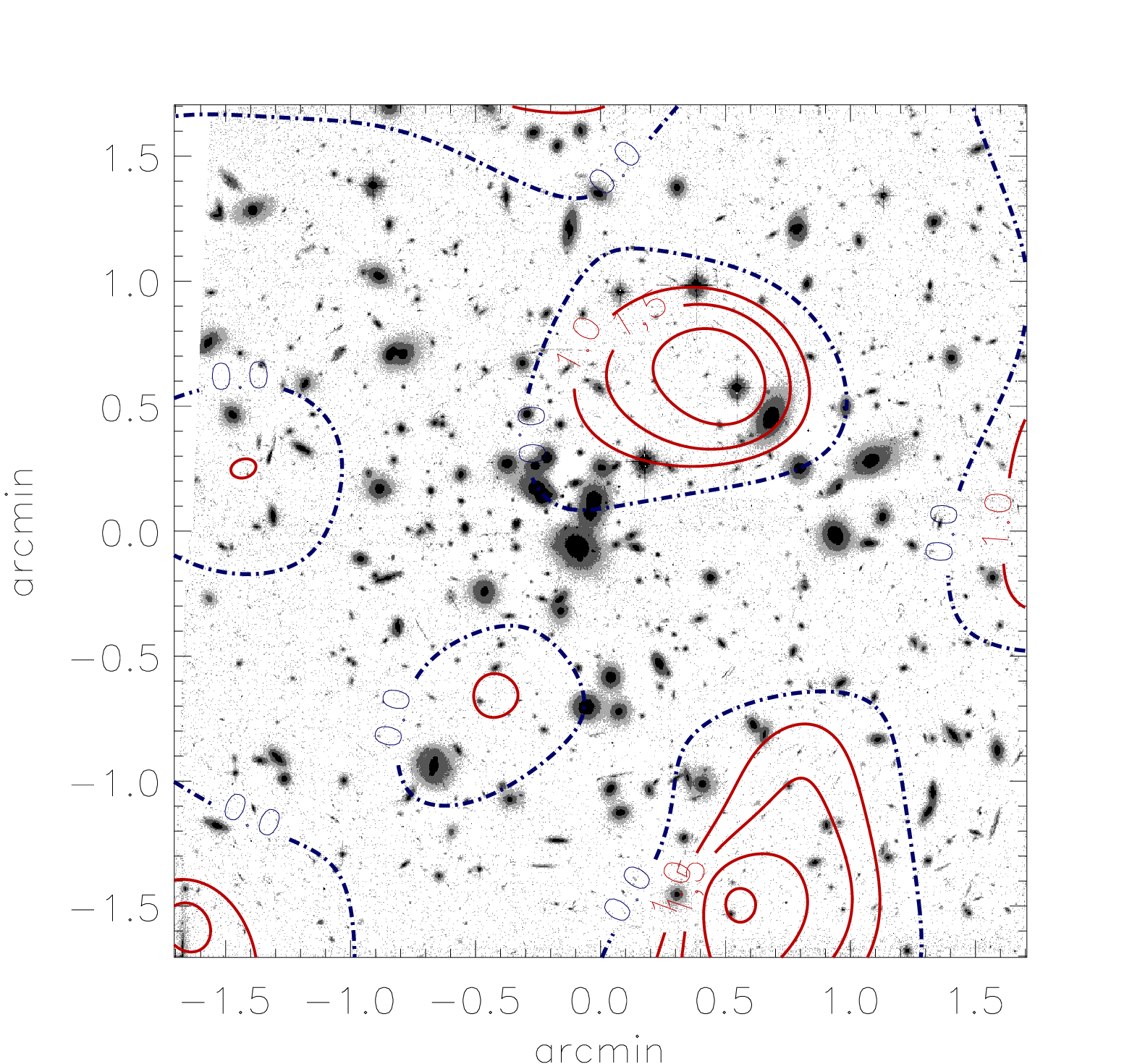}\vspace{15pt}
\includegraphics[width=0.25\textwidth]{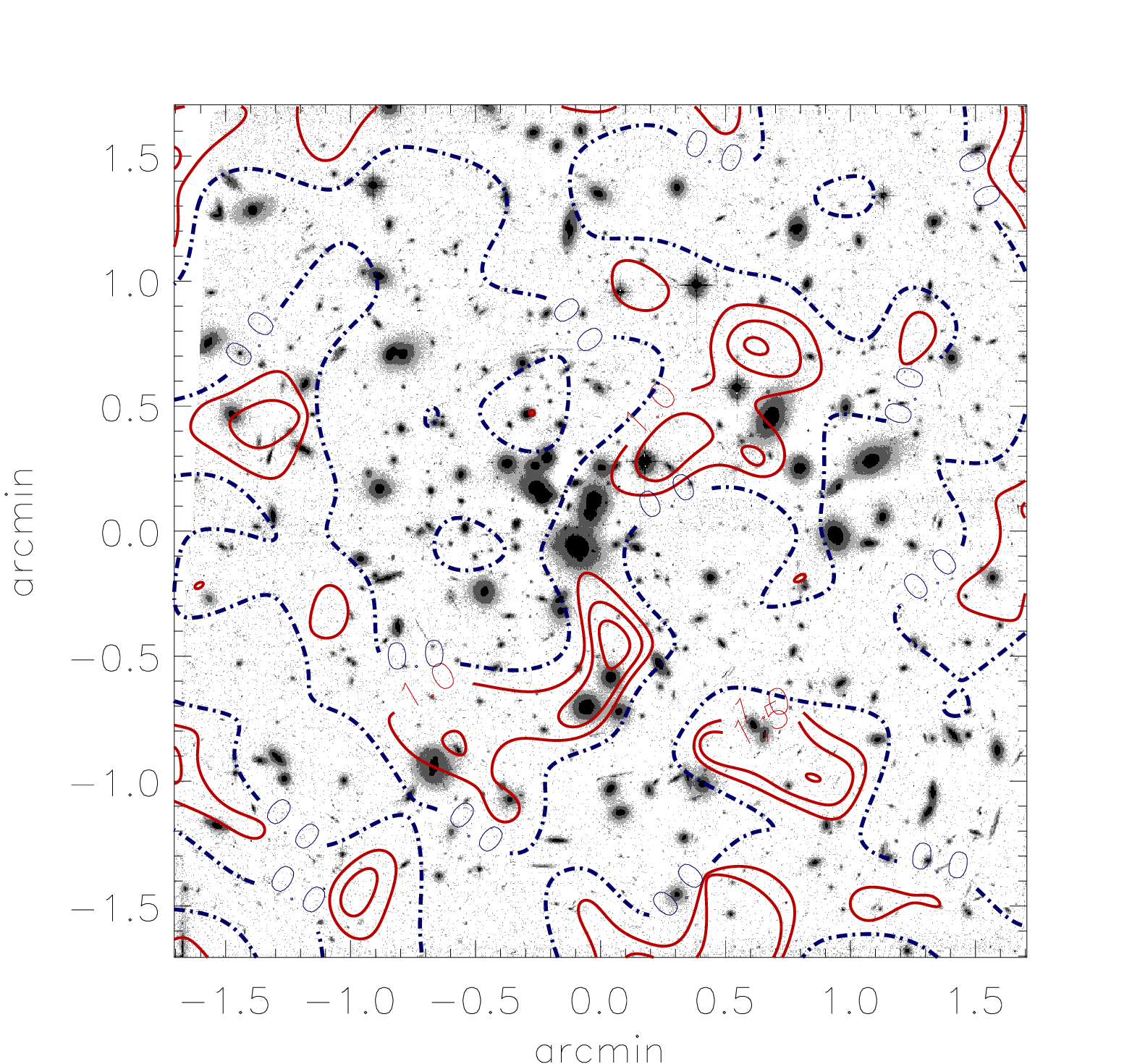}\includegraphics[width=0.25\textwidth]{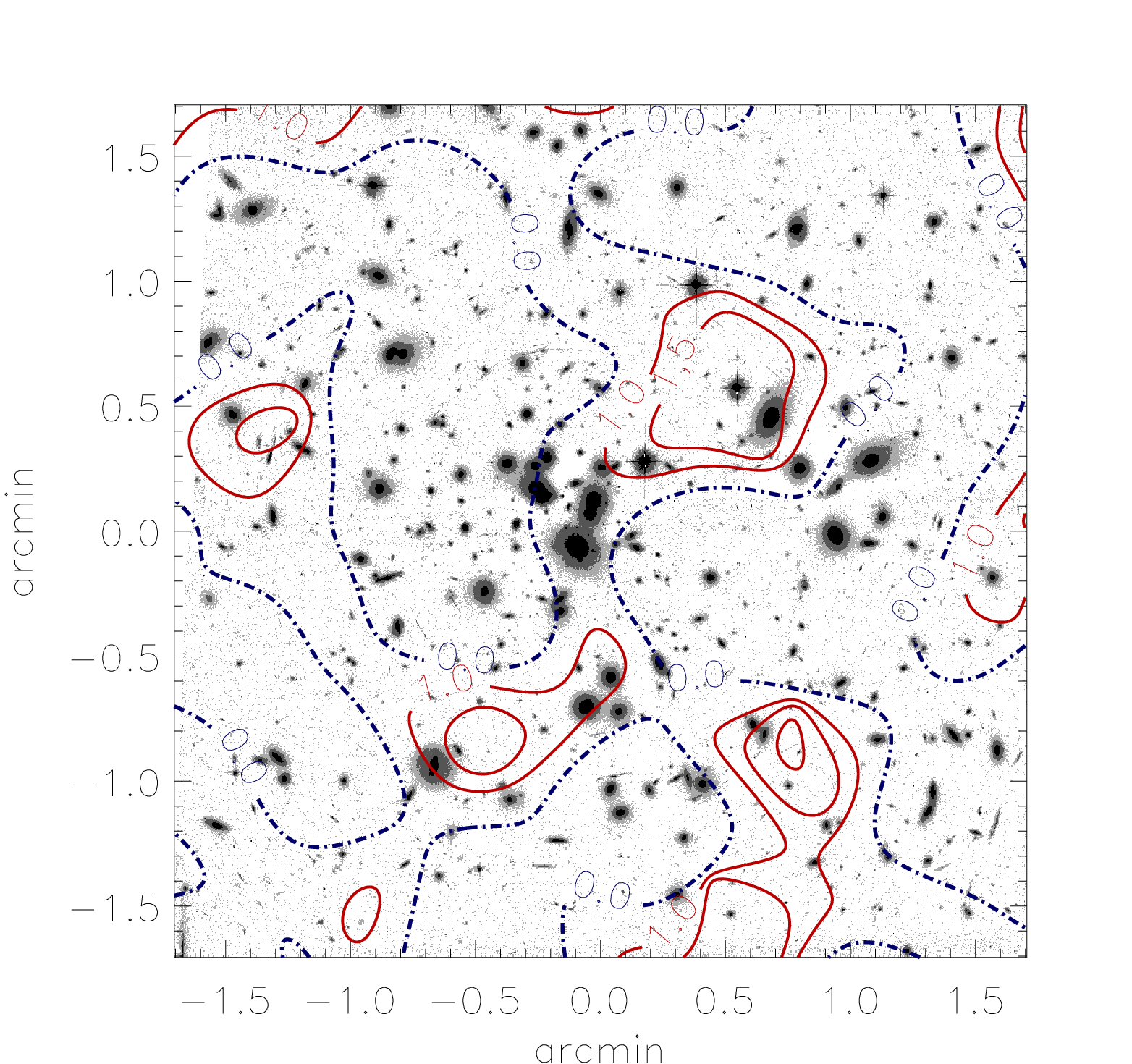}\includegraphics[width=0.25\textwidth]{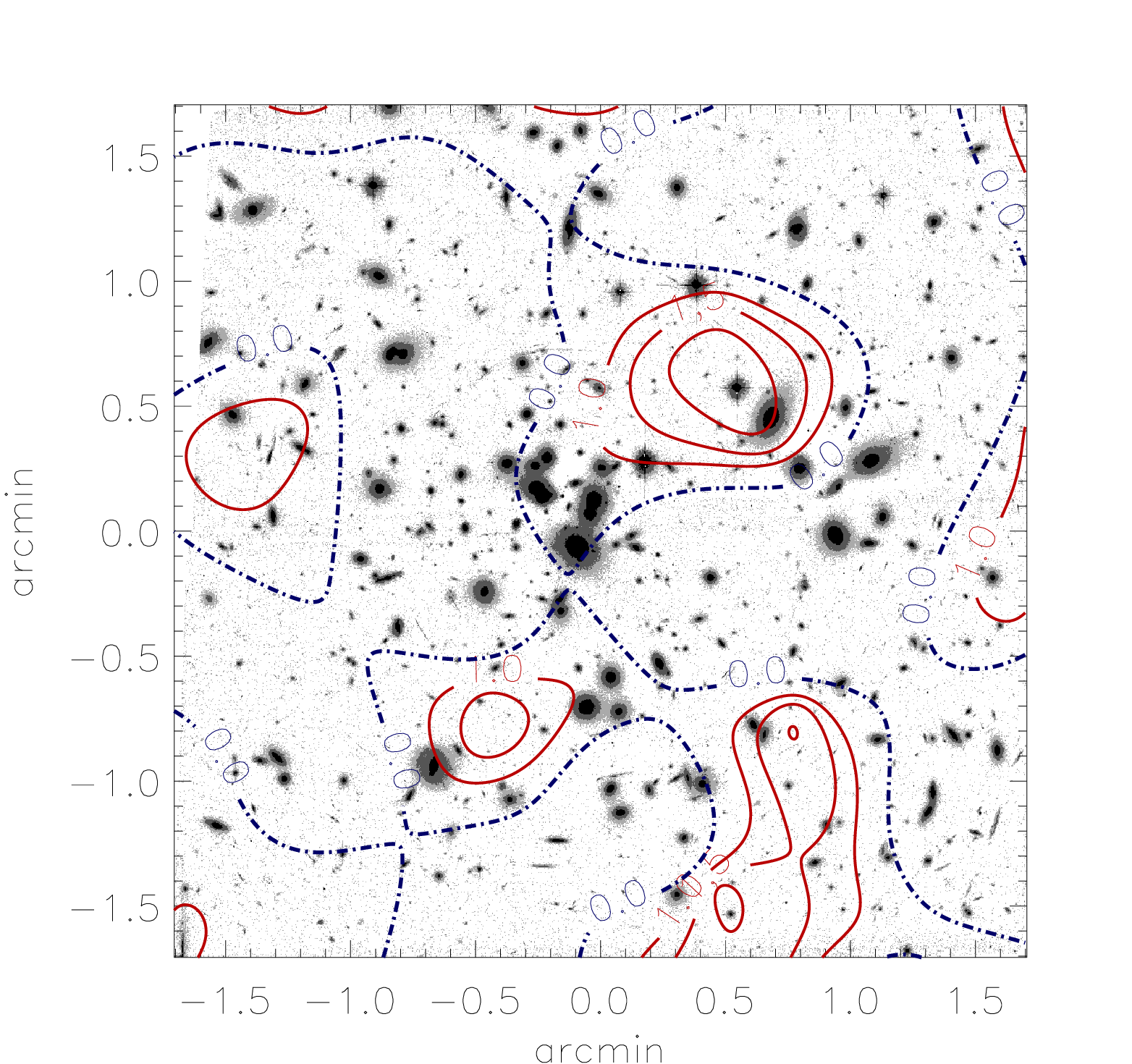}\includegraphics[width=0.25\textwidth]{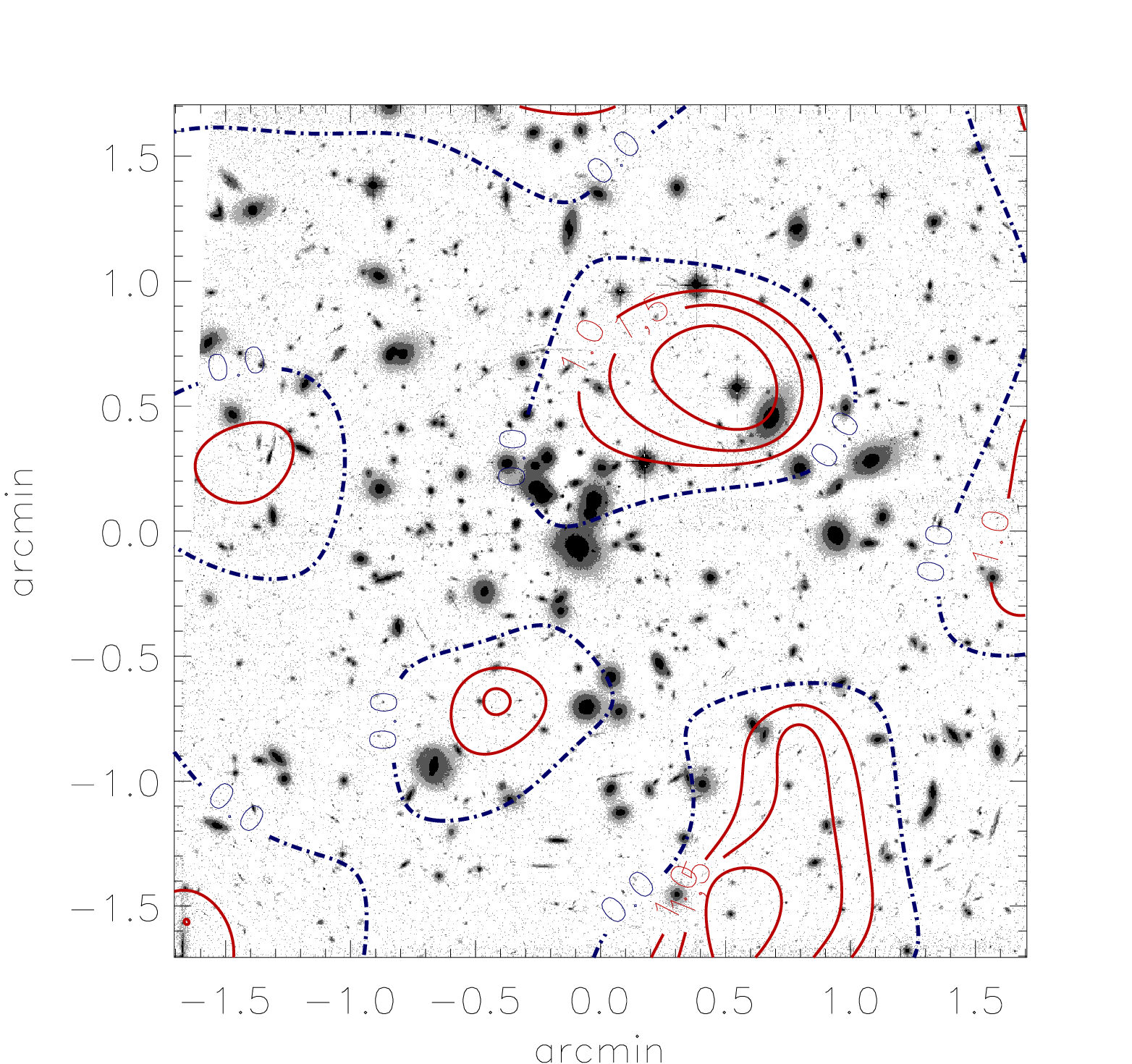}\vspace{15pt}
\includegraphics[width=0.25\textwidth]{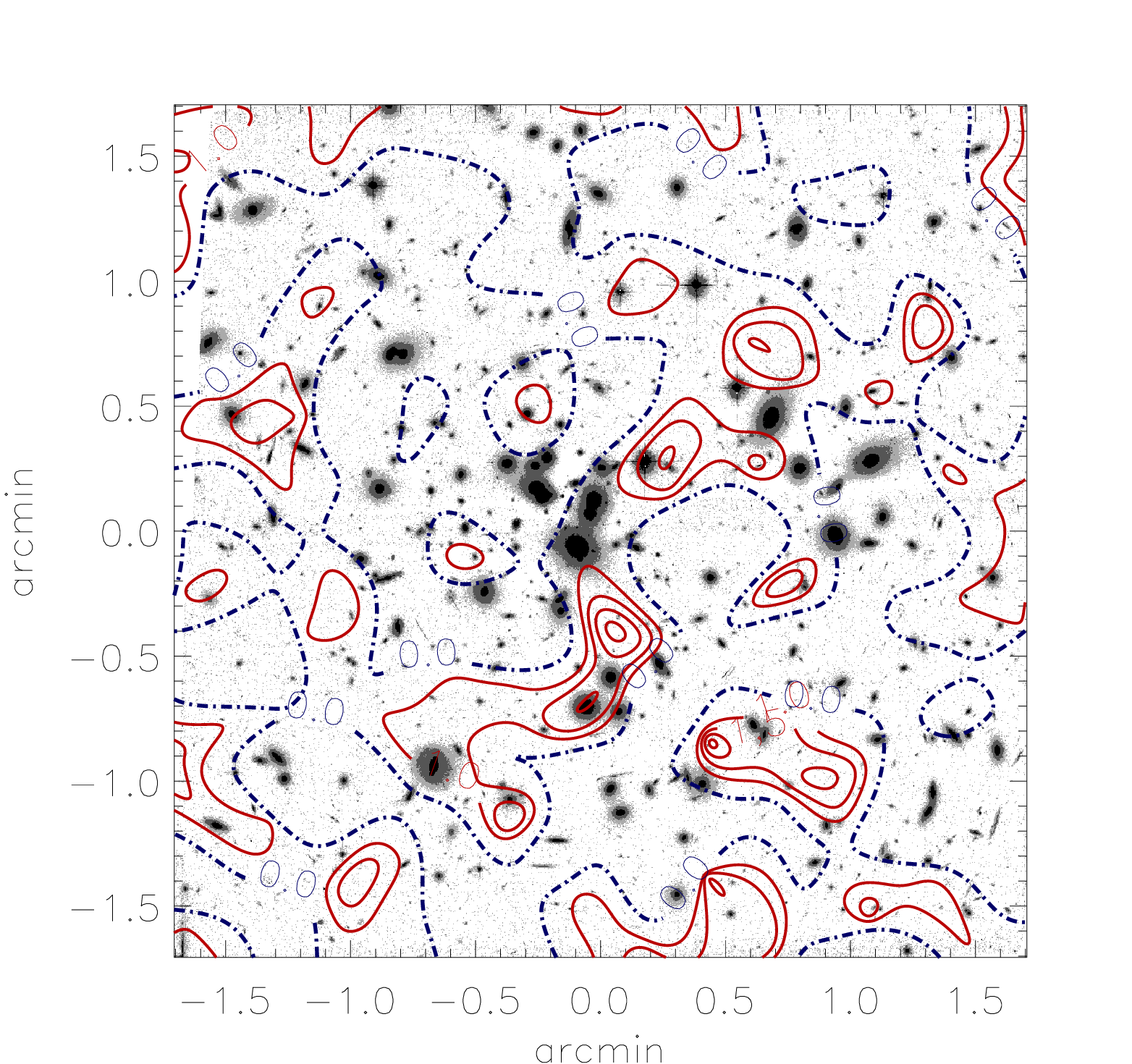}\includegraphics[width=0.25\textwidth]{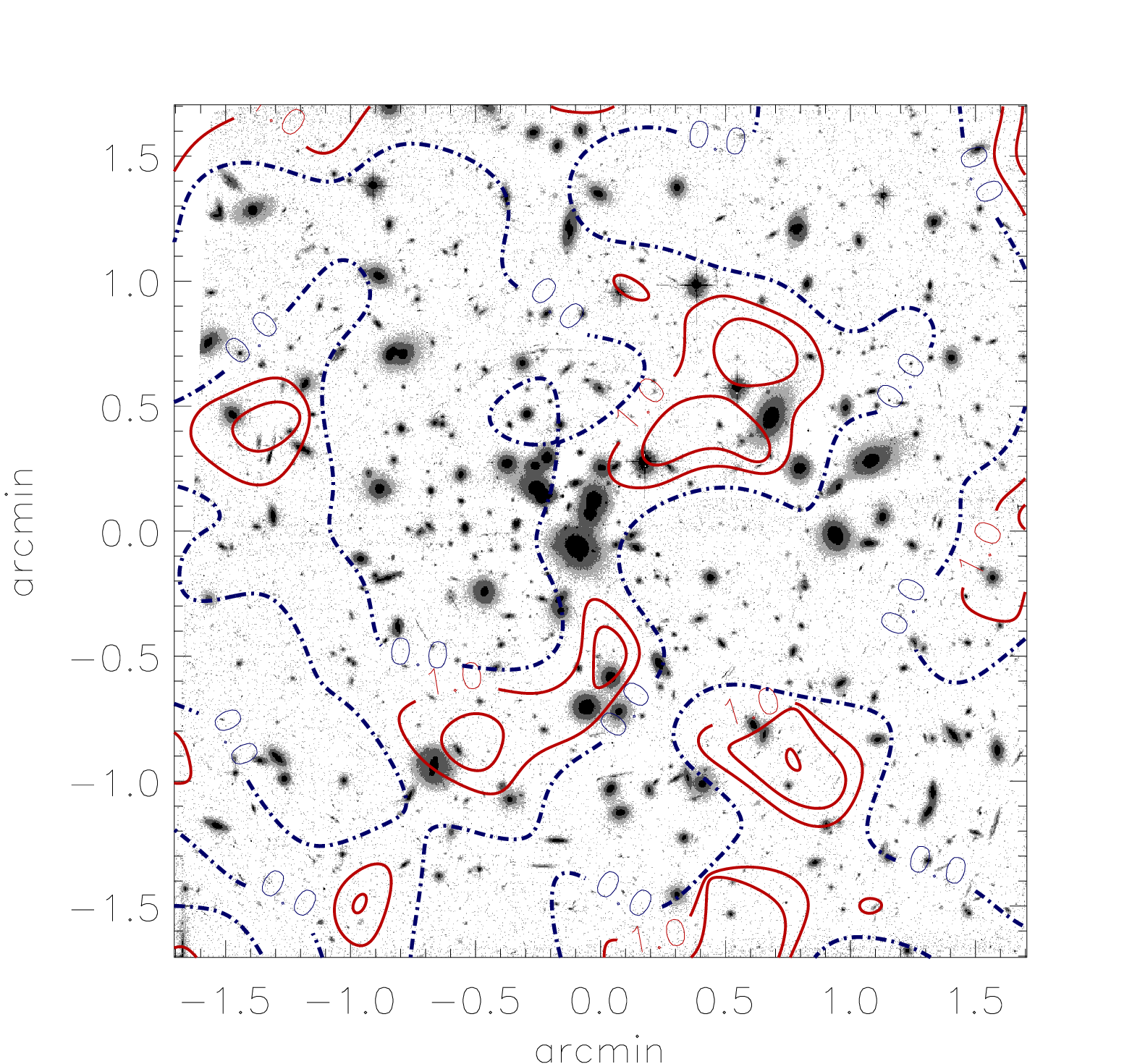}\includegraphics[width=0.25\textwidth]{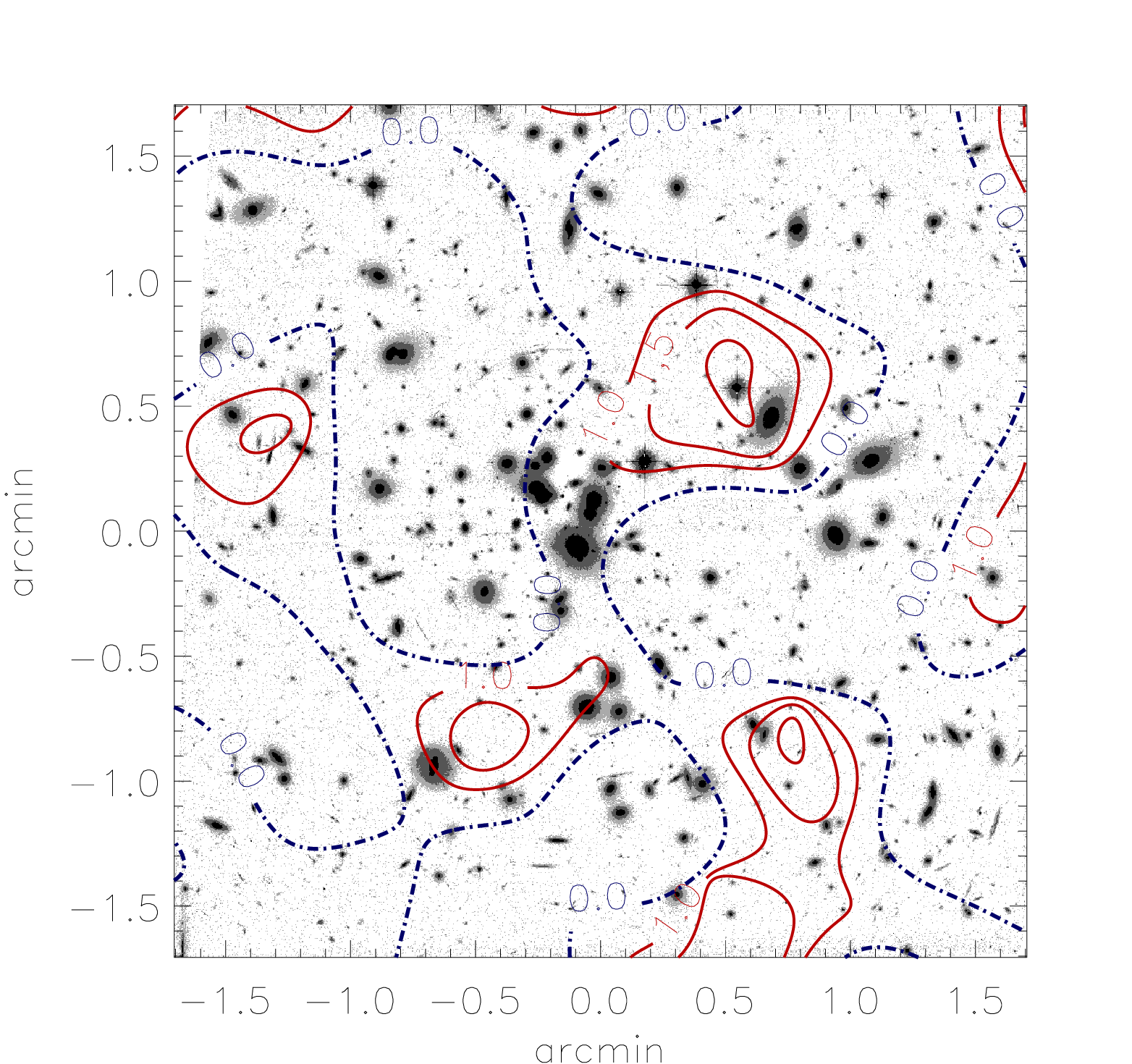}\includegraphics[width=0.25\textwidth]{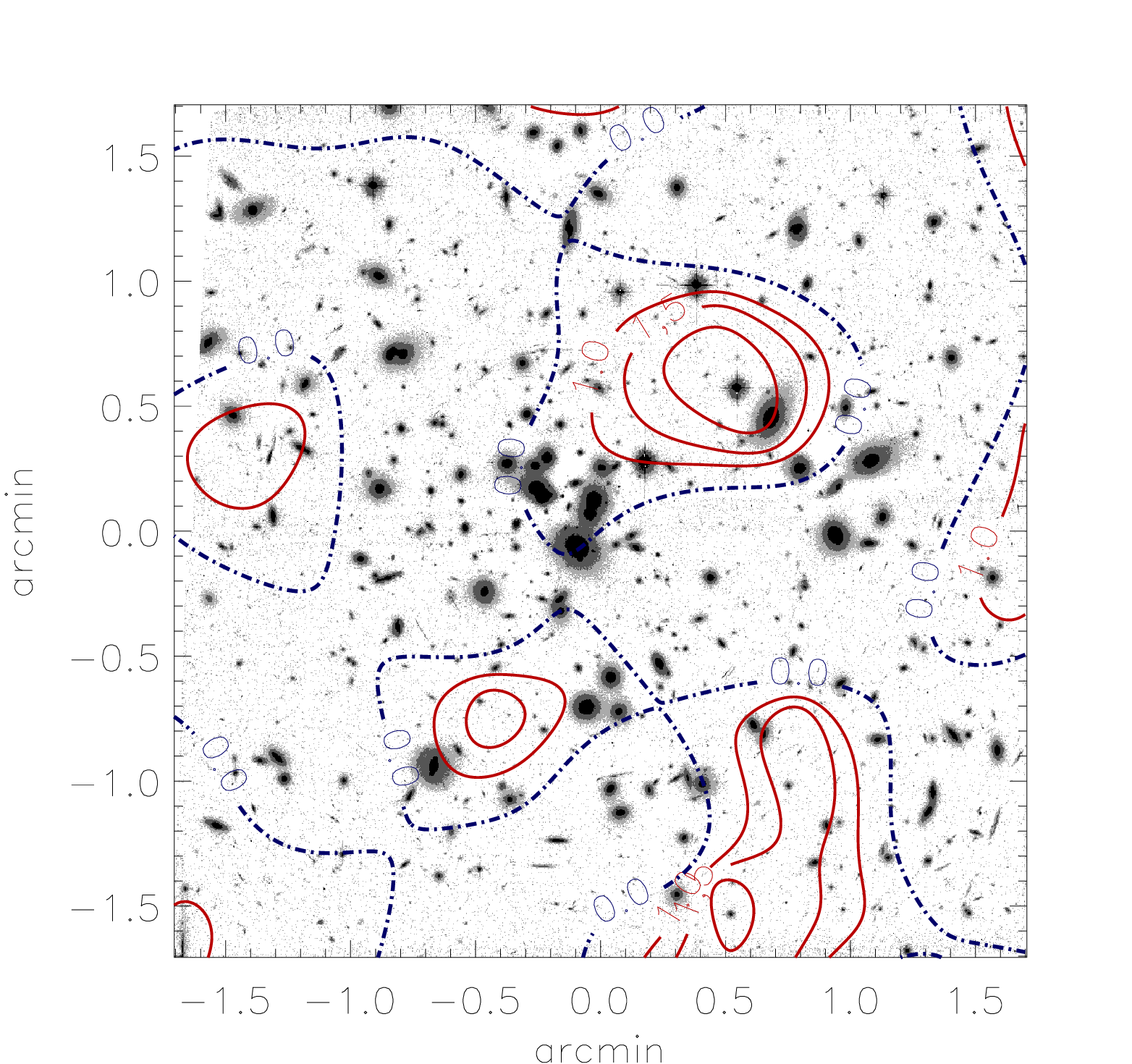}\vspace{15pt}
\caption[Flexion Aperture Mass Reconstructions of Abell 1689]{The figure above shows flexion aperture mass reconstructions for Abell 1689 using a polynomial flexion filter function with $\ell=3$ (top row), $\ell=5$ (second row), $\ell=7$ (third row) and $\ell=10$ (bottom row), and an aperture radius of $R=45''$ (first column), $R=60''$ (second column), $R=75''$ (third column) and $R=90''$ (fourth column). \label{fg:snfmap}}
\end{figure*}

The signal to noise contours of the 16 $\fmap$ reconstructions are plotted in Figure\,\ref{fg:snfmap}, overlaid on an image of the cluster. Here again, the contours are plotted at levels ${\cal S}=[1.0,1.5,2.0,2.5]$. The peak signal to noise value for each of the structures detected vary with each reconstruction, generally being reduced slightly as the aperture size is increased at constant $l$, but is consistently found to be above ${\cal S}_{\rm peak}\sim 2$ for the larger, dominant structures. 

Several interesting features arise in the reconstructions. Firstly, as expected, the reconstructions using smaller apertures reveal structures on finer scales. Indeed, as expected, when the filter polynomial order is increased for a fixed aperture size, a similar effect is seen. For example, remarkable similarity is seen in the signal to noise plots for $\ell=5,\ R=45^{\prime\prime}$ and $\ell=10,\ R=60^{\prime\prime}$. This is an important consistency check of the method. The $\fmap$ reconstructions are not independent as the filter functions themselves are not orthogonal. Therefore, any large inconsistencies between the maps, particularly those whose filter functions peak at the same angular scale, are indicative of persistent systematic or computational errors.

\begin{figure}
\includegraphics[width=0.5\textwidth]{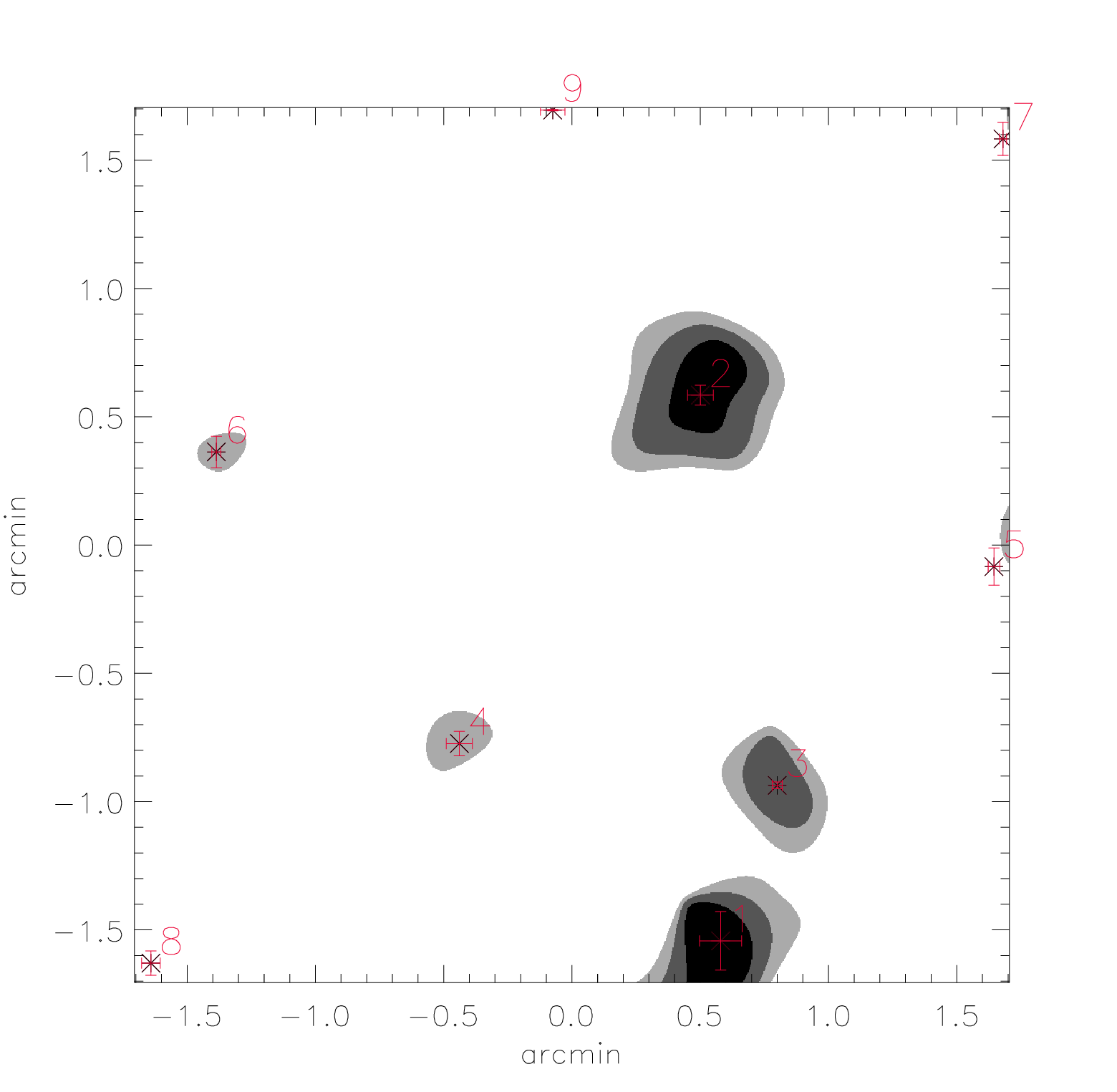}
\caption{The average signal to noise map of the 16 $\fmap$ reconstructions, showing all peaks identified with a signal to noise greater than 1. Overplotted aon the map are the locations of each of the 9 peaks identified, labels for each peak, and the $1\sigma$ dispersion in the location of the peaks across the 16 reconstructions\label{fg:agg}}
\end{figure}

It is clear that there are several prominent, persistent features across the 16 maps. These structures can be seen to break up into smaller substructures in the signal to noise maps as the aperture size is reduced or the polynomial order increased. In order to determine the most significant persistent structures in the $\fmap$ reconstructions, we consider the aggregate map, plotted in Figure\,\ref{fg:agg}. 9 peaks are identified in total; for each, the signal-to-noise weighted centroid is highlighted in the figure, and the peaks are labelled in order of decreasing signal to noise in the aggregate map.

We now consider the peak detections in the individual $\fmap$ reconstructions. The signal to noise maps for each reconstruction are cross-matched with the average map, and where a peak has been broken down into smaller sub-peaks, the sub-peaks are identified by considering all peaks in the individual map that are located within the ${\cal S}=1$ contour of the peak in the average map. 

\begin{figure*}
\center
\includegraphics[width=0.25\textwidth]{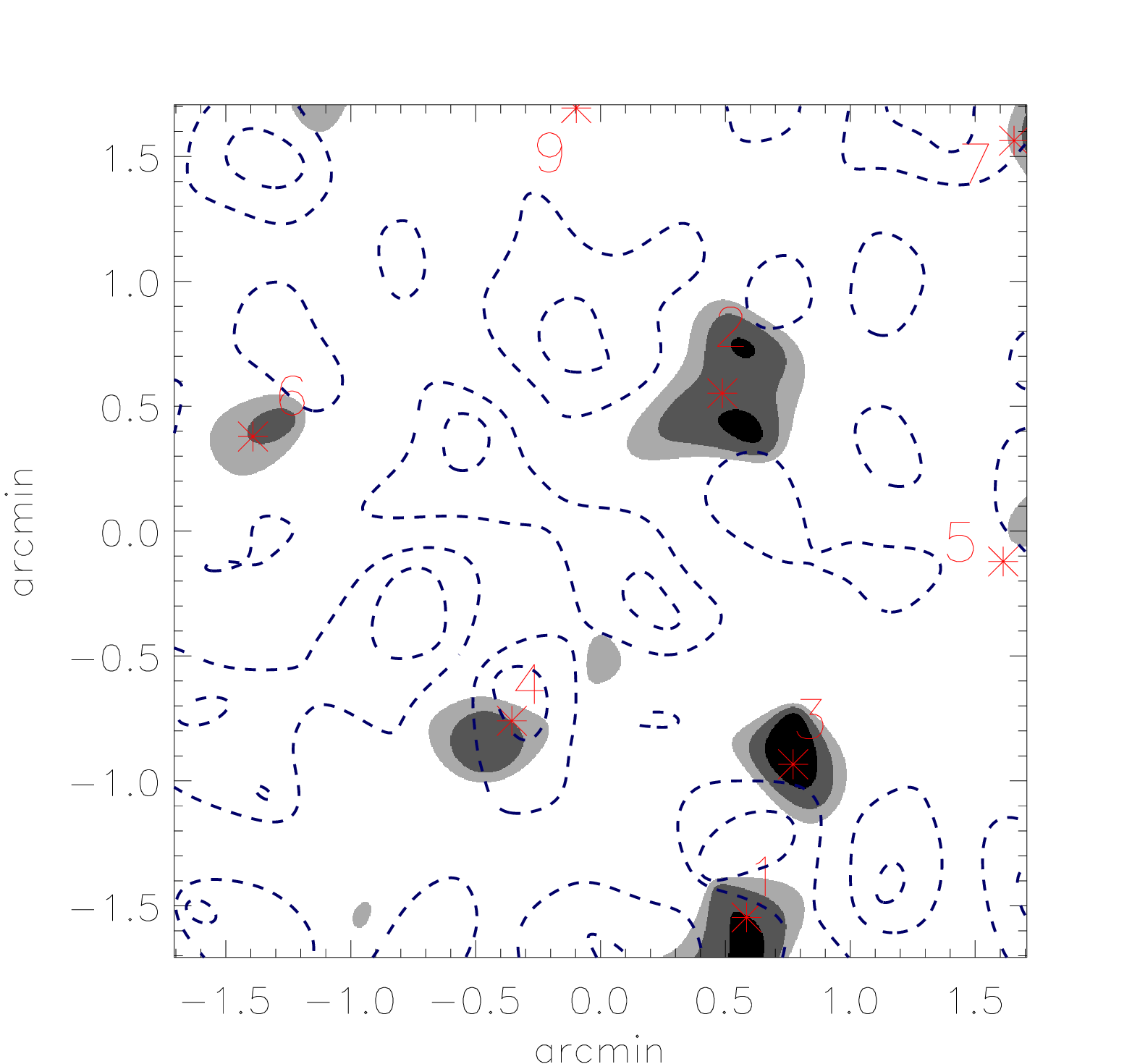}\includegraphics[width=0.25\textwidth]{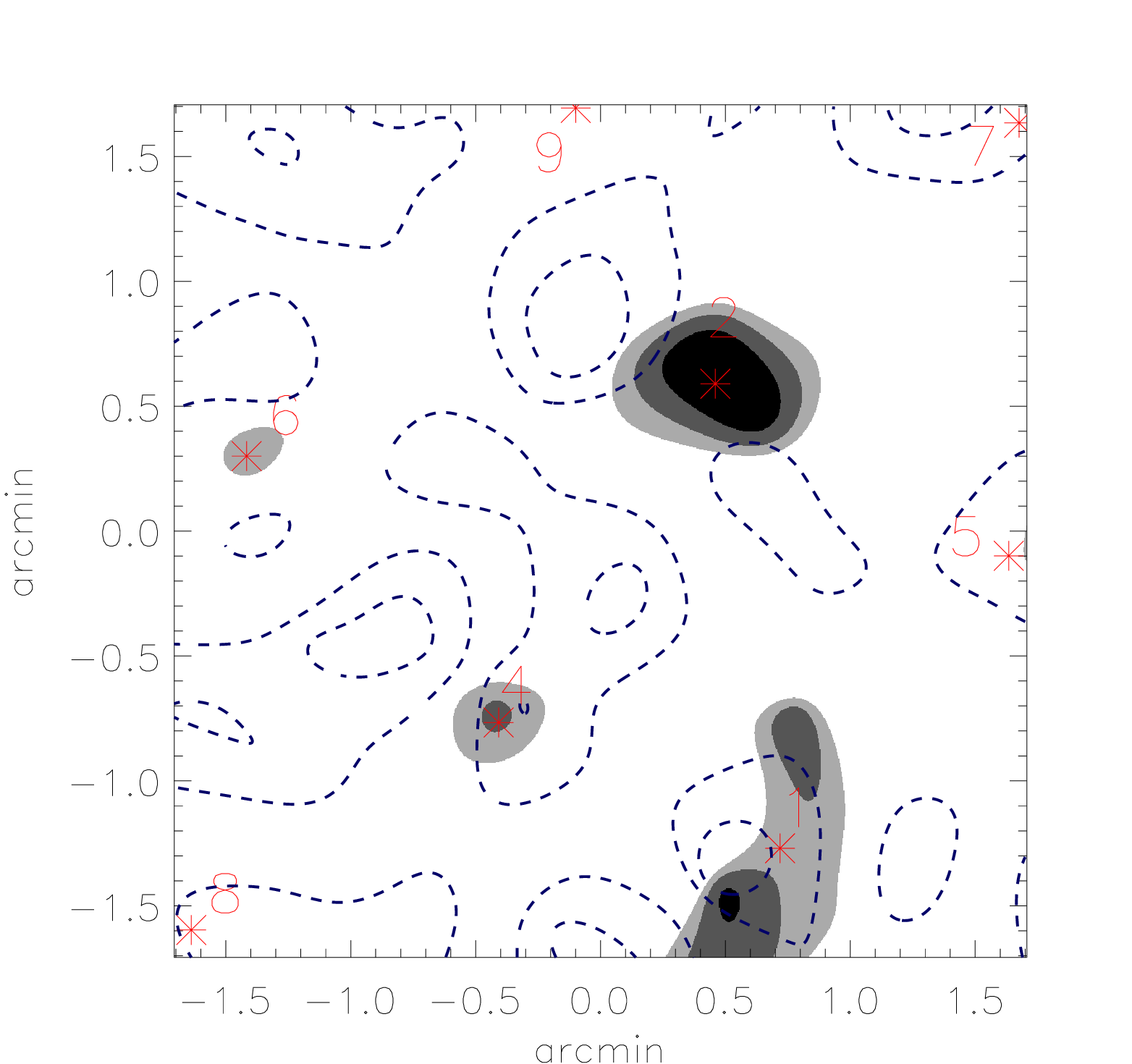}\includegraphics[width=0.25\textwidth]{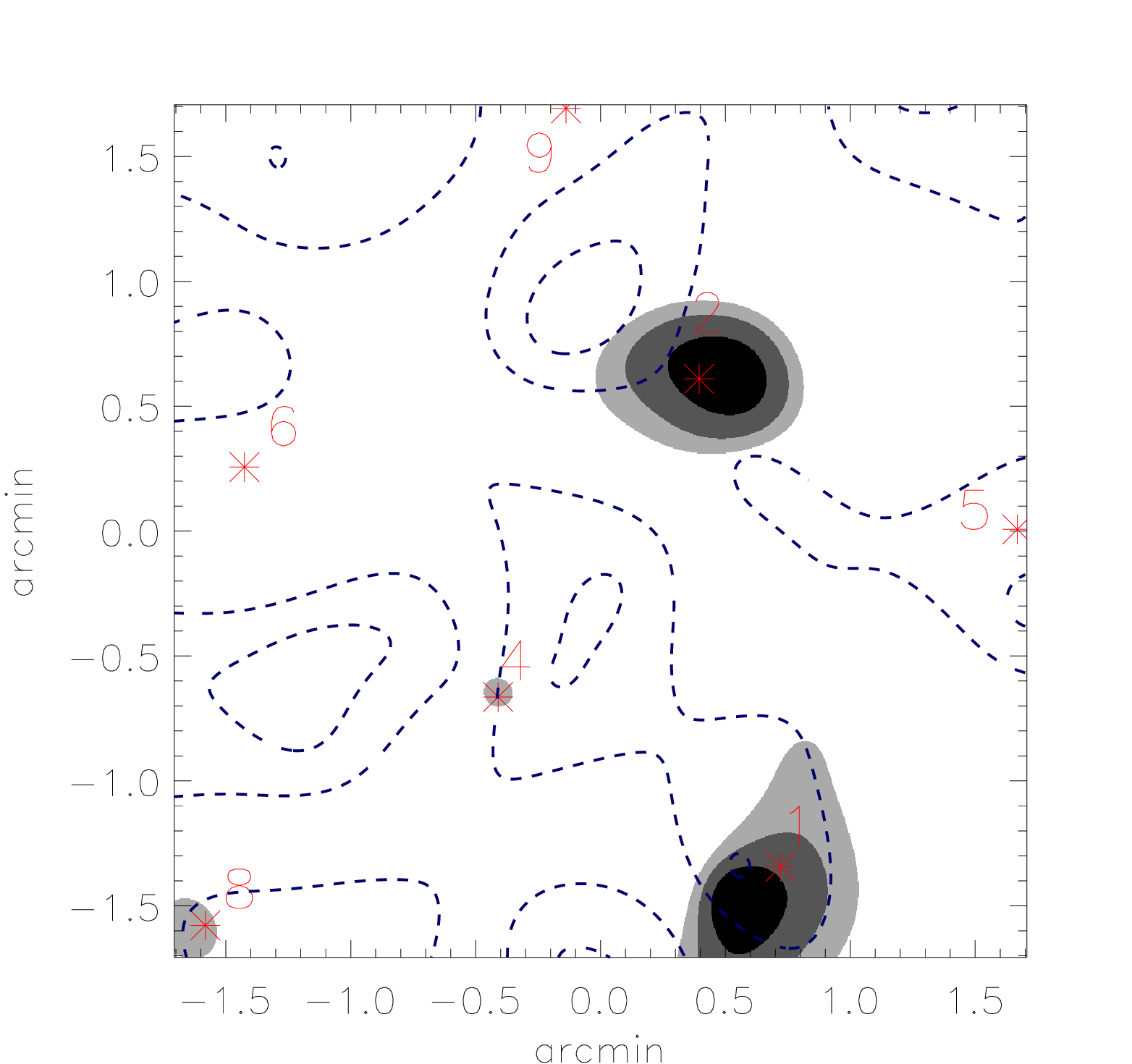}\includegraphics[width=0.25\textwidth]{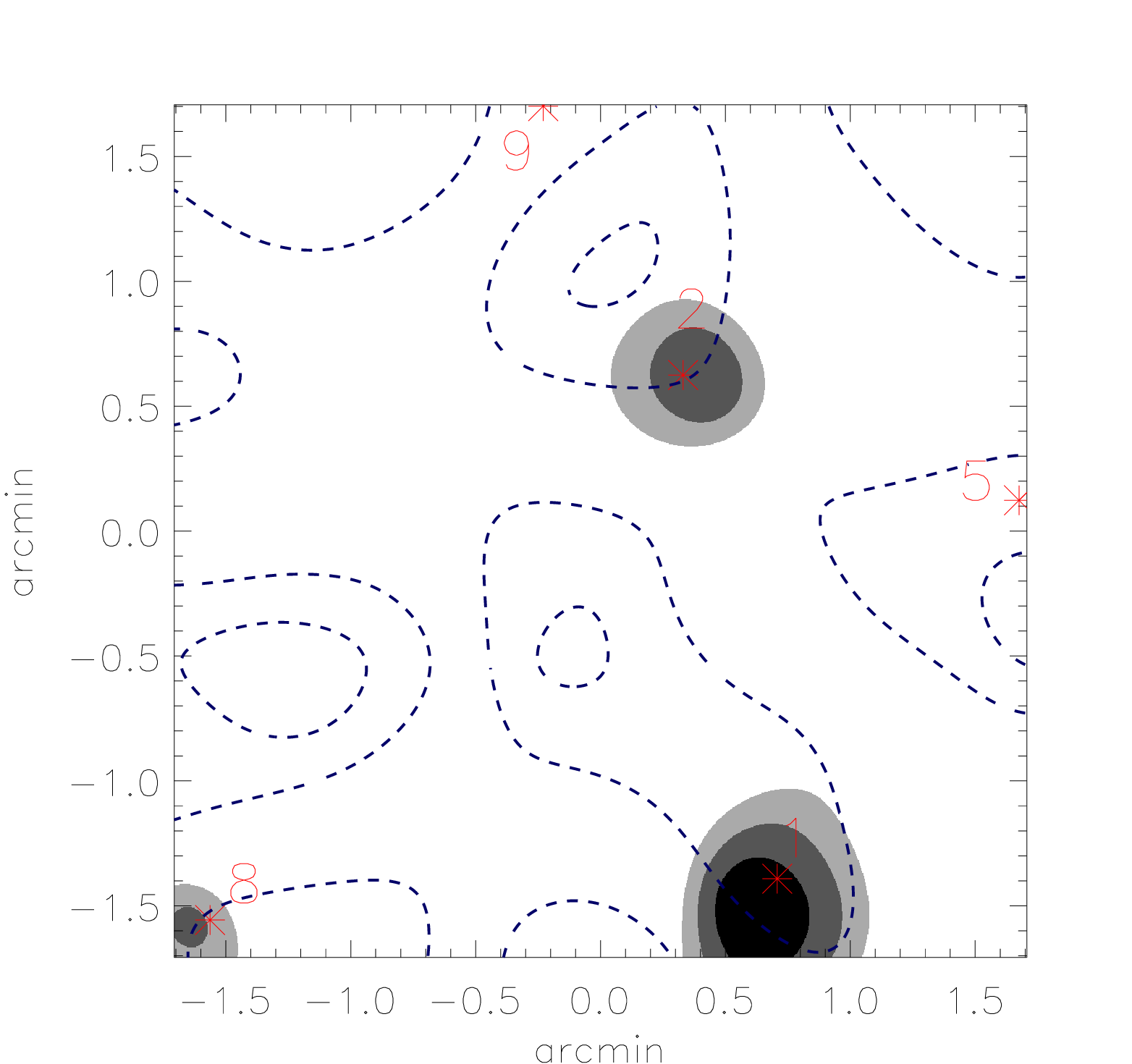}\vspace{15pt}
\includegraphics[width=0.25\textwidth]{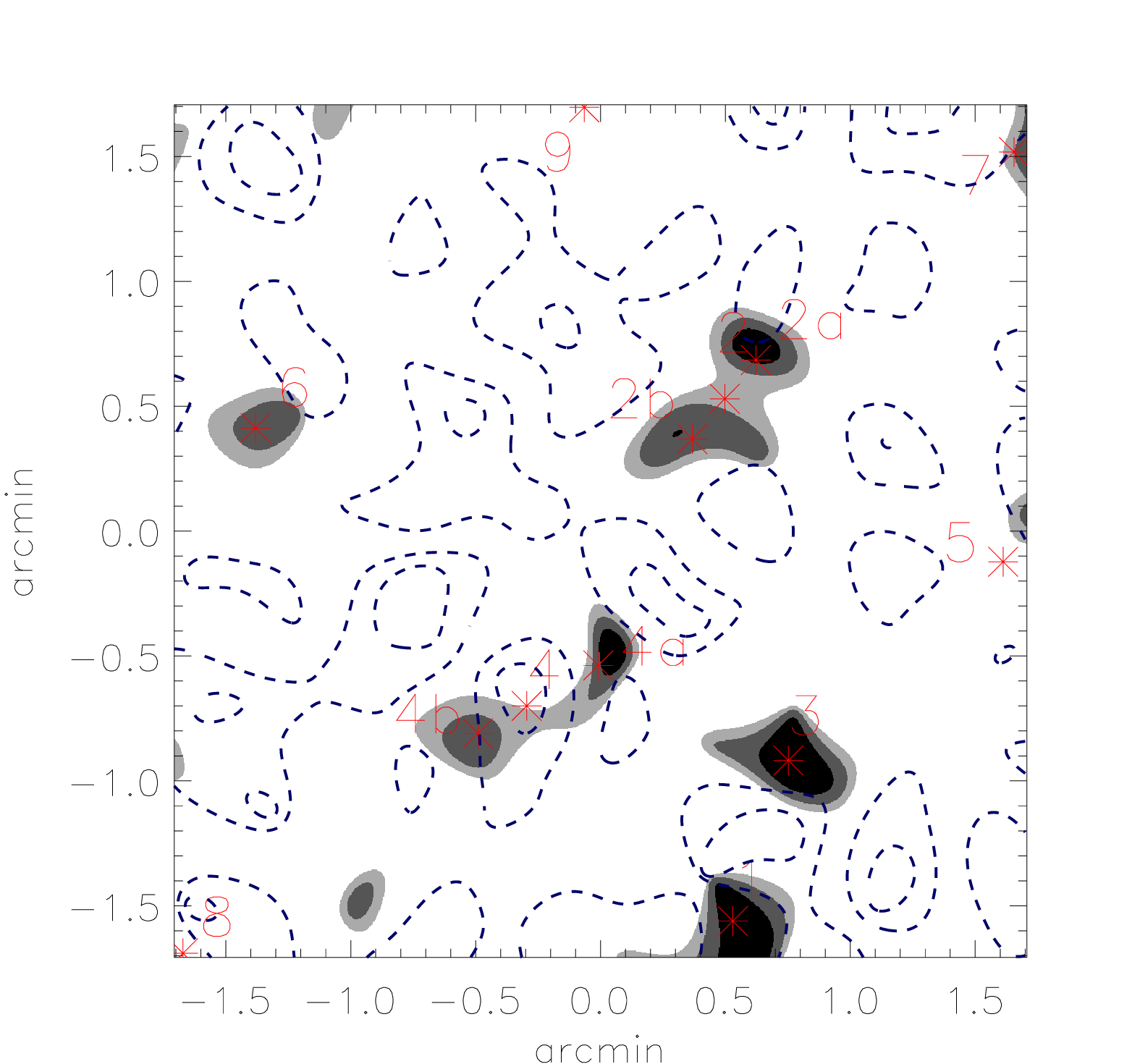}\includegraphics[width=0.25\textwidth]{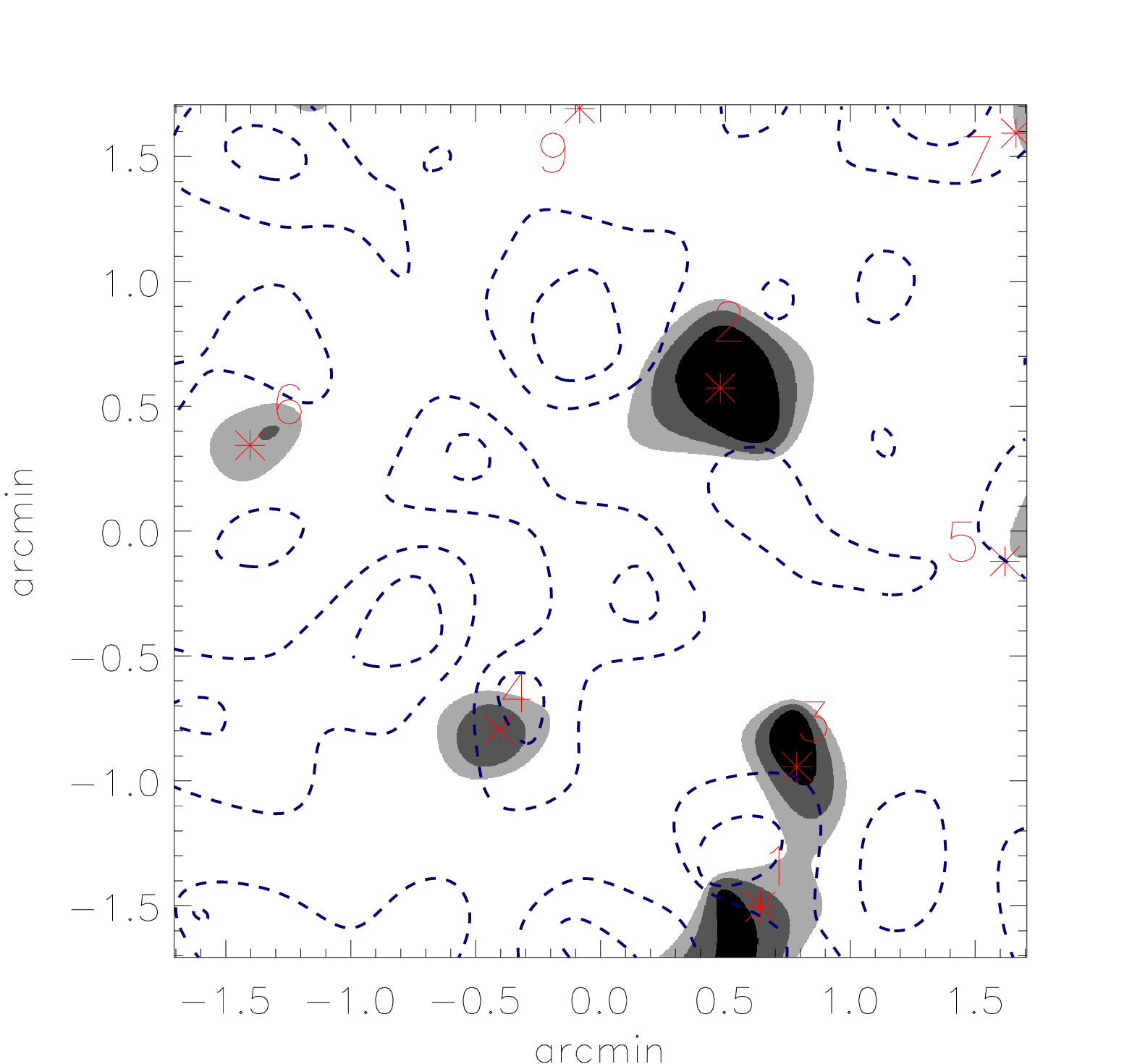}\includegraphics[width=0.25\textwidth]{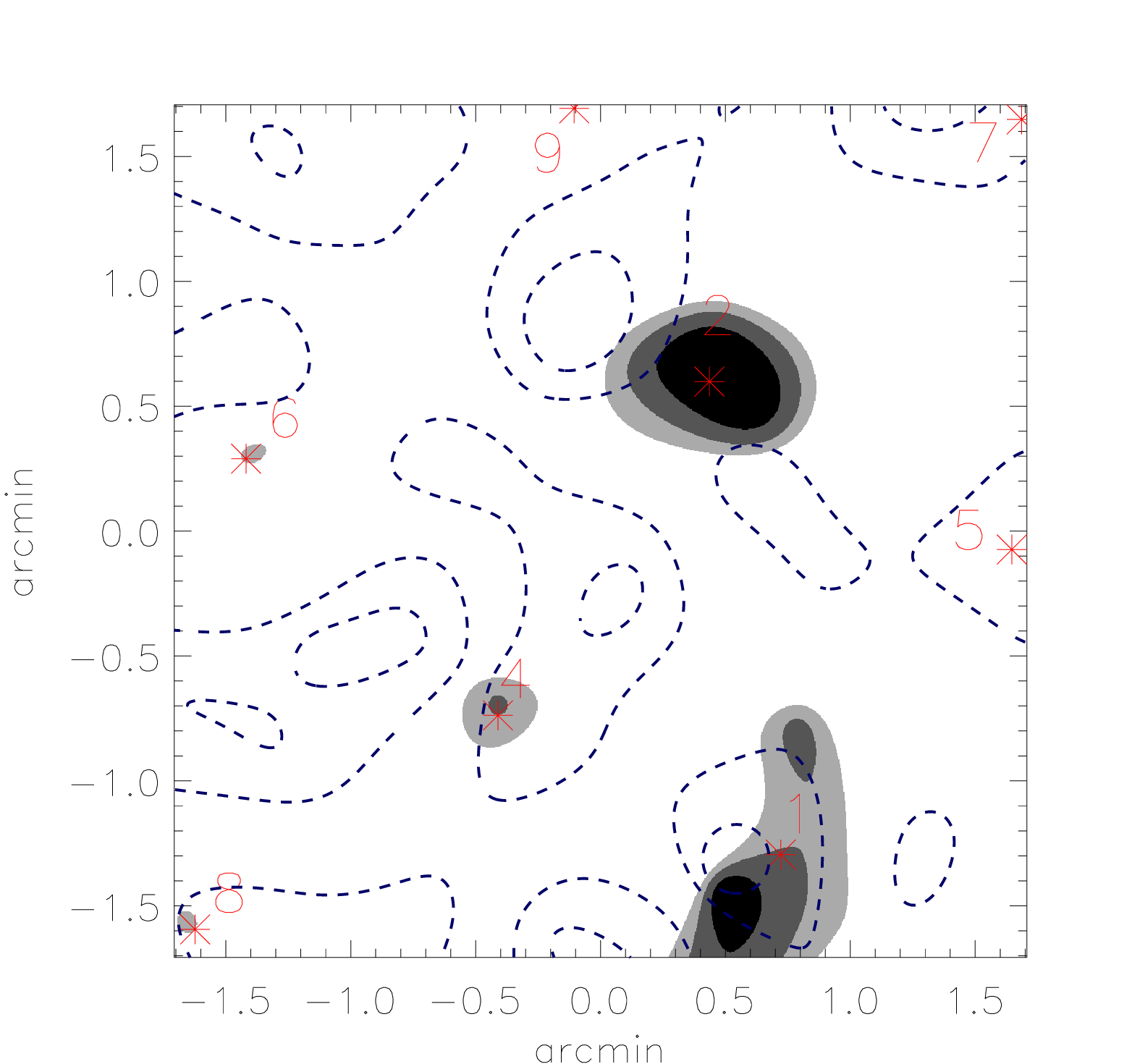}\includegraphics[width=0.25\textwidth]{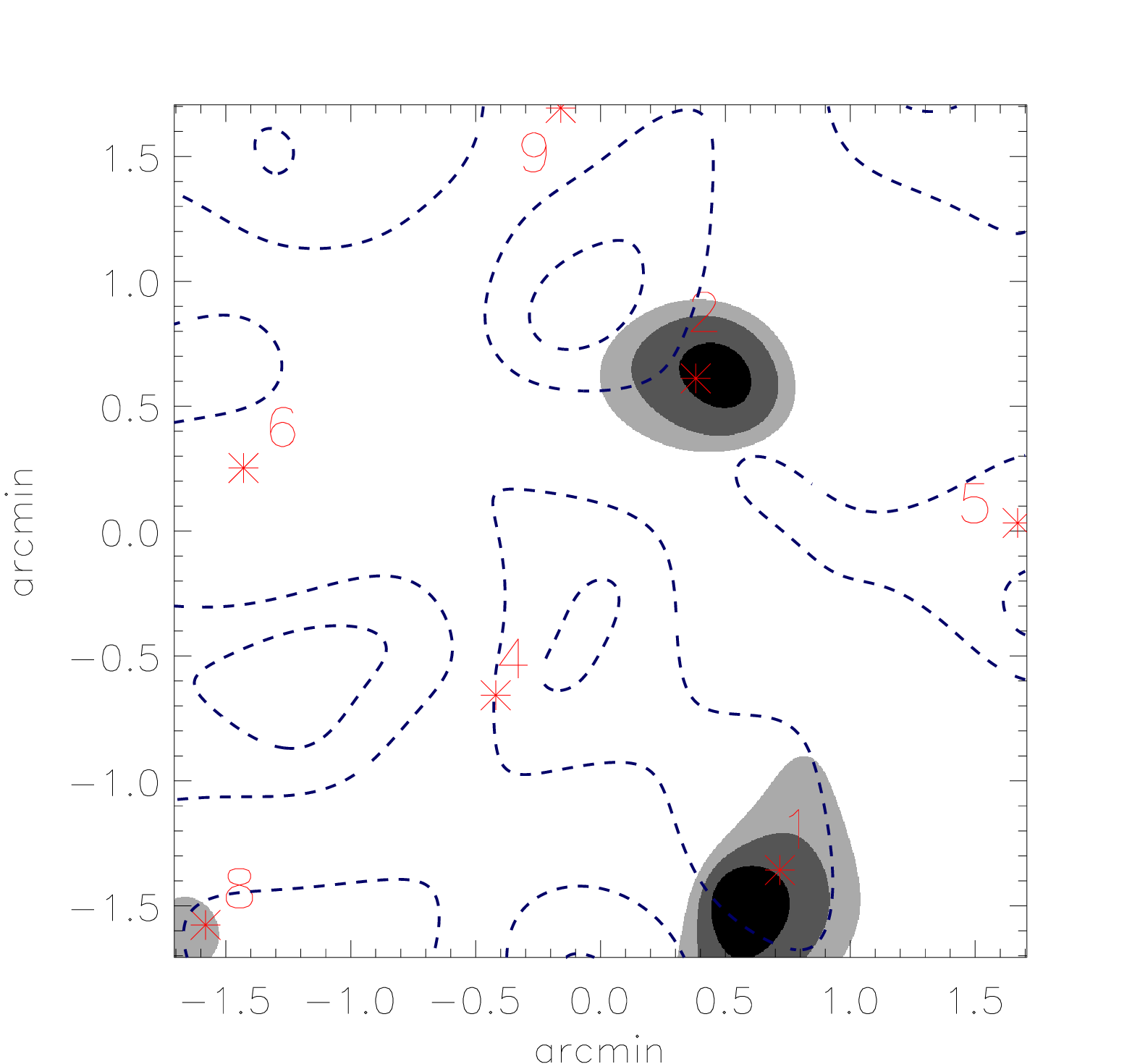}\vspace{15pt}
\includegraphics[width=0.25\textwidth]{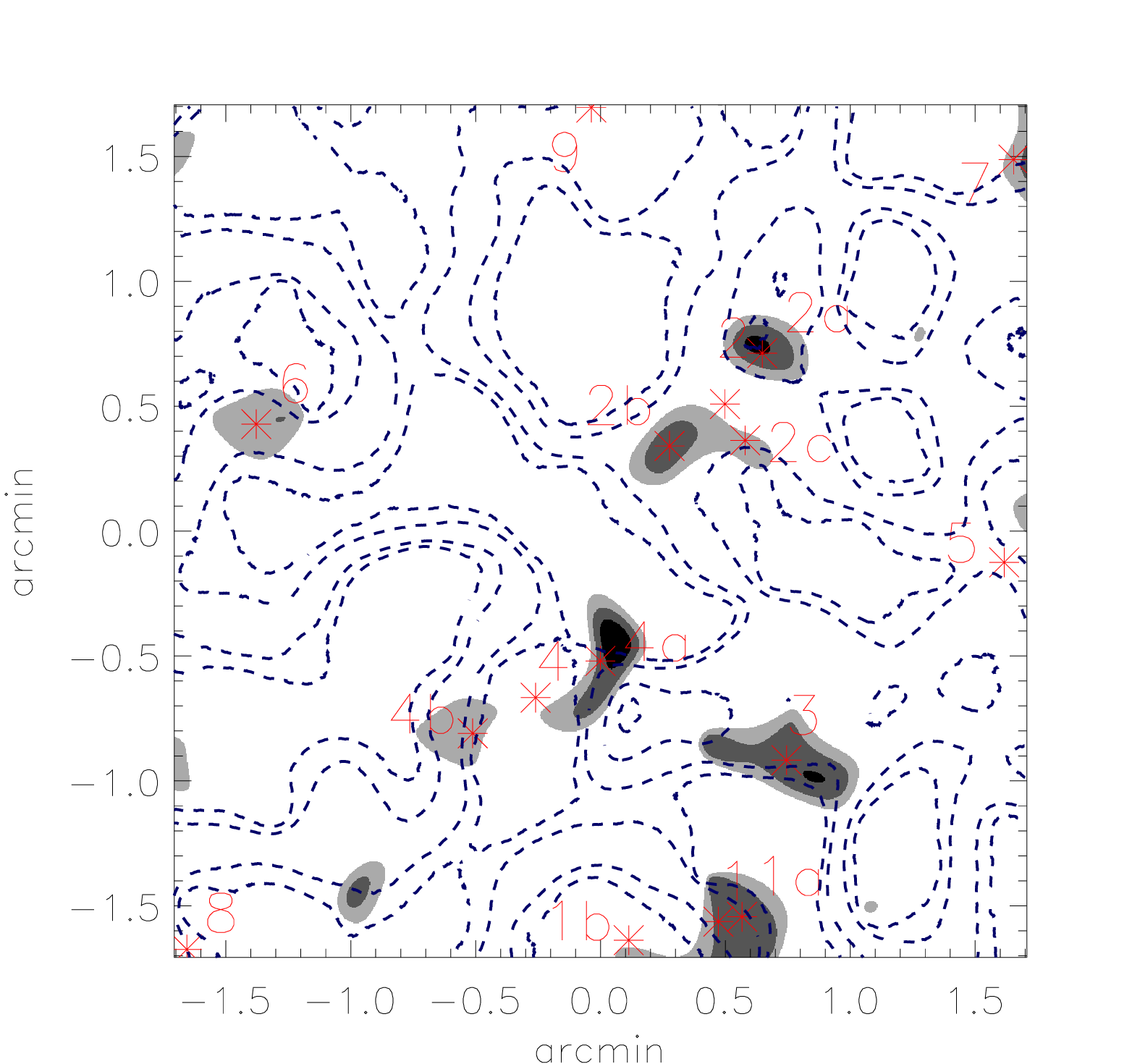}\includegraphics[width=0.25\textwidth]{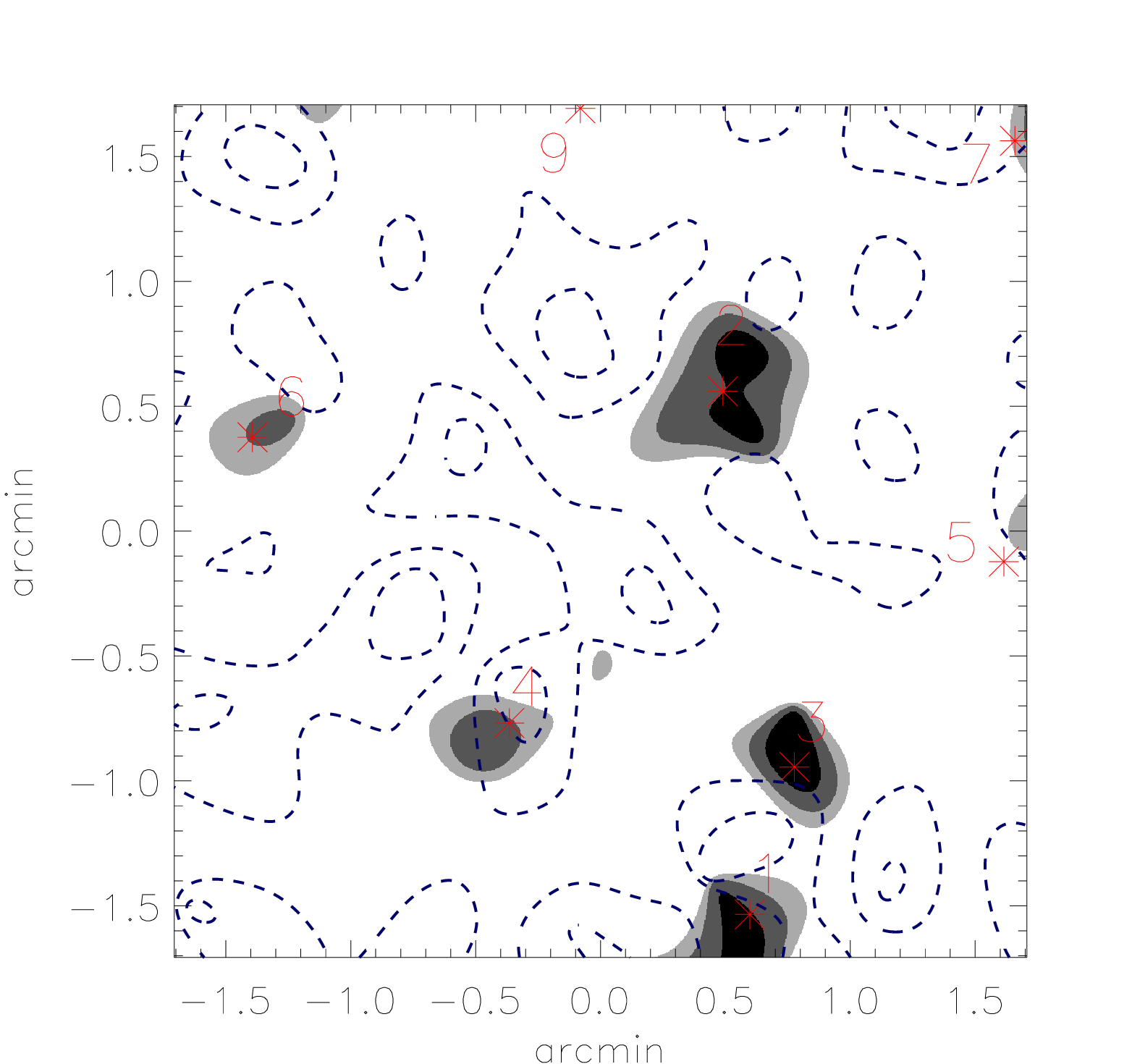}\includegraphics[width=0.25\textwidth]{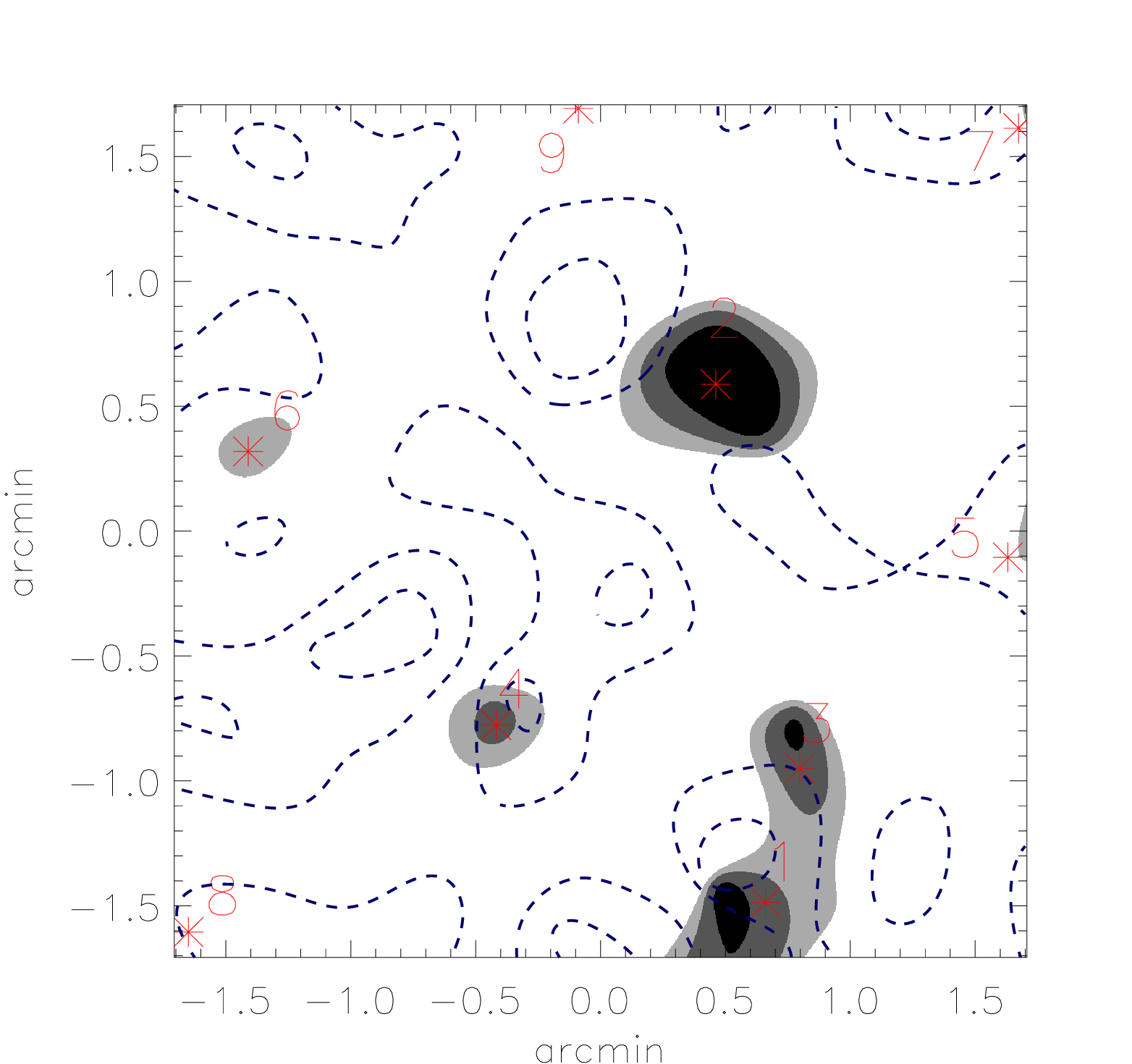}\includegraphics[width=0.25\textwidth]{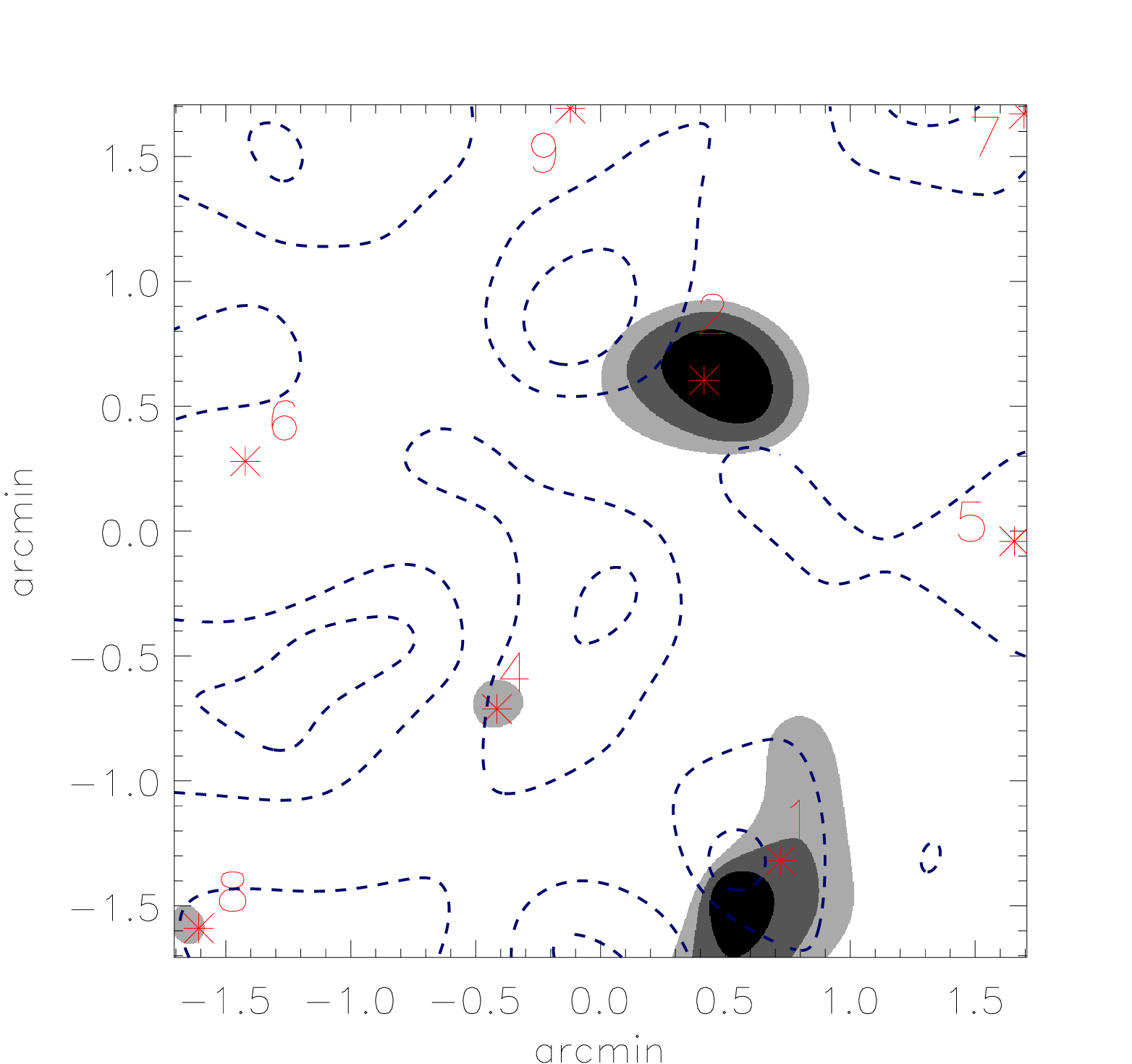}\vspace{15pt}
\includegraphics[width=0.25\textwidth]{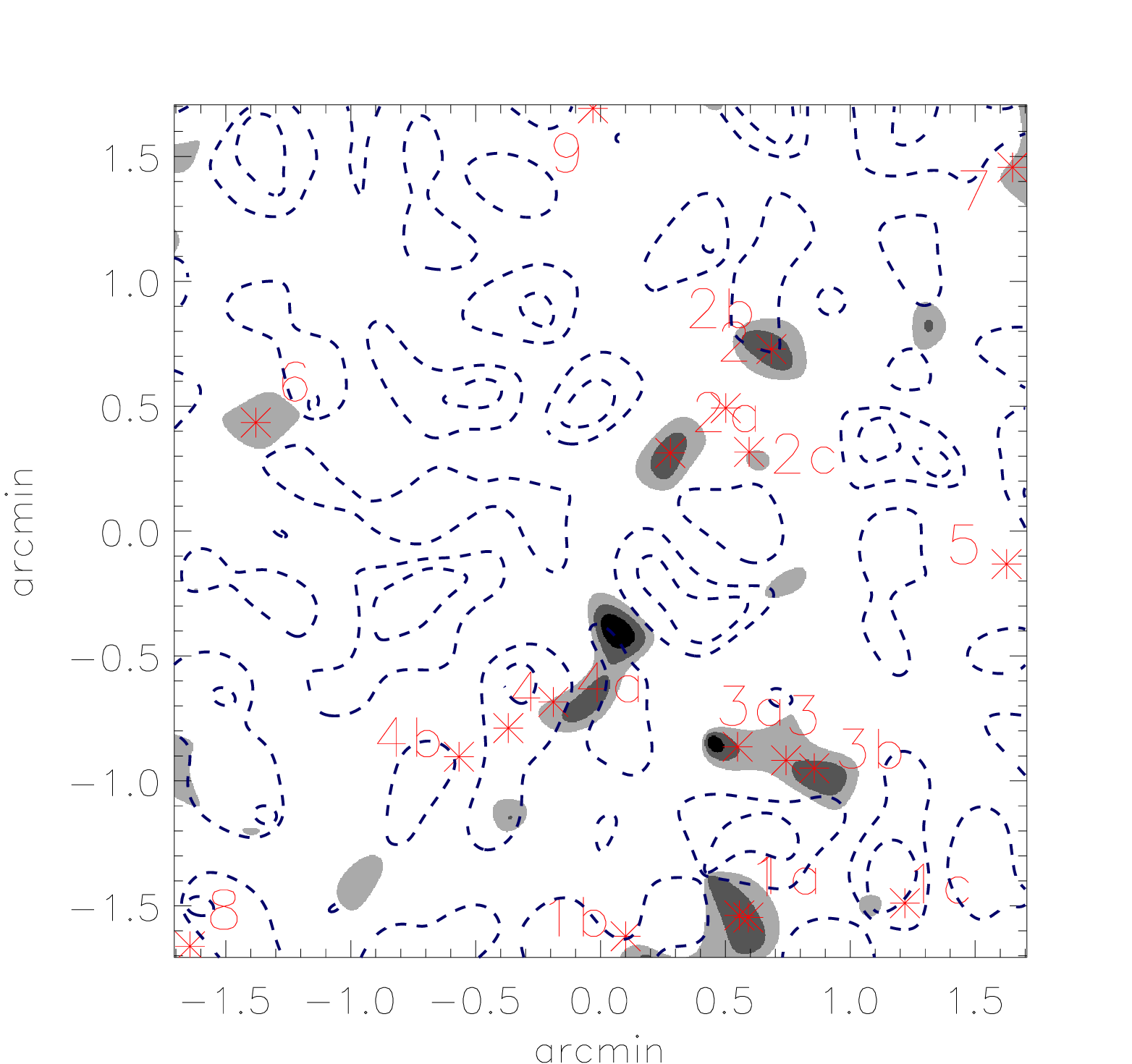}\includegraphics[width=0.25\textwidth]{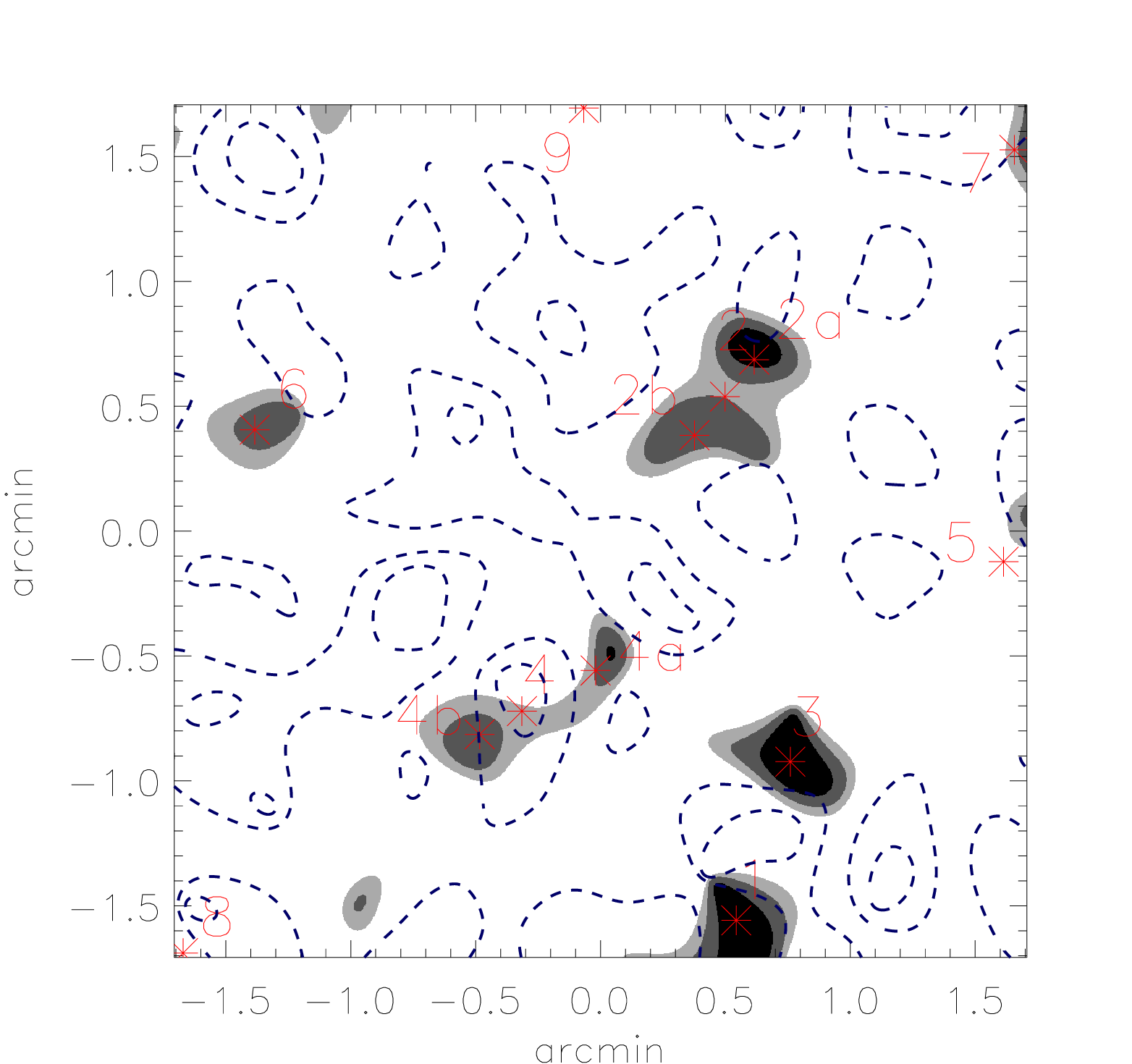}\includegraphics[width=0.25\textwidth]{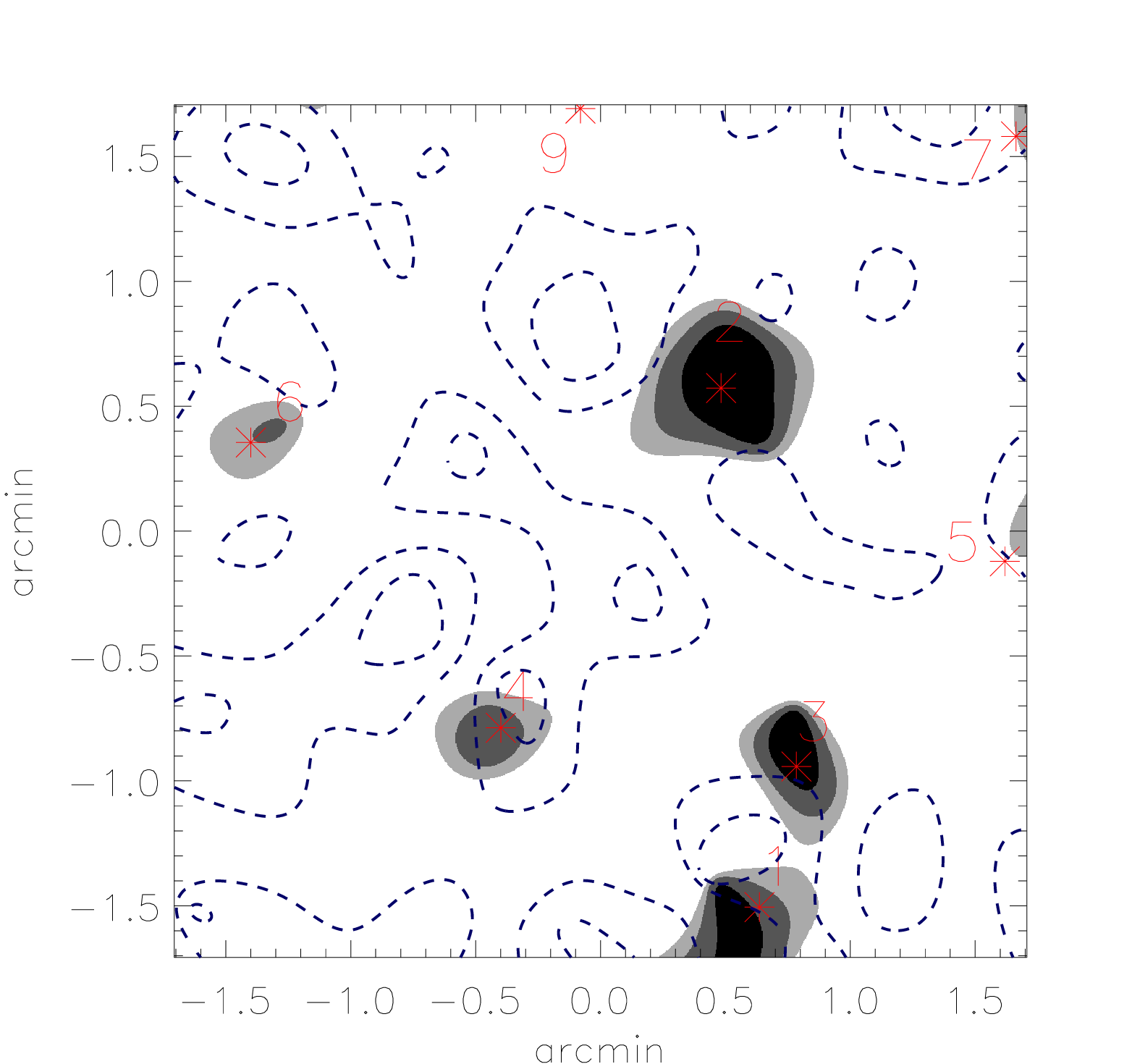}\includegraphics[width=0.25\textwidth]{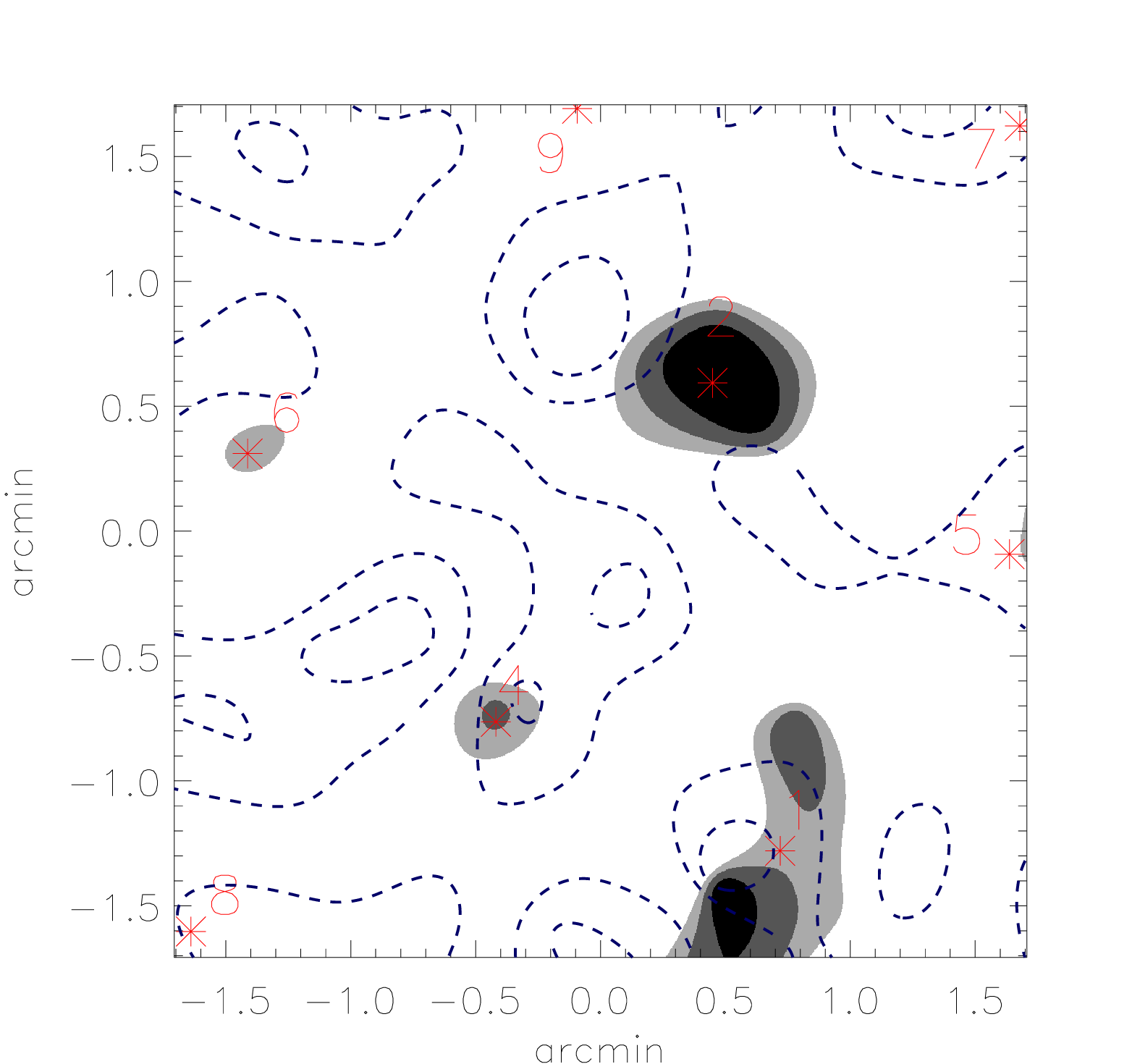}\vspace{15pt}
\caption{The figure above shows the detected peaks in $\fmap$ reconstructions of Abell 1689 using a polynomial flexion filter function with $\ell=3$ (top row), $\ell=5$ (second row), $\ell=7$ (third row) and $\ell=10$ (bottom row), and an aperture radius of $R=45''$ (first column), $R=60''$ (second column), $R=75''$ (third column) and $R=90''$ (fourth column). Peaks and their subpeaks are labelled in each reconstruction at the location of the signal-to-noise weighted centroid of the peak. Dashed contours show the absolute value of the B-mode signal, plotted at levels of 1$\sigma$ and 2$\sigma$.\label{fg:fmap_peaks}}
\end{figure*}

In addition, a signal-to-noise weighted centroid is computed for the collection of sub-peaks corresponding to the primary peak in the average map. Figure\,\ref{fg:fmap_peaks} shows the peaks detected in each $\fmap$ realisation, appropriately labelled at the centroid of each peak or sub-peak. Overlaid on each map are contours showing the absolute value of the B-mode map for each reconstruction, normalised by the noise in that reconstruction. Contours are plotted for $|{\fmap}_{,\rm B}|=[1\sigma,\ 2\sigma]$.

The $1\sigma$ dispersion in the locations of each of the peaks are overplotted in Figure\,\ref{fg:agg}, and provide an indication of the extent to which the locations of the peaks are offset from one another in the 16 $\fmap$ reconstructions. Offsets in the peaks can arise due to noise in the reconstruction, or as a result of a peak breaking up in to smaller sub-peaks; a small dispersion in the peak locations is indicative of a robust detection of the position a given peak.

Of the 9 peaks identified, four (peaks 5, 7, 8, and 9) are located at the edge of the field. The B-mode maps appear to show some structure around the edges of the field, thus we cannot consider these to be true detections. Even if they are indeed real features, due to their locations it is impractical to attempt to compare their radial profiles with those of the template models, given that they are located so close to the edge of the field; a complete circular average around the centre of each peak is impossible to obtain given the data. 

In addition, significant B-modes were detected in all $\fmap$ reconstructions of peak $4$, as is evident from Figure\,\ref{fg:fmap_peaks}. This peak can therefore also be considered to arise out of noise, and is excluded from further analysis. Therefore, we restrict our study to peaks $[1,3],\ 2,$ and $6$. Each peak is considered separately below. Note that peaks 1 and 3 are considered together, as it is likely that they form part of the same structure. Indeed, in many of the large $R$, low $l$ reconstructions, peak 3 is undetected.

\subsection{Constraints on mass profiles}

\subsubsection{Peaks 1 \& 3}

Peak 1 is detected in all 16 $\fmap$ reconstructions, whilst peak 3 appears in only 9 reconstructions, forming part of peak 1 for the remainder. It therefore seems likely that peak 3 is a slightly lower-mass substructure located in the neighbourhood of peak 1 that becomes blended with peak 1 as the aperture radius is increased. Both peak 1 and peak 3 are seen to break up into smaller parts in the high $\ell$, low $R$ aperture mass reconstructions, which indicates a complex structure that is unlikely to be well-modelled by a spherical halo. Nonetheless, Leonard \& King (2010) found that the peak signal and zero signal radius of the flexion aperture mass statistic could be used to estimate the masses of haloes even when they possess a moderate ellipticity.

In the instances where peak 3 is clearly resolved, it does not show prominent B-modes within the error bars of the $\fmap$ reconstruction in each case. This implies that it is, indeed, a real mass concentration within the cluster. Peak 1 is found to show very prominent B-modes in the 7 reconstructions in which peak 3 is not resolved, which suggests that in these reconstructions, the signal from peak 3 might be a strong contaminant to the signal. Peak 1 shows prominent B-modes in a further 3 reconstructions, all corresponding to low-$\ell$, large-$R$ filters. These B-modes most likely result from edge effects, as peak 1 is located close to the image boundary. 

Excluding those profiles with substantial B-modes, we are left with 6 reconstructions of peak 1 and 9 of peak 3, and we can attempt to constrain the mass distribution associated with these peaks from the remaining reconstructions.

\begin{figure}
\center
\includegraphics[width=0.45\textwidth]{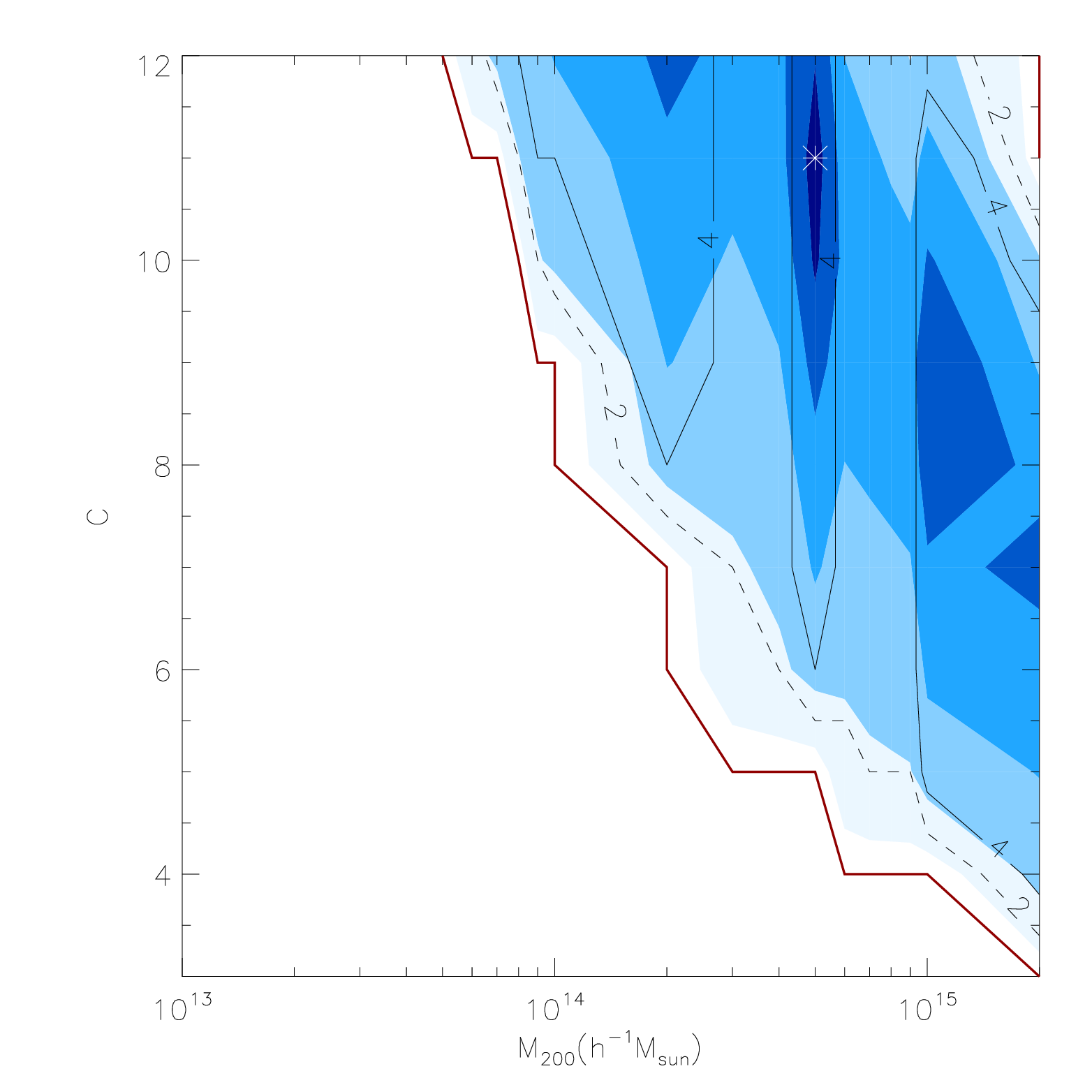}
\includegraphics[width=0.45\textwidth]{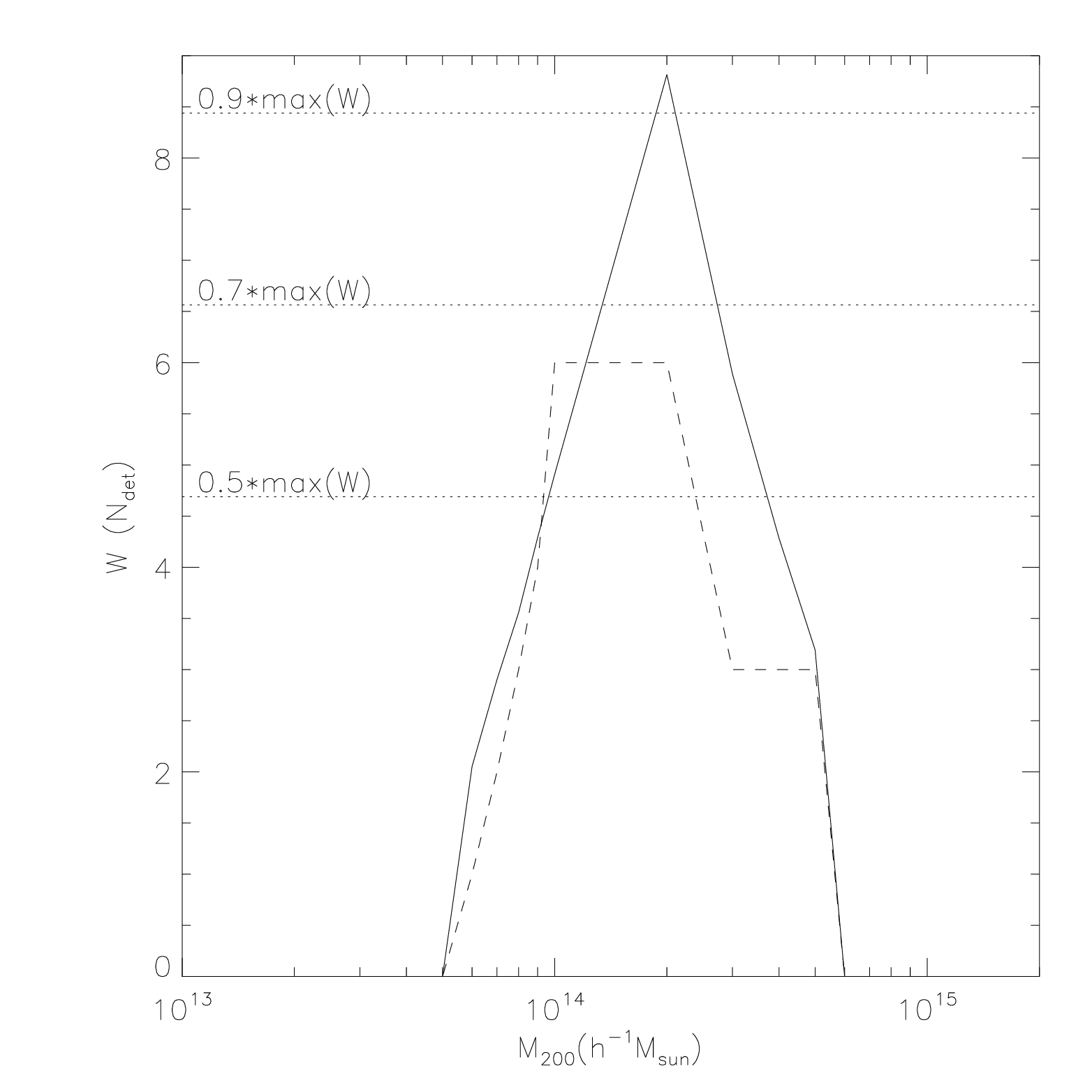}
\caption{The pseudo-likelihood distributions for NFW (top panel) and SIS (bottom panel) models for peak 1 generated as described in the text. In the top panel, filled contours indicate ${\cal W}$  at contour levels equal to $[0.1, 0.3, 0.5, 0.7,0.9]\times{\cal W}_{\rm max}$, with the model corresponding to the peak ${\cal W}$ in the NFW distribution denoted by a white asterisk. The solid red curve shows ${\cal W}=0$. Corresponding levels are shown in the bottom panel as dotted horizontal lines. The dotted and dashed contours in the top panel indicate $N_{\rm det}$, represented in the bottom panel as the dashed curve.  \label{fg:pk1prob}}
\end{figure}

We first consider the ${\cal W}$ distribution of peak 1. Figure\,\ref{fg:pk1prob} shows the measured ${\cal W}$ distribution for the set of NFW template models (top panel) and SIS models (bottom panel). In the NFW case, contours are plotted in $M_{200}-c$ space at levels equal to $[0.5,0.7,0.9]\times$ the maximum ${\cal W}$ found amongst all (both NFW and SIS) models. Dotted and dashed contours indicate $N_{\rm det}$ for each model. In the SIS case, the solid curve shows the ${\cal W}$ distribution as a function of mass, and the dashed curve shows $N_{\rm det}$. Dotted horizontal curves show the ${\cal W}$ values corresponding to the contours in the NFW plot. 

The best-fit NFW model to the data is one with $M_{200}=5\times10^{14}h^{-1}M_\odot$ and $c=11$ (${\cal W}_{\rm max}=9.38,\ N_{\rm det}=6$), and the best-fit SIS has $M_{200}=2\times10^{14}h^{-1}M_\odot$ (${\cal W}_{\rm max}=8.82,\ N_{\rm det}=6$). Thus, the data seem to favour the NFW model, and Figure\,\ref{fg:pk1prob} shows a rather narrow range of masses favoured by the data. 

It is important to note that the range of allowed masses is similar in both the NFW and SIS cases. This is an important consistency check, as one would not expect the signal from low mass SIS profiles to be comparable to that from high mass NFW profiles, and vice versa. Despite the presence of significant structure within each peak at small aperture radii and high filter polynomial order, and the large error bars in the measurements of $m_{\rm peak}$ and $R_0$, we are able to eliminate large regions of parameter space as incompatible with the data, and isolate a small range of best-fit models for the data.

We now consider peak 3, and the results are shown in Figure\,\ref{fg:pk3prob} below. The best-fit NFW model to the data is one with $M_{200}=2\times10^{14}h^{-1}M_\odot$ and $c=10$ (${\cal W}_{\rm max}=19.79,\ N_{\rm det}=9$), and the best-fit SIS has $M_{200}=2\times10^{14}h^{-1}M_\odot$ (${\cal W}_{\rm max}=12.13,\ N_{\rm det}=7$). Here, again, the data seem to favour the NFW model, and Figure\,\ref{fg:pk3prob} shows a narrow range of allowed masses. 

\begin{figure}
\includegraphics[width=0.45\textwidth]{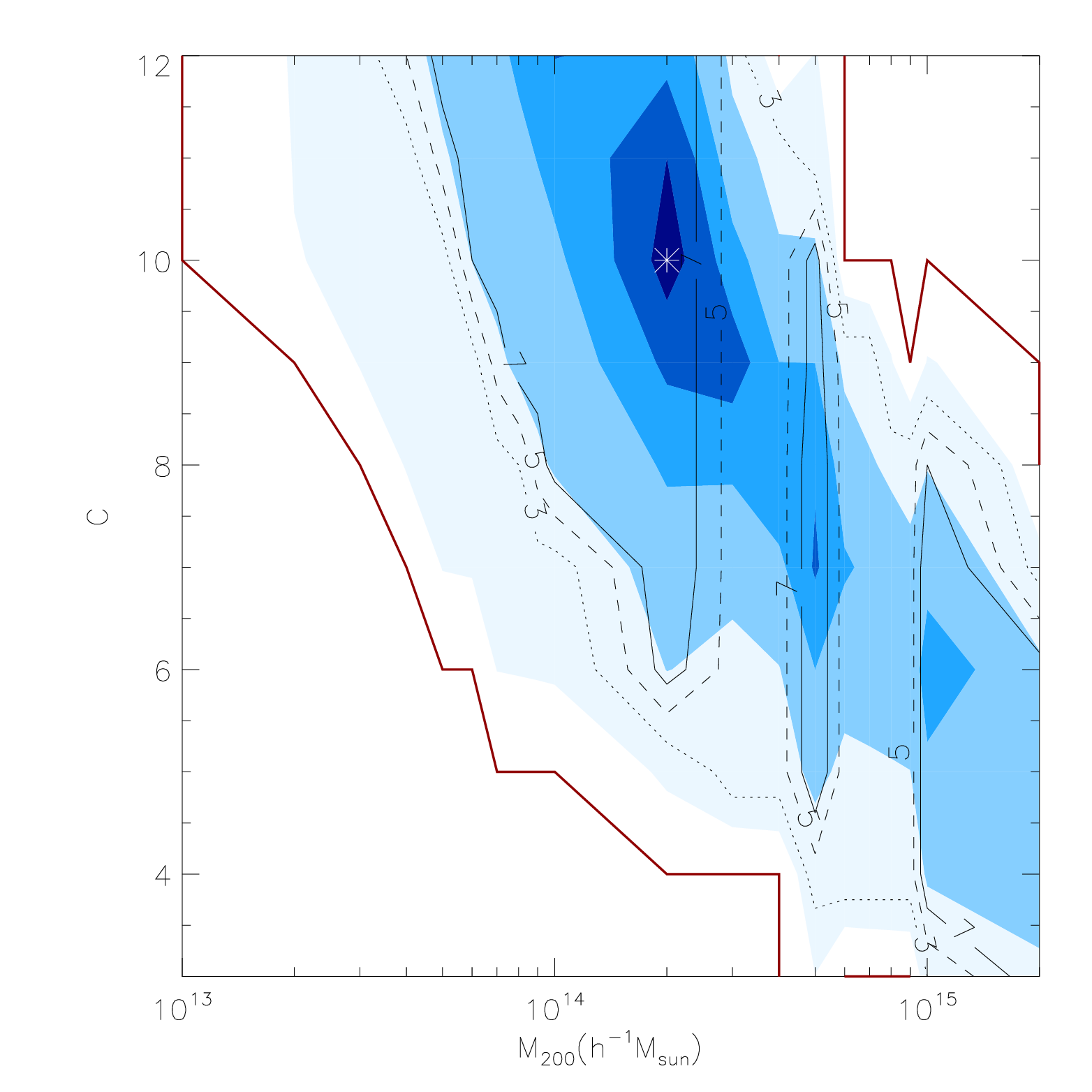}
\includegraphics[width=0.45\textwidth]{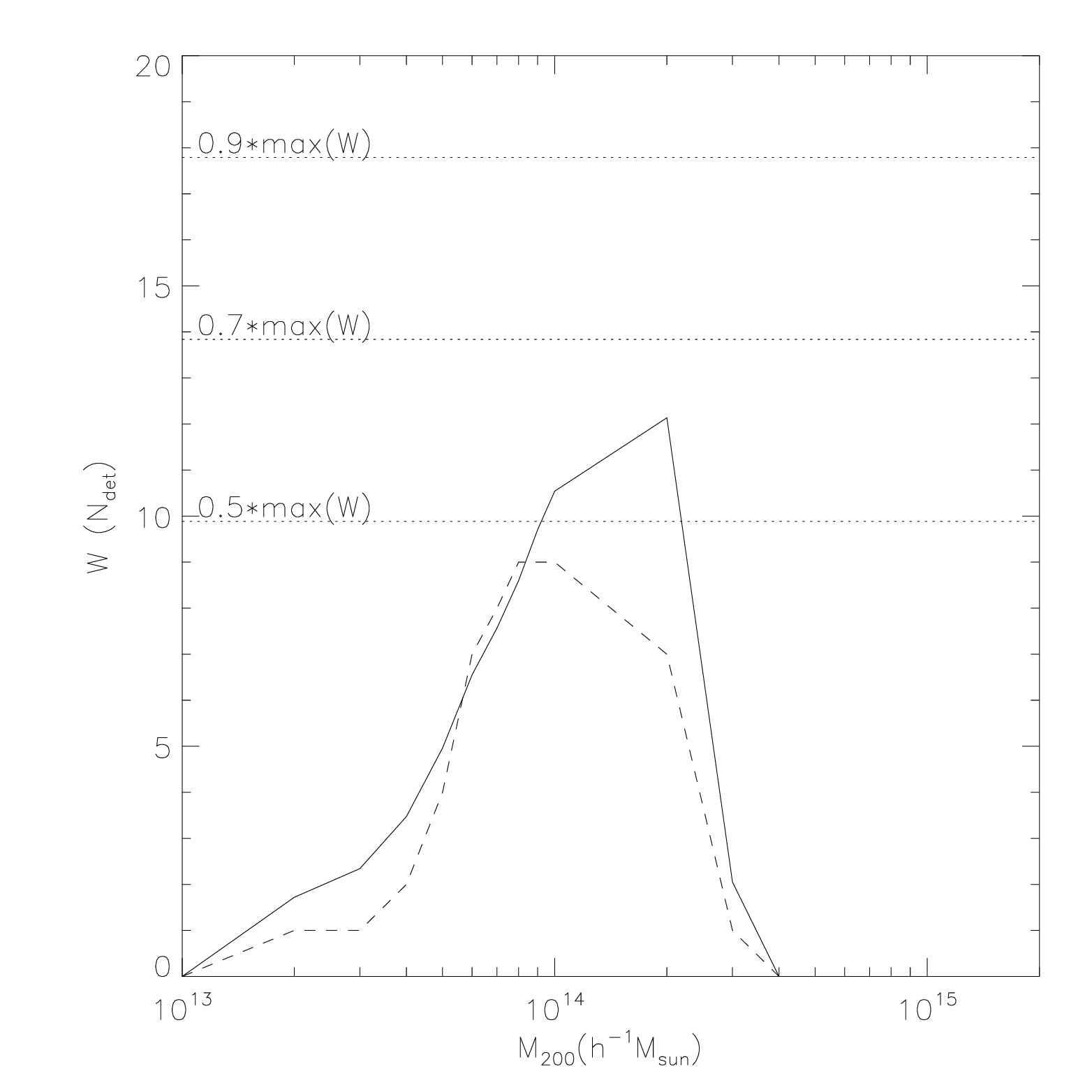}
\caption{The pseudo-likelihood distributions for NFW (top panel) and SIS (bottom panel) models for peak 3. In the top panel, filled contours indicate ${\cal W}$  at contour levels equal to $[0.1, 0.3, 0.5, 0.7,0.9]\times{\cal W}_{\rm max}$, with the model corresponding to the peak ${\cal W}$ in the NFW distribution denoted by a white asterisk. The solid red curve shows ${\cal W}=0$. Corresponding levels are shown in the bottom panel as dotted horizontal lines. The dotted and dashed contours in the top panel indicate $N_{\rm det}$, represented in the bottom panel as the dashed curve.  \label{fg:pk3prob}}
\end{figure}

As expected, the peak 3 data favour a lower-mass halo than those of peak 1, and imply an overall mass for the substructure system of $M_{200}\sim 5-7\times10^{14}h^{-1}M_\odot$, consisting of two prominent sub-clumps. It is clear that the use of a non-circular model and the presence of wider-field data are needed to constrain the mass distribution more accurately.

\subsubsection{Peak 2}

\begin{figure}
\includegraphics[width=0.45\textwidth]{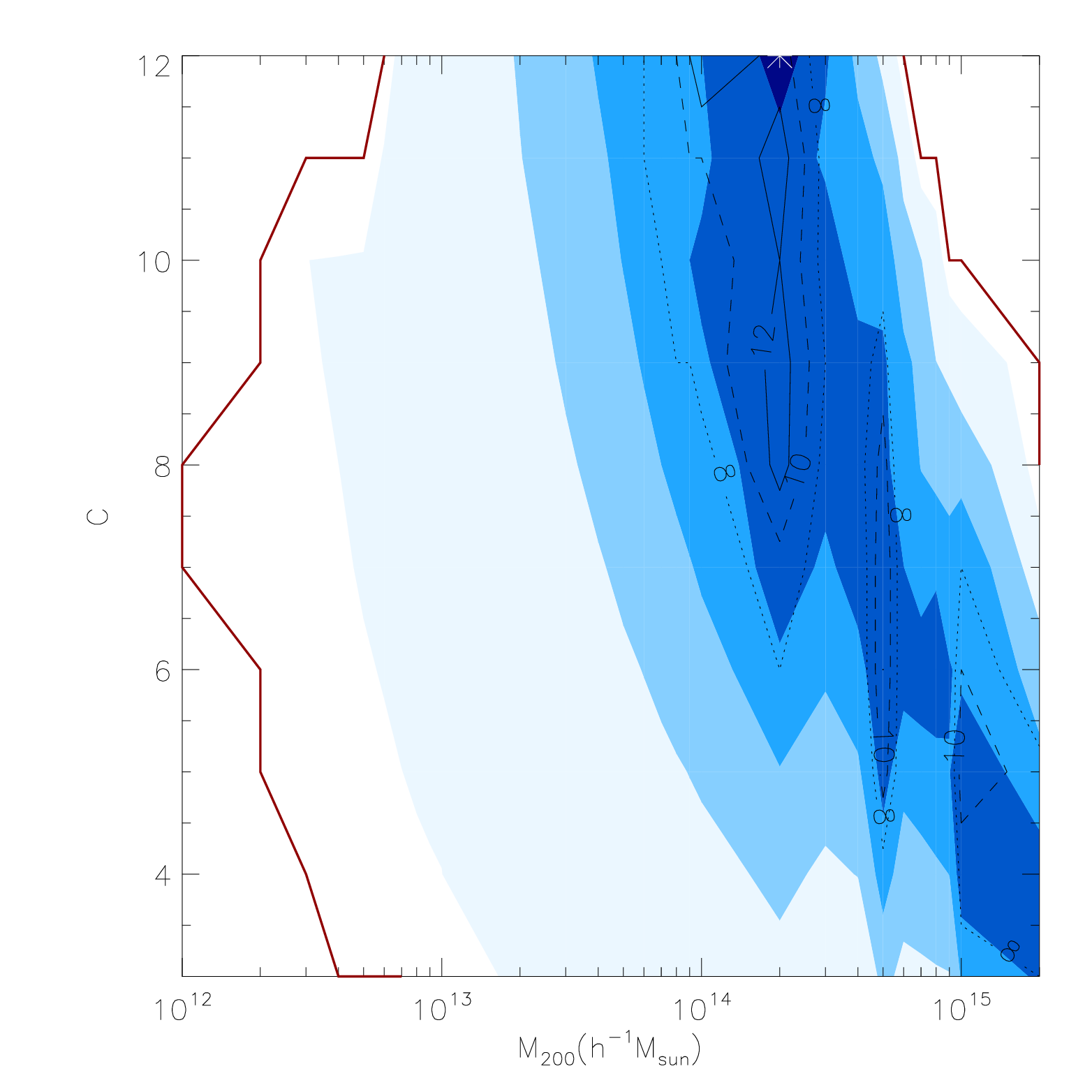}
\includegraphics[width=0.45\textwidth]{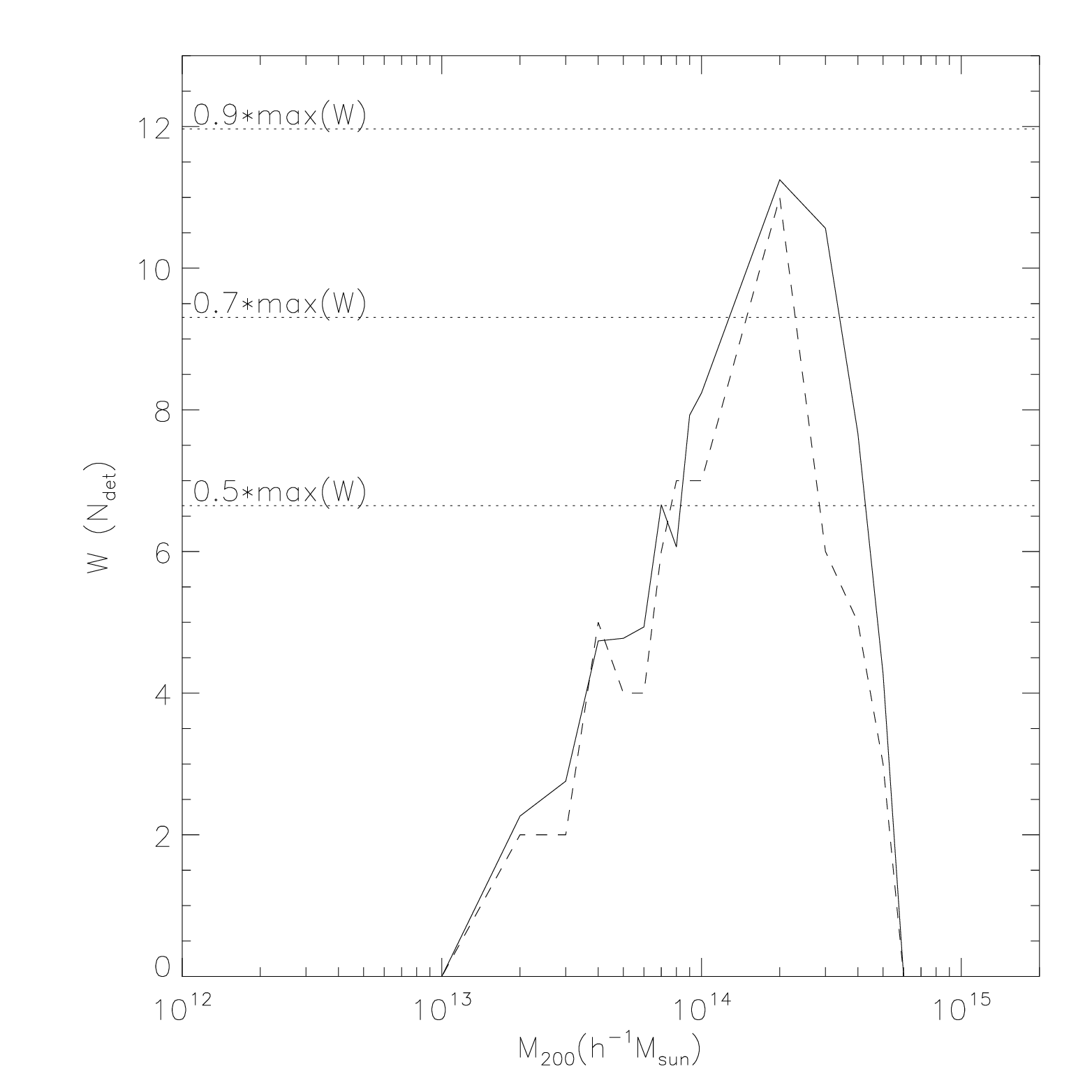}
\caption{The pseudo-likelihood distributions for NFW (top panel) and SIS (bottom panel) models for peak 2. In the top panel, filled contours indicate ${\cal W}$ at contour levels equal to $[0.1, 0.3, 0.5, 0.7,0.9]\times{\cal W}_{\rm max}$, with the model corresponding to the peak ${\cal W}$ in the NFW distribution denoted by a white asterisk. The solid red curve shows ${\cal W}=0$. Corresponding levels are shown in the bottom panel as dotted horizontal lines. The dotted and dashed contours in the top panel indicate $N_{\rm det}$, represented in the bottom panel as the dashed curve.  \label{fg:pk2prob}}
\end{figure}

As before, but to a lesser extent, in the case of peak 2 we find a rather complicated structure, with the peak breaking down into 2-3 sub-peaks in four of the $\fmap$ reconstructions. However, a B-mode signal appears in only one reconstruction ($\ell=3,\ R=90^{\prime\prime}$). Excluding this reconstruction, and treating peak 2 as a single halo, we compute the ${\cal W}$ pseudo-probability distribution as before. The results are shown in Figure\,\ref{fg:pk2prob}. 

For this peak, we find a best-fit NFW halo with $M_{200}=2\times10^{14}h^{-1}M_\odot$ and $c=12$, with ${\cal W}_{\rm max}=13.30$ and $N_{\rm det}=11$. Consideration of the $N_{\rm det}$ contours in this case suggests that a lower concentration parameter (perhaps as low as 8) may be compatible for the same $M_{\rm 200}$. The best fit SIS model yields the same $M_{200}$ and $N_{\rm det}$, with ${\cal W}_{\rm max}=11.25$. Note that we see a tail in the NFW probability distribution towards high-mass, low-$c$ models, which does not appear in the SIS distribution. This indicates that substructure may play a role in overestimating the zero-signal contour, thus giving rise to the bias. It is important to note, however, that this does not affect our overall mass estimate, even when the constraint on $R_0$ is removed. 

\subsubsection{Peak 6}

\begin{figure}
\includegraphics[width=0.45\textwidth]{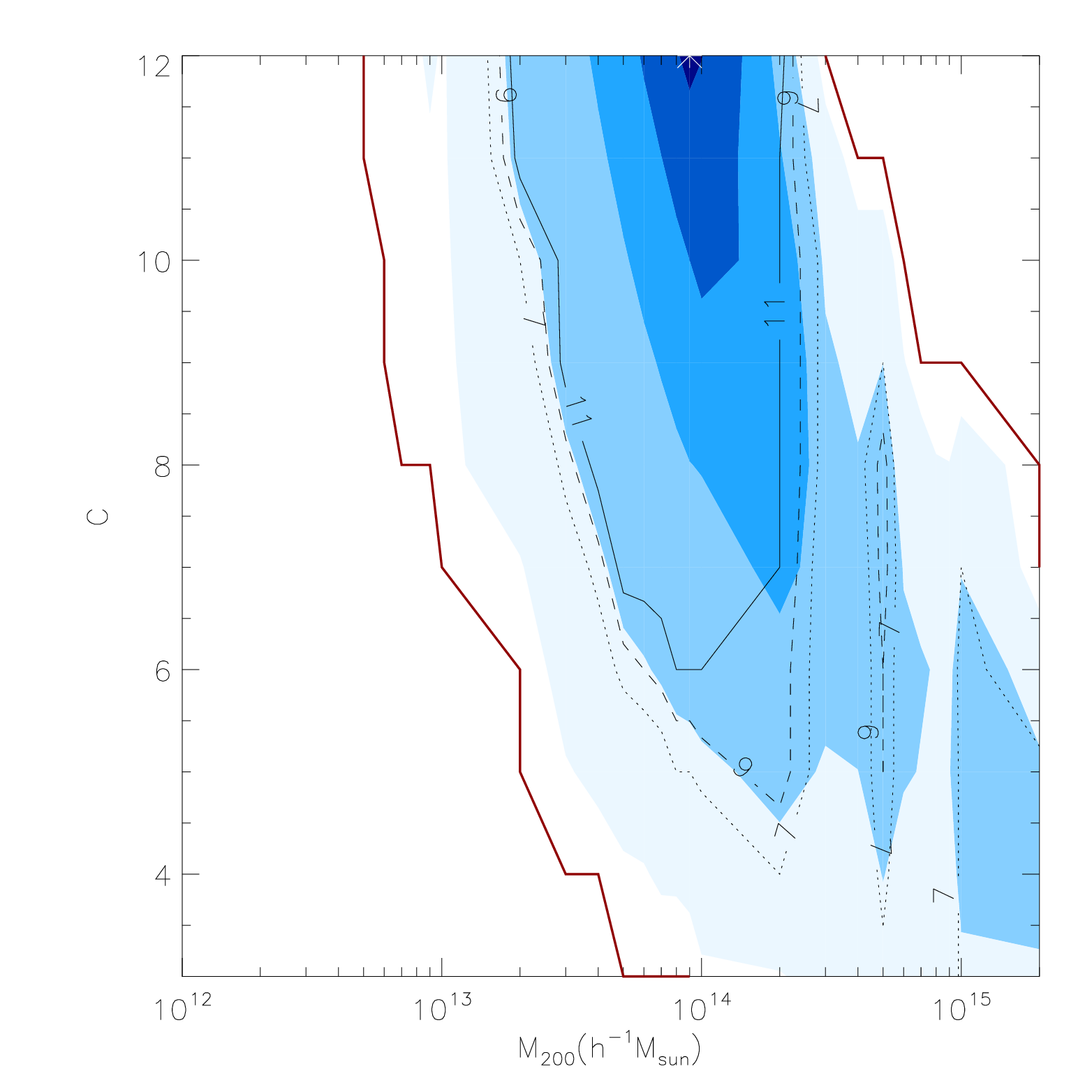}
\includegraphics[width=0.45\textwidth]{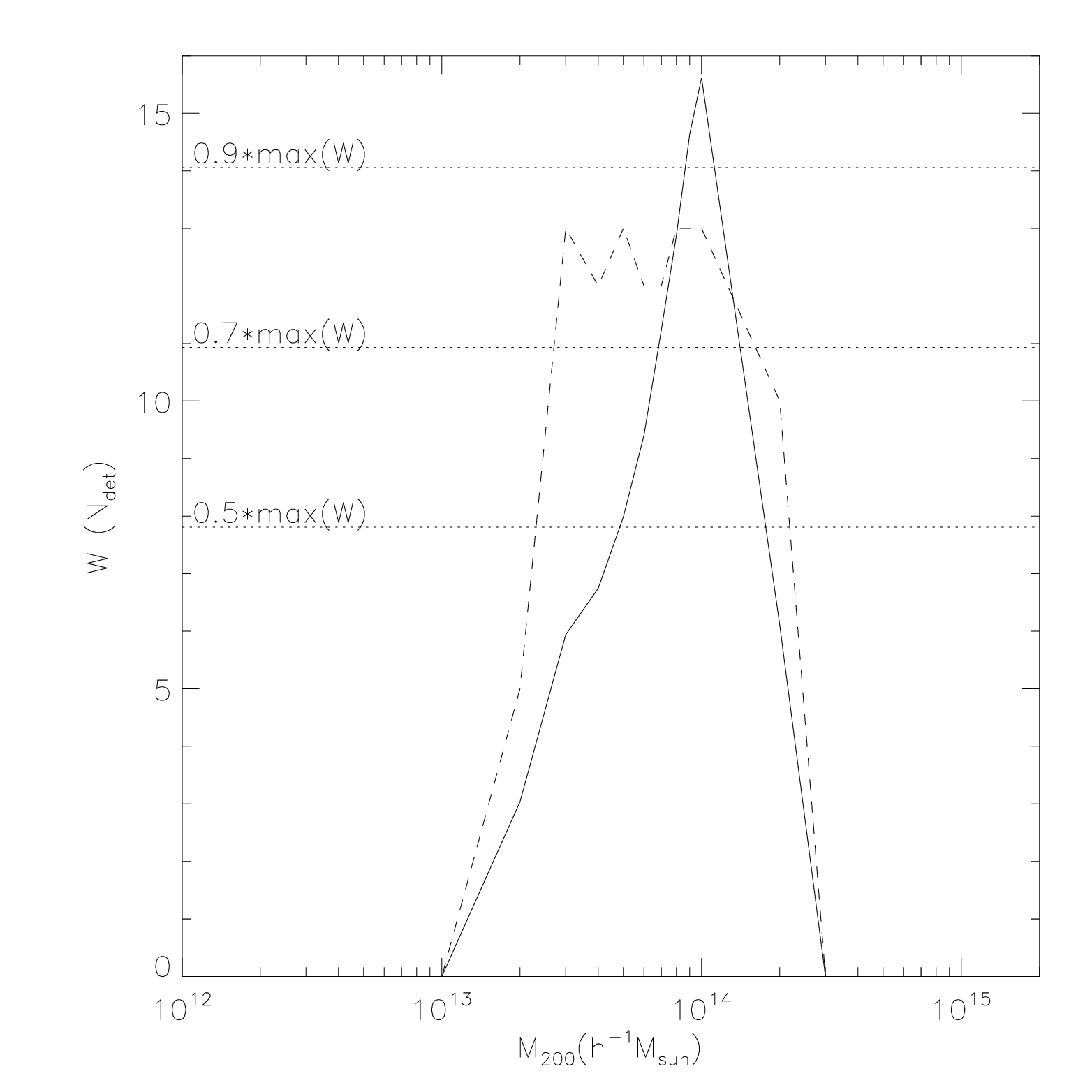}
\caption{The pseudo-likelihood distributions for NFW (top panel) and SIS (bottom panel) models for peak 6. In the top panel, filled contours indicate ${\cal W}$  at contour levels equal to $[0.1, 0.3, 0.5, 0.7,0.9]\times{\cal W}_{\rm max}$, with the model corresponding to the peak ${\cal W}$ in the NFW distribution denoted by a white asterisk. The solid red curve shows ${\cal W}=0$. Corresponding levels are shown in the bottom panel as dotted horizontal lines. The dotted and dashed contours in the top panel indicate $N_{\rm det}$, represented in the bottom panel as the dashed curve.  \label{fg:pk6prob}}
\end{figure}

This peak is the lowest signal to noise peak considered here, and does not appear in the $\ell=3$, $R=90^{\prime\prime}$ reconstruction, indicating it is likely to be a relatively low-mass structure with a small physical size (or high concentration parameter). In contrast to the other peaks, peak 6 shows no complicated structure, and is likely to be well-fit by a circular model. Furthermore, it shows no B-mode signal in any of the 15 $\fmap$ reconstructions in which it is detected. The pseudo-probability distribution for this peak is shown in Figure\,\ref{fg:pk6prob}.

We find that this peak is best fit by an SIS model with $M_{200}=1\times10^{14}h^{-1}M_\odot$ (${\cal W}_{\rm max}=15.62,\ N_{\rm det}=13$. The best-fit NFW model has $M_{200}=9\times10^{13}h^{-1}M_\odot,\ c=12$, with only slightly lower ${\cal W}_{\rm max}=14.66$ and $N_{\rm det}=12$. Thus, a highly concentrated halo with mass $\sim 10^{14}h^{-1}M_\odot$ is clearly implied with a fairly narrow spread of allowed masses, and a clear preference to high concentration parameters (perhaps in excess of the maximum $c=12$ sampled by our template models). 

\subsection{Comparison of best-fit models to data}

In order to demonstrate how well our `best-fit' models, as identified by the ${\cal W}$ statistic, actually fit the data, it is helpful to compare directly the measured radial profile with the best-fit SIS and NFW model for each peak.

\begin{figure*}
\center
\includegraphics[width=0.25\textwidth]{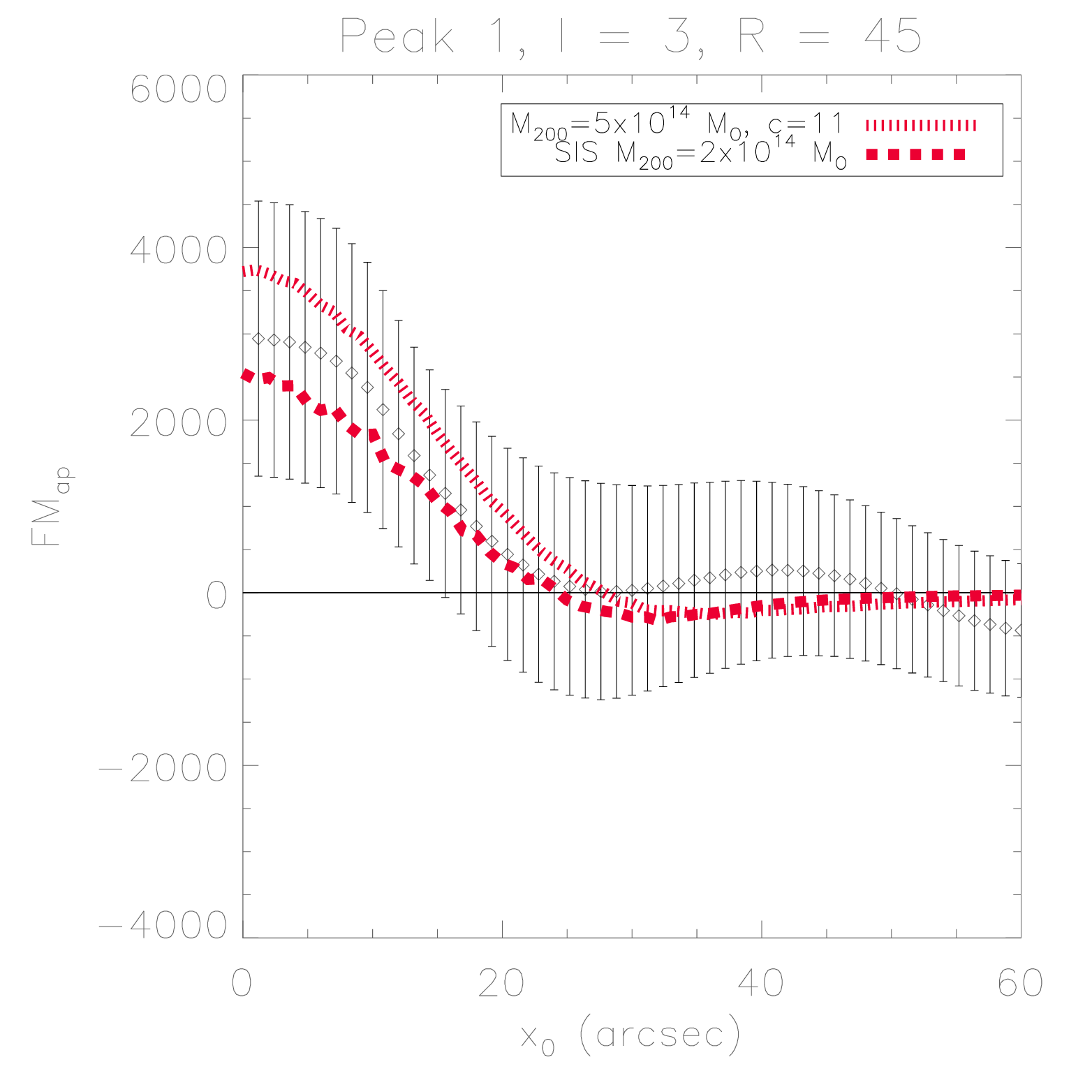}\includegraphics[width=0.25\textwidth]{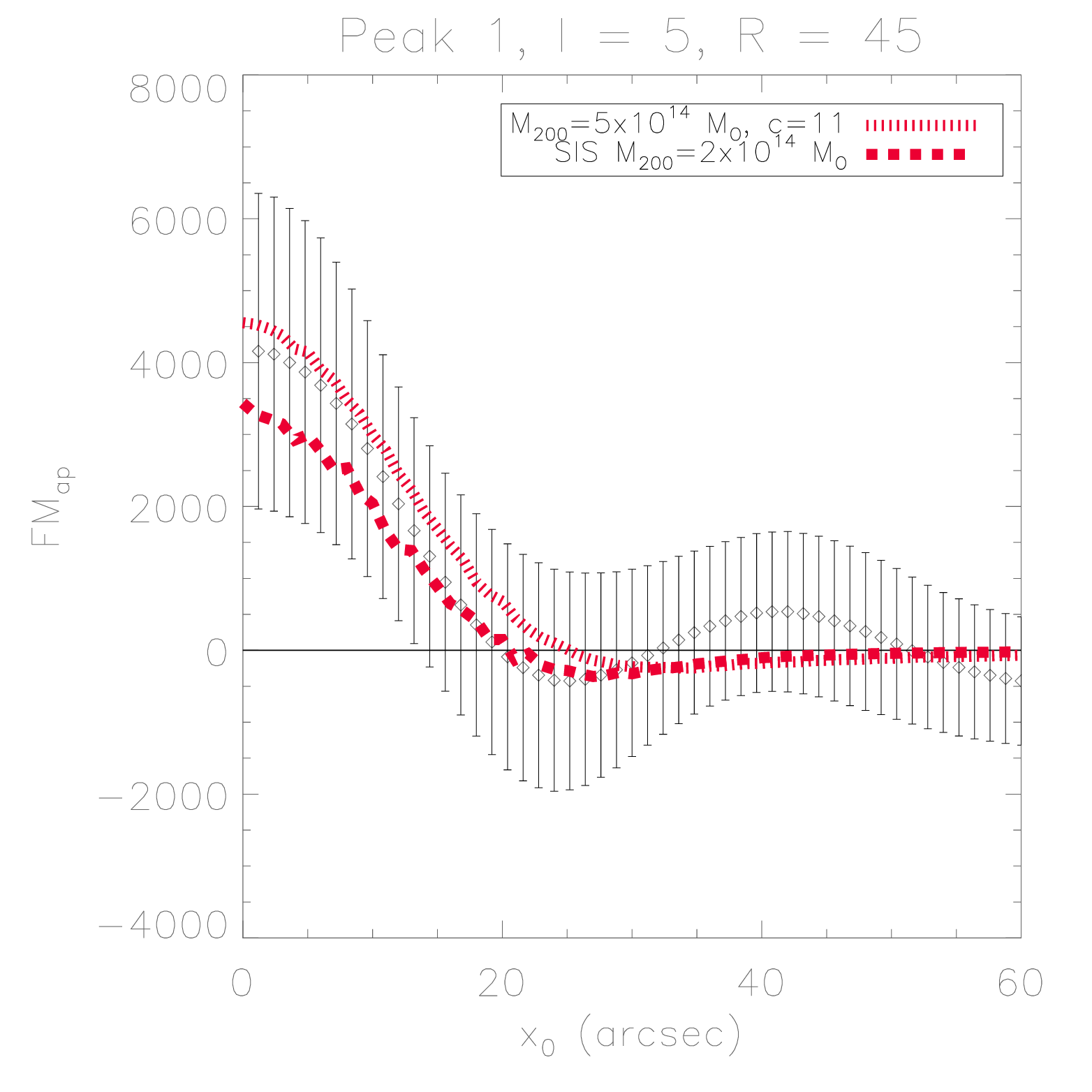}\includegraphics[width=0.25\textwidth]{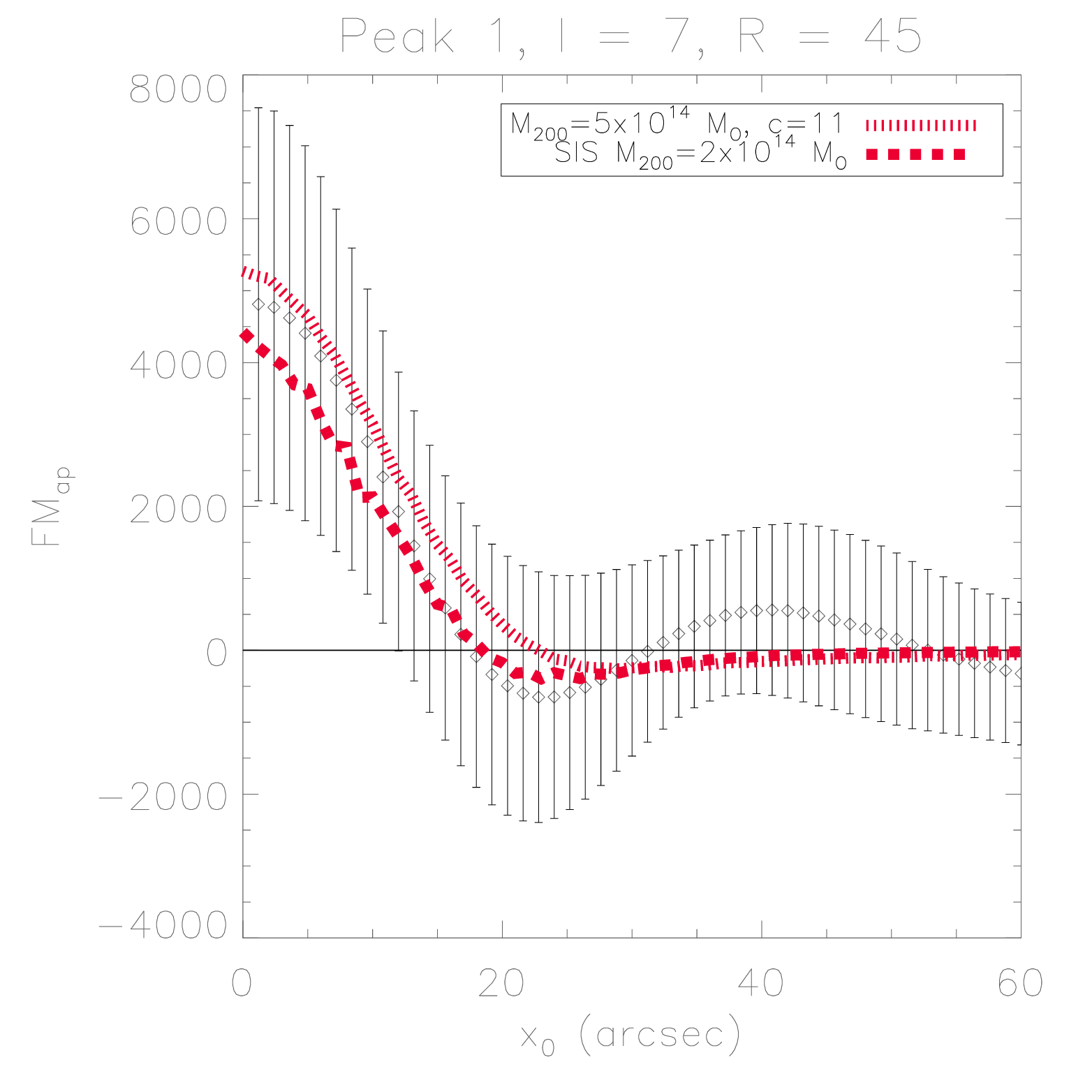}\includegraphics[width=0.25\textwidth]{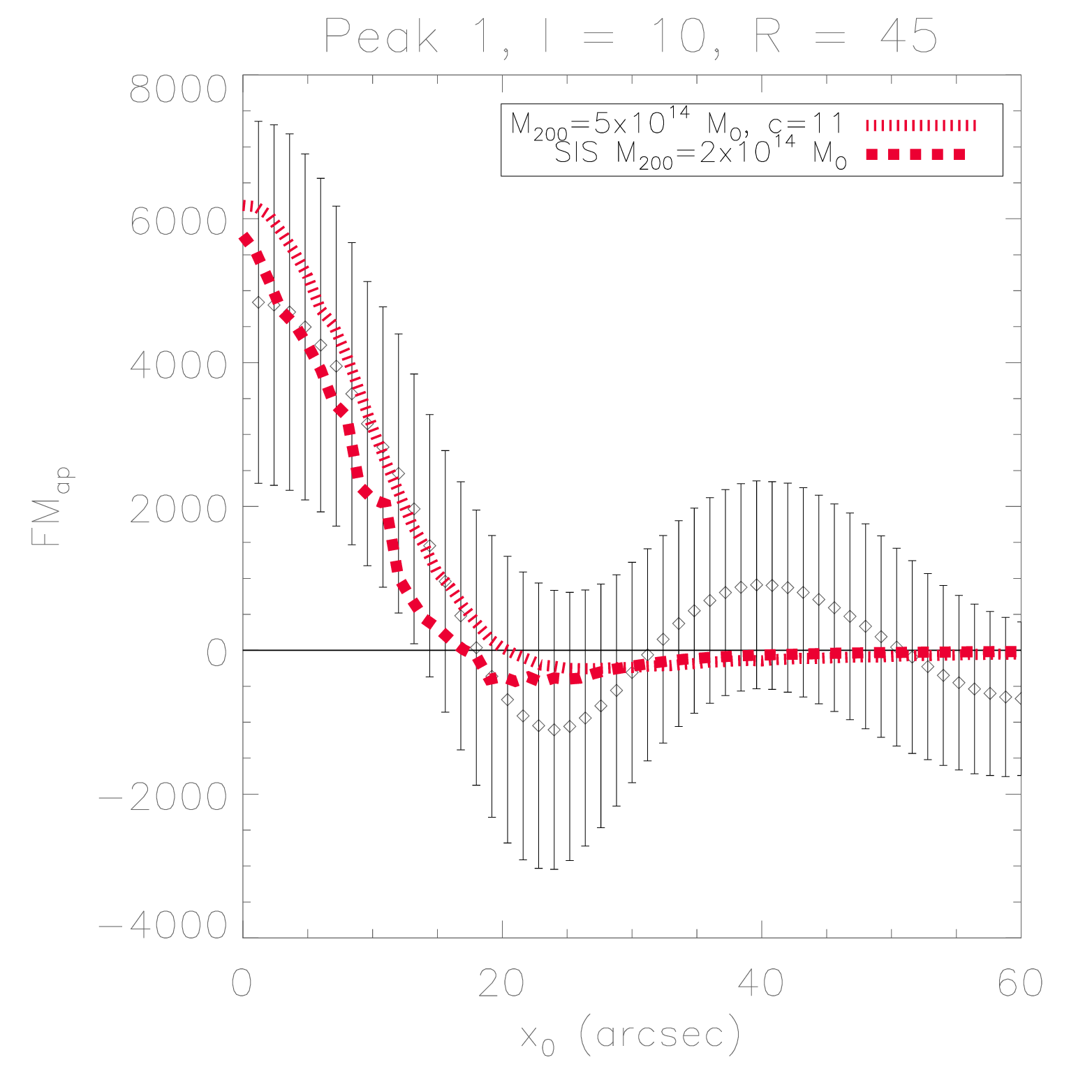}\vspace{15pt}
\includegraphics[width=0.25\textwidth]{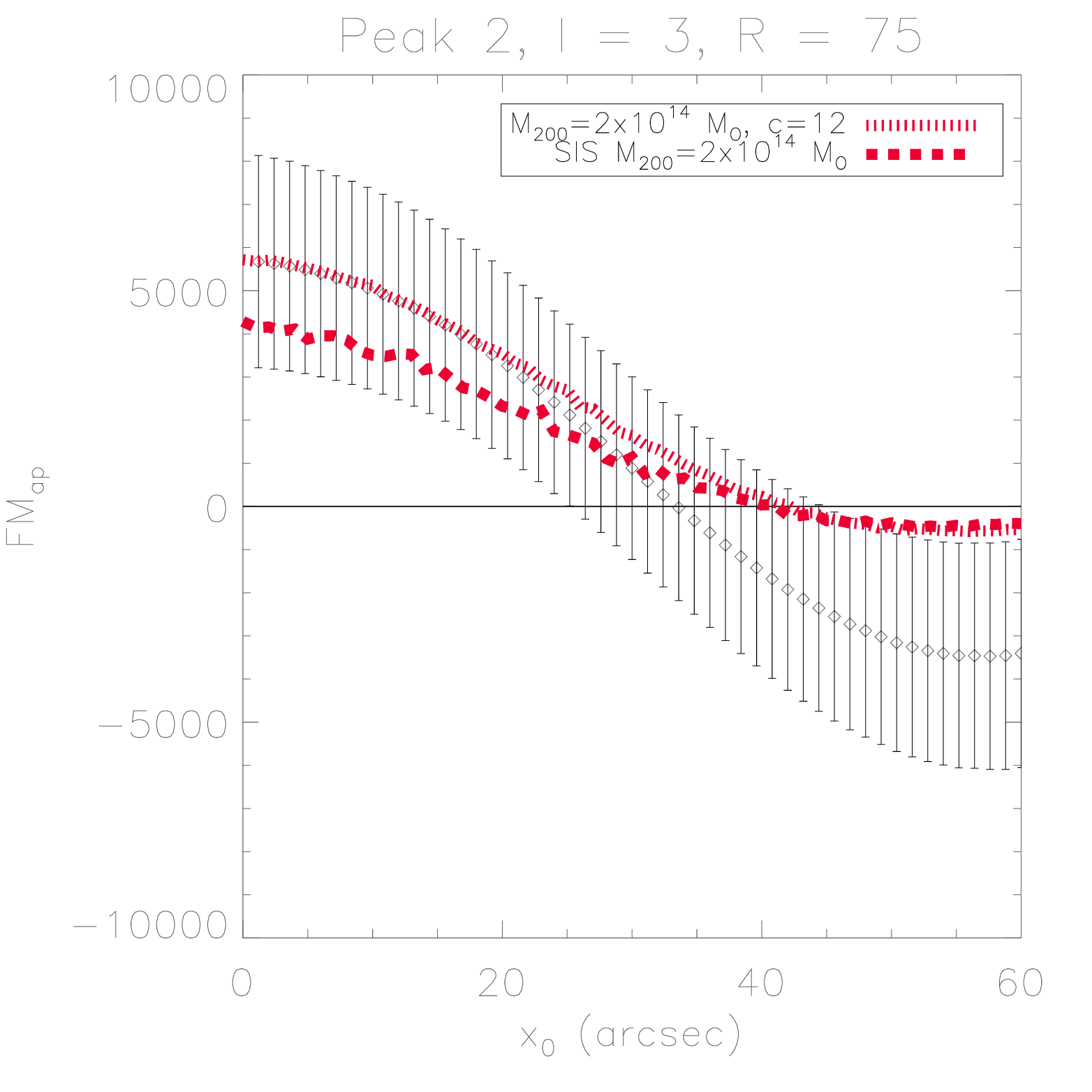}\includegraphics[width=0.25\textwidth]{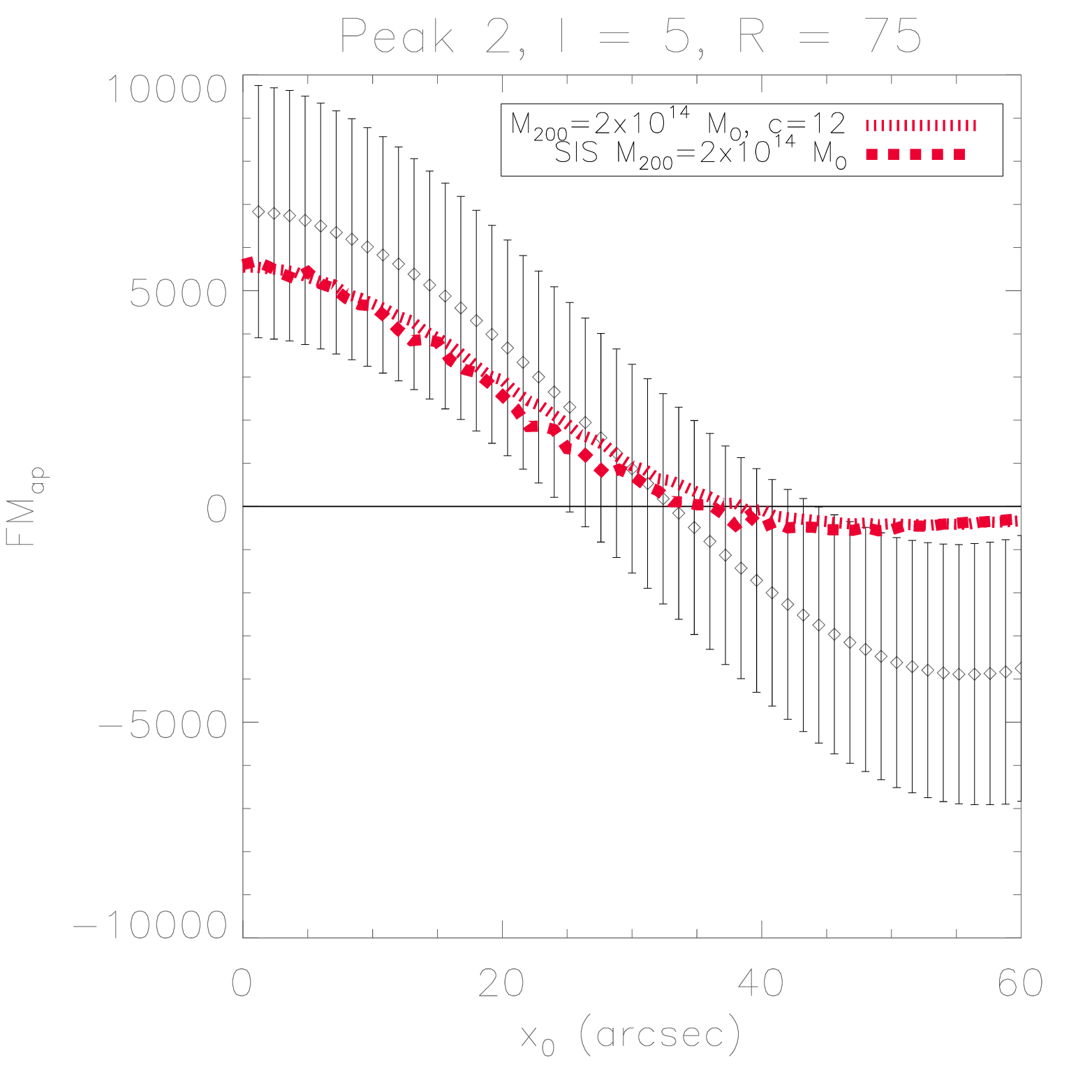}\includegraphics[width=0.25\textwidth]{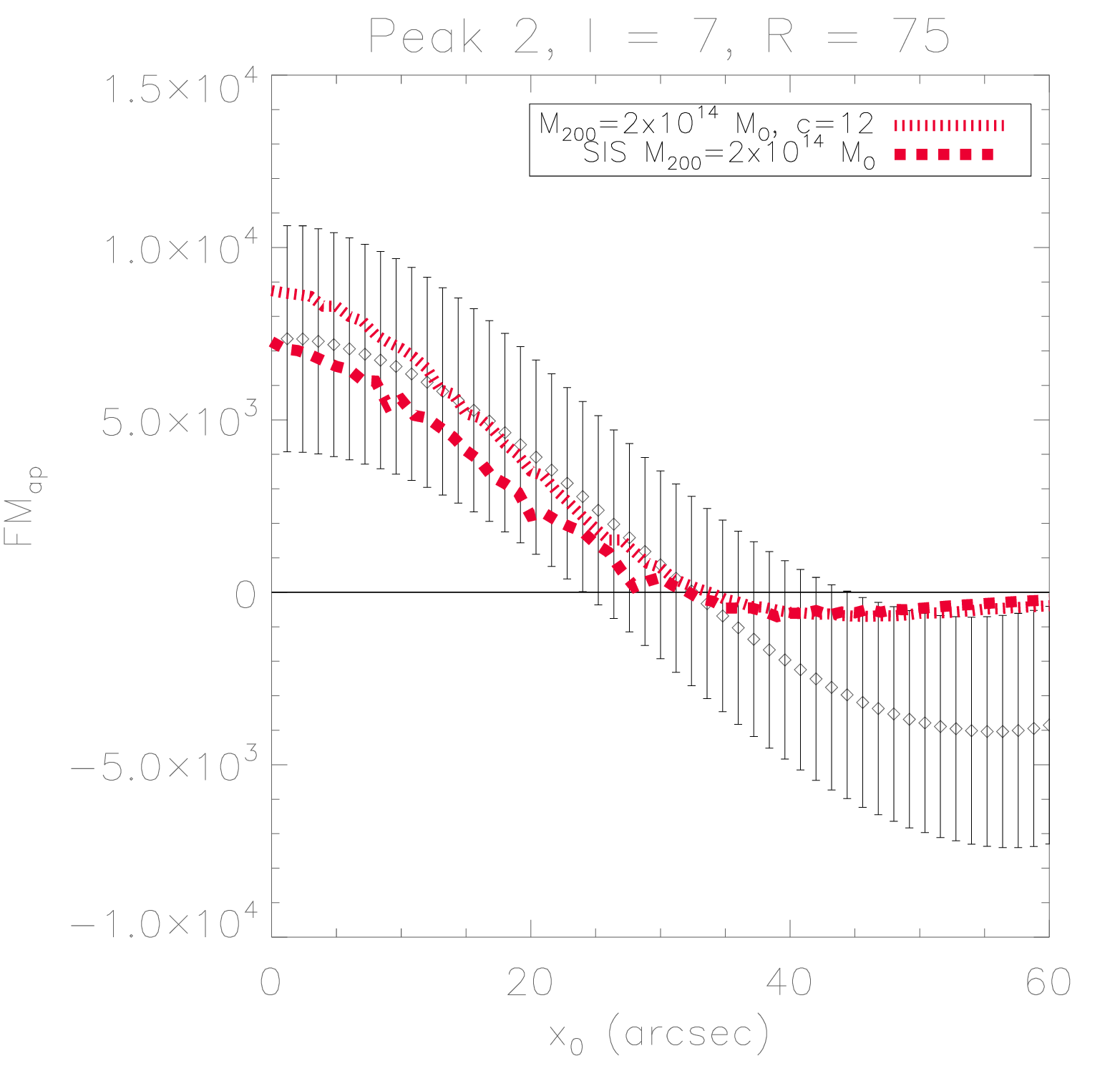}\includegraphics[width=0.25\textwidth]{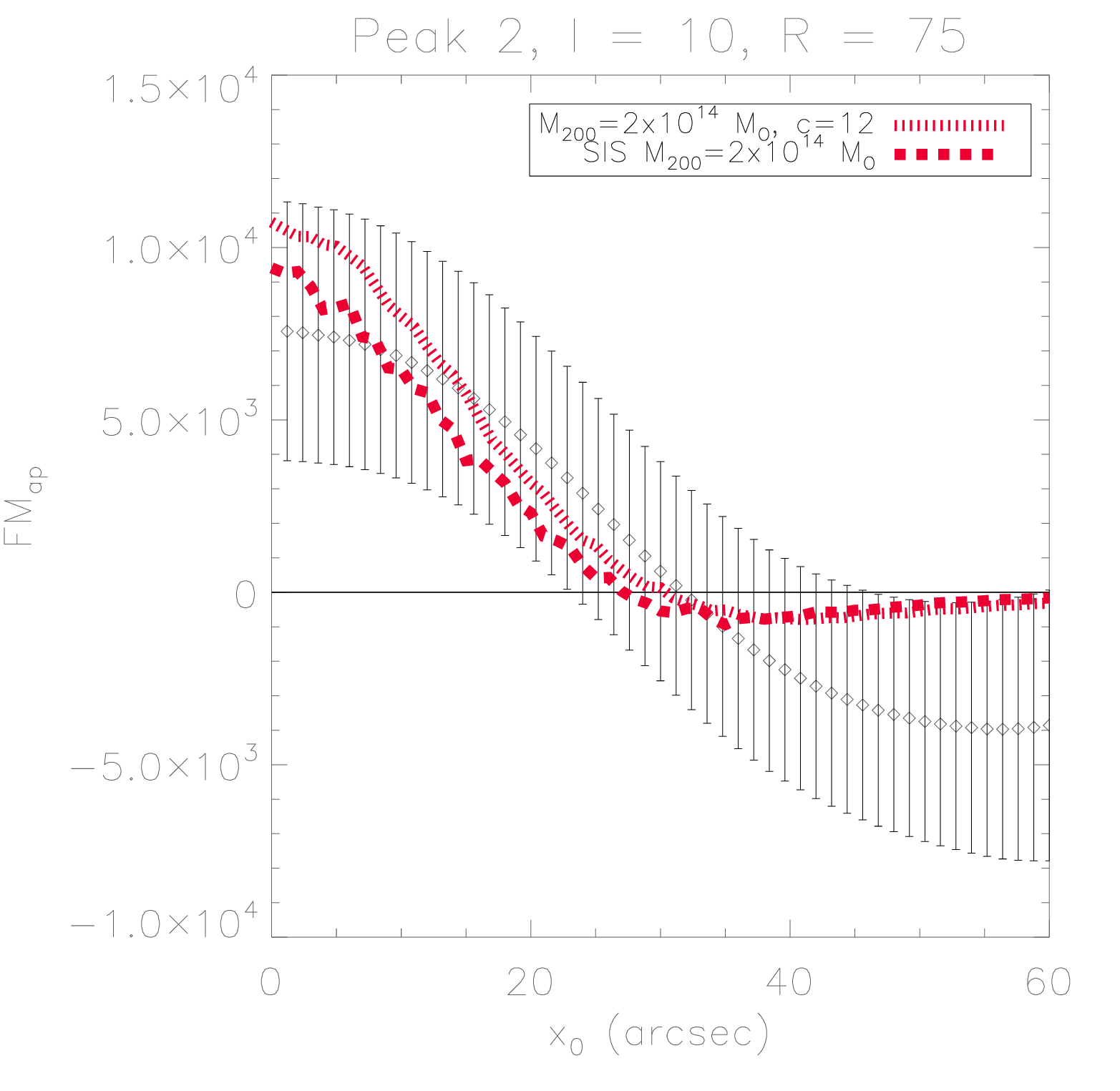}\vspace{15pt}
\includegraphics[width=0.25\textwidth]{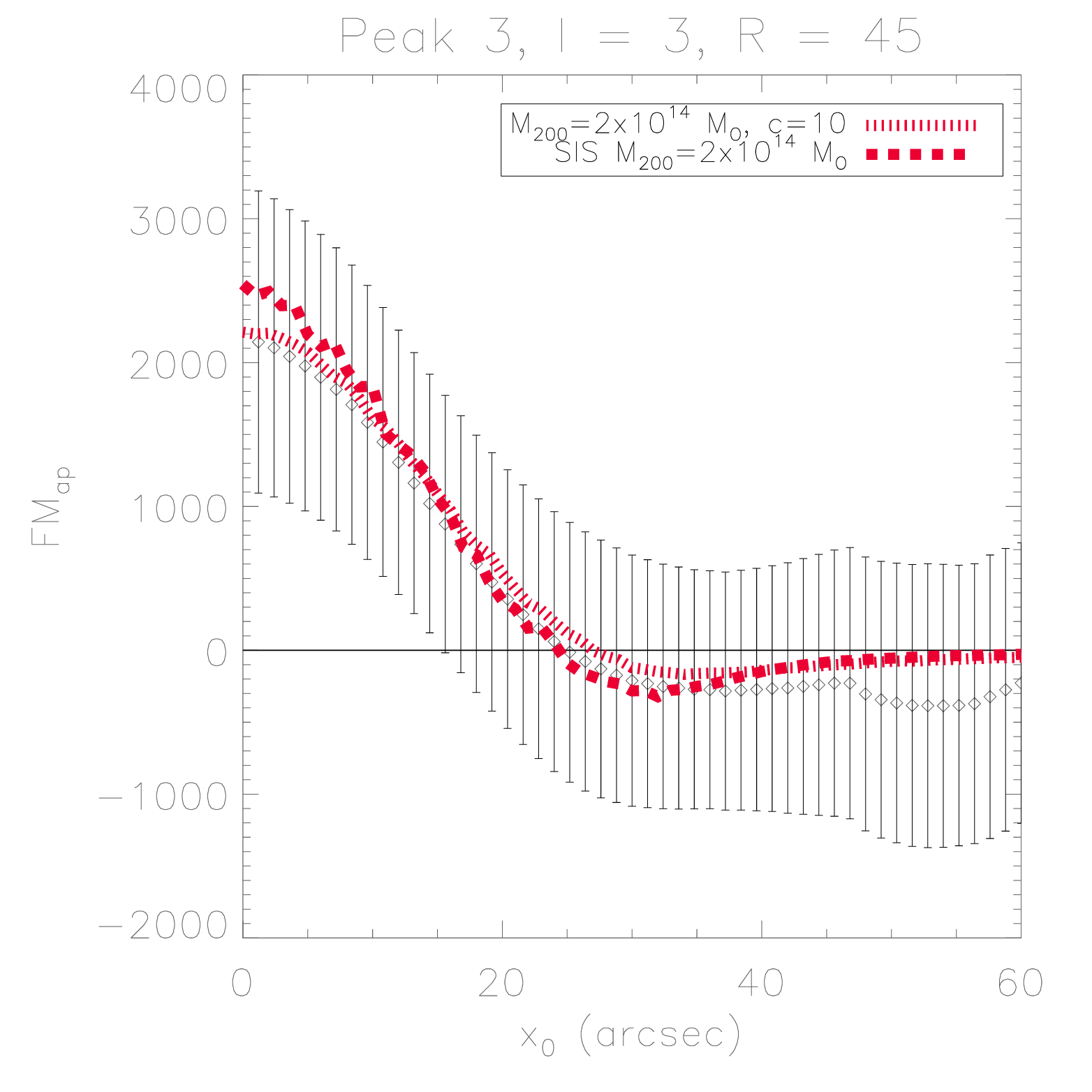}\includegraphics[width=0.25\textwidth]{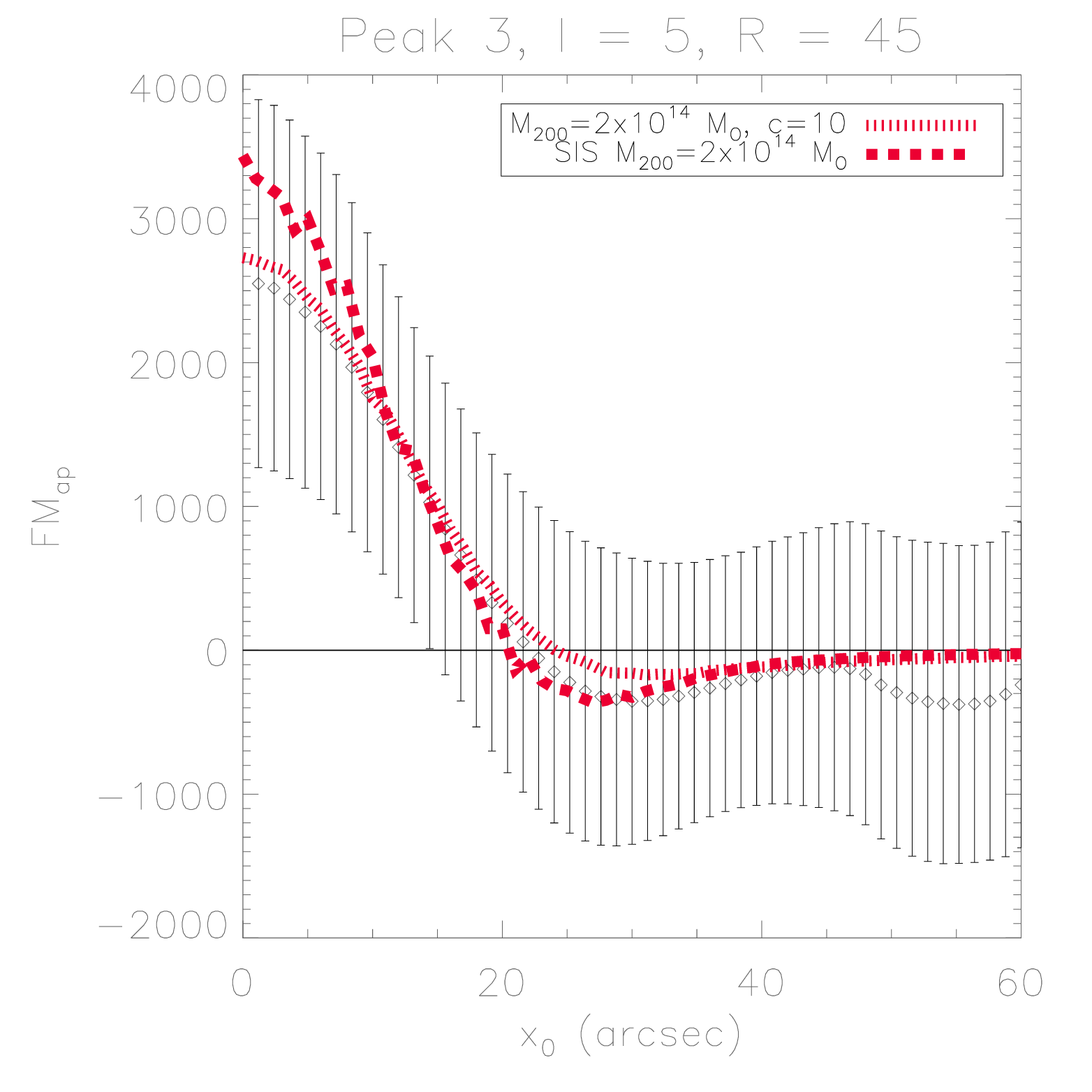}\includegraphics[width=0.25\textwidth]{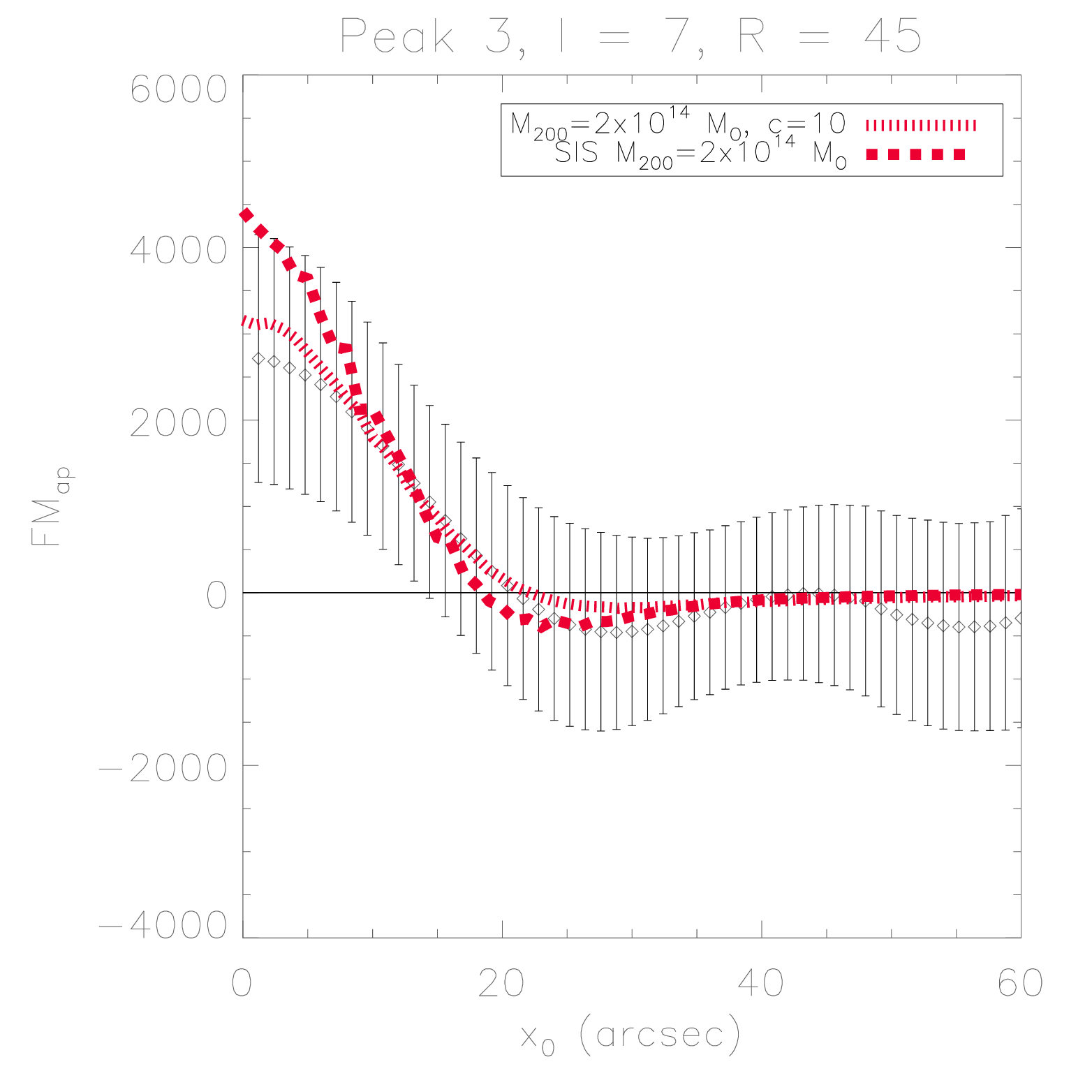}\includegraphics[width=0.25\textwidth]{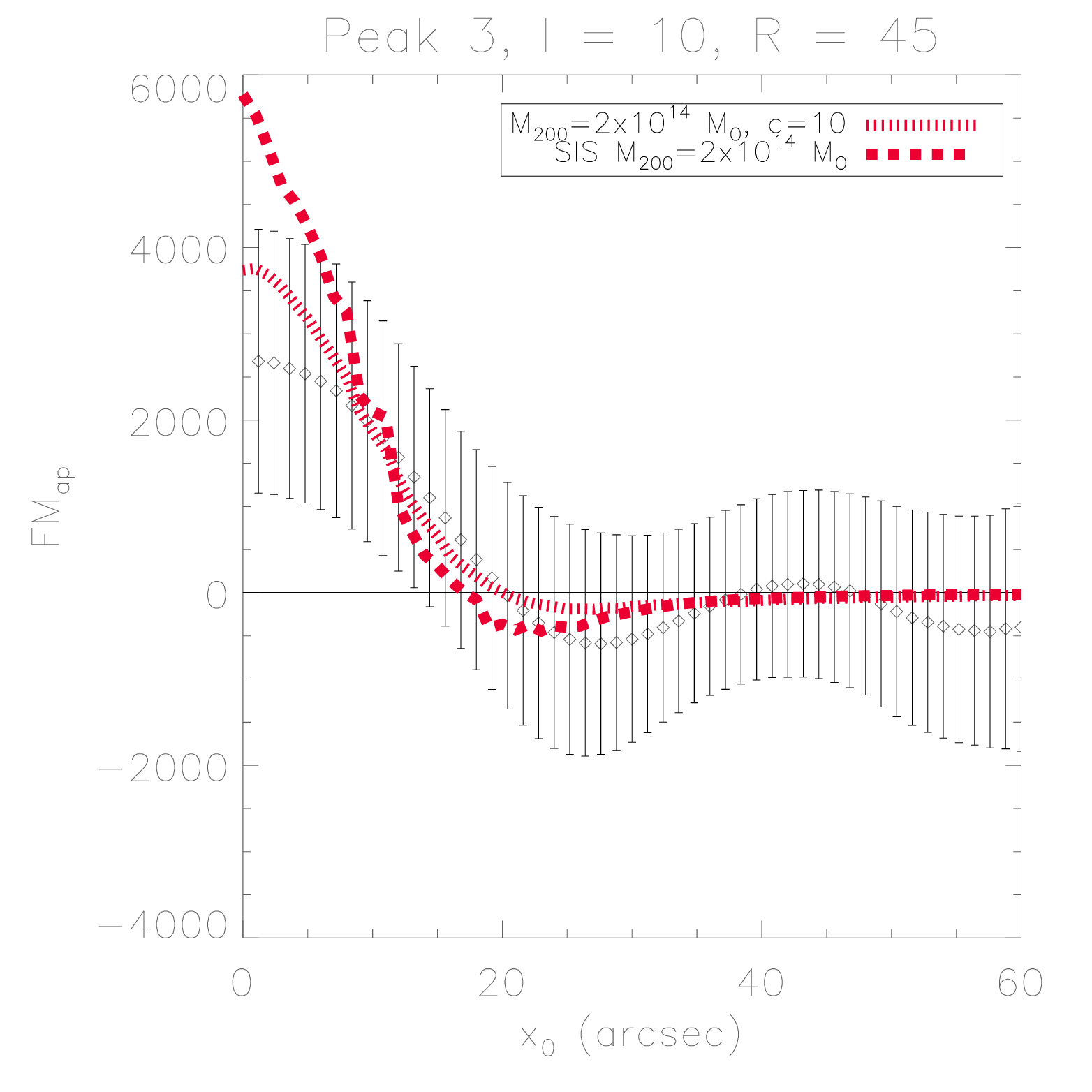}\vspace{15pt}
\includegraphics[width=0.25\textwidth]{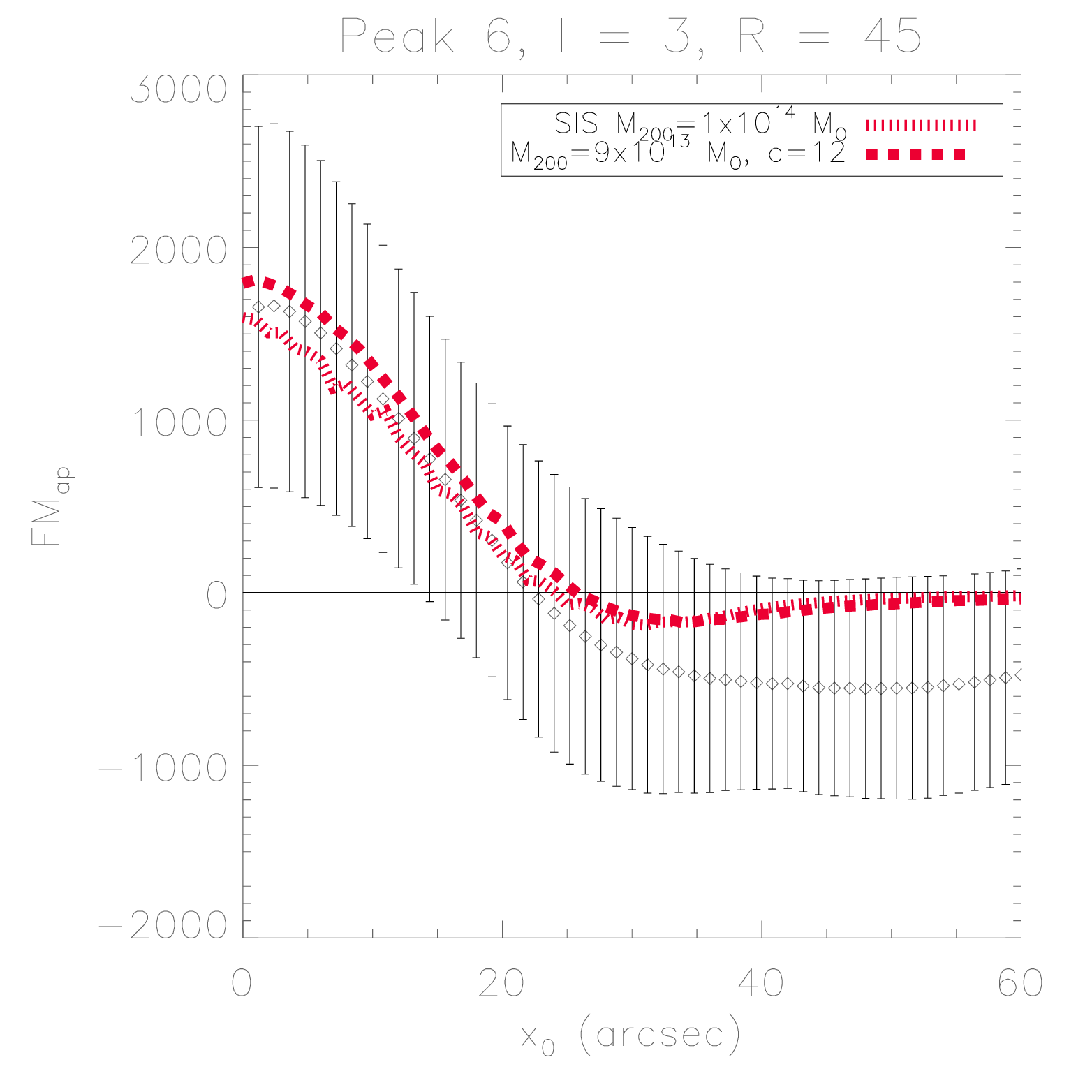}\includegraphics[width=0.25\textwidth]{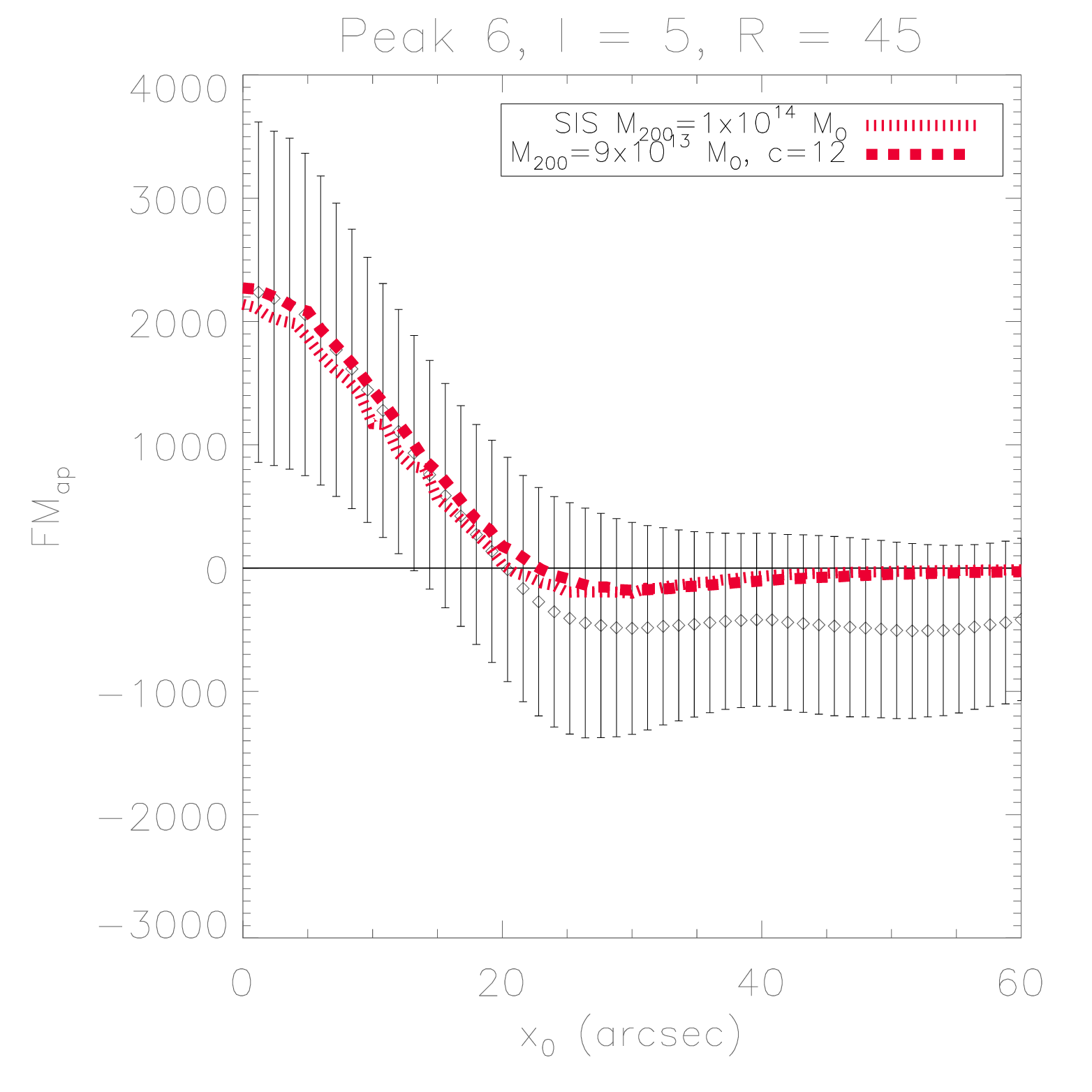}\includegraphics[width=0.25\textwidth]{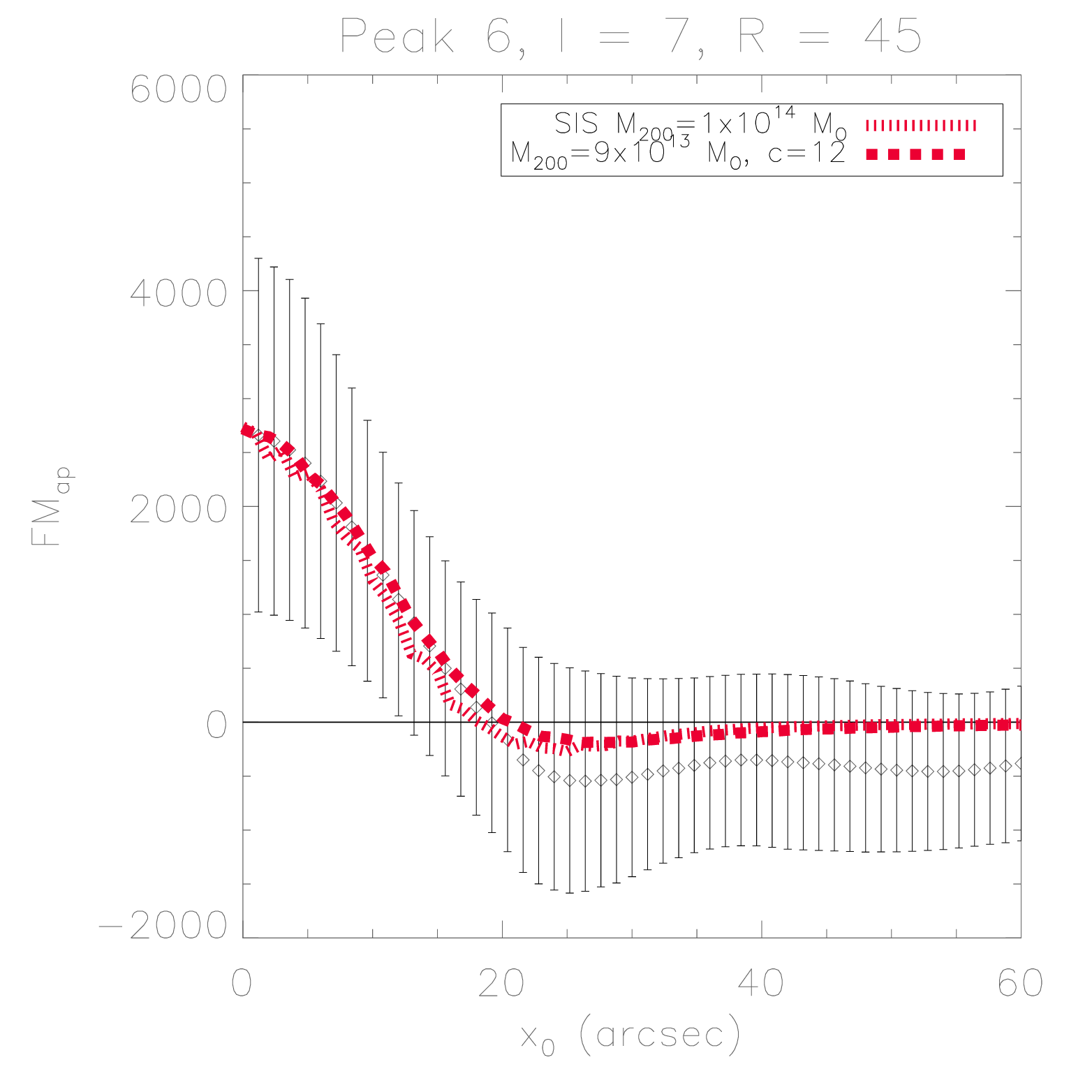}\includegraphics[width=0.25\textwidth]{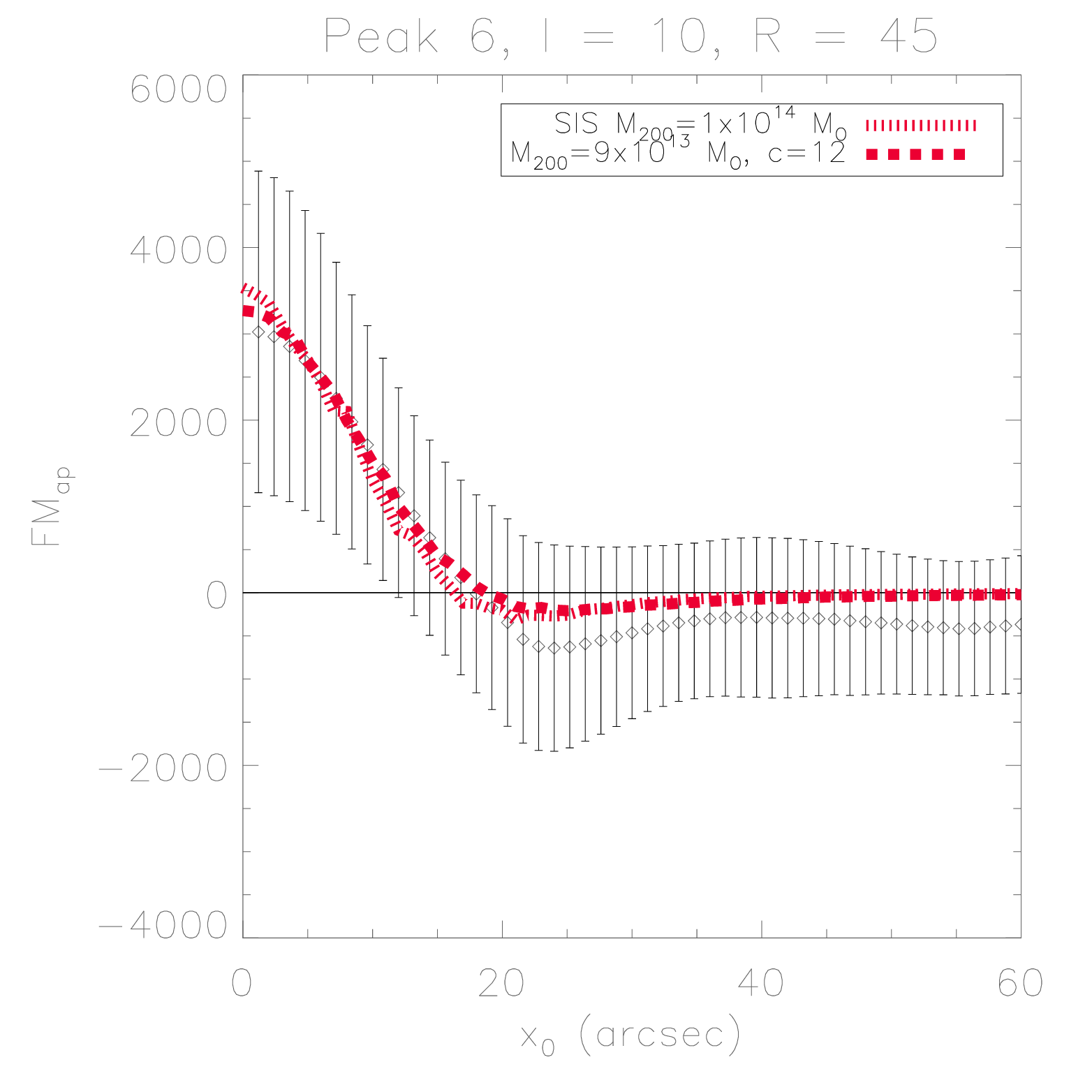}\vspace{15pt}
\caption{The figure shows a representative sample of radial profiles for the four detected peaks. Each row represents a single peak, $\ell$ increases left to right, and $R$ is fixed in each row. the best-fit models are overplotted as dashed red lines.\label{fg:radprofs}}
\end{figure*}

As a representative sample of our results, we plot in Figure\,\ref{fg:radprofs} the radial profiles for all peaks for $\ell = [3, 5, 7, 10]$ and $R=45^{\prime\prime}$, with the exception of peak 2, for which we use $R=75^{\prime\prime}$. 
This is because this peak breaks down into smaller sub-peaks in the $R=45^{\prime\prime}$ reconstructions. In these cases, the radial profile strongly under-estimates the peak signal and central regions of the profile, as a result of the fact that the centroid used does not correspond to the peak signal-to-noise for any of the sub-clumps. As a result, these reconstructions have few (if any) models compatible with their values of $m_{\rm peak}$ and $R_0$, and are not included in the calculation of the ${\cal W}$ statistic. It is for this reason that $N_{\rm det}$ is substantially lower than the total number of reconstructions of this peak. 

For clarity, Table\,\ref{tab:props} below details the properties of the 4 peaks discussed above in each of the 16 $\fmap$ reconstructions. 

It is clear from Figure\,\ref{fg:radprofs} that the best-fit models not only fit individual radial profiles well, but also track the evolution of the profile under variation of $\ell$ or $R$ consistently, indicating that the method is very effective at model-selection.

\begin{table*}
\centering
\begin{tabular}{c c c c c c l}
\hline
Peak & $\ell$ & $R$ (arcsec) & ${\cal S}$ at centroid &B-modes? & Fits model? & Comments\\
\hline\hline
1 & 3 & 45 & 1.85 & \ding{55} & \checkmark & \\
   &     & 60 & 1.58 & \checkmark & n/a & blended with peak 3\\
   &     & 75 & 2.06 &\checkmark & n/a & blended with peak 3\\
   & & 90 & 2.39 &\checkmark & n/a & blended with peak 3\\
   & 5 & 45 & 1.90 &\ding{55}  & \checkmark & \\
   & & 60 & 1.72 &\checkmark & n/a & \\
   & & 75 & 1.71 &\checkmark & n/a& blended with peak 3\\
   & & 90 & 2.12 &\checkmark & n/a & blended with peak 3\\
   & 7 & 45 & 1.76 &\ding{55} & \checkmark & 2 sub-clumps, centroid coincident with clump $a$\\
   & & 60 & 1.84 &\ding{55} & \checkmark & \\
   & & 75 & 1.79 &\checkmark & n/a & \\
   & & 90 & 1.85 &\checkmark & n/a & blended with peak 3\\
   & 10 & 45 & 1.92 &\ding{55} & \checkmark & 3 sub-clumps, centroid coincident with clump $a$\\
   & & 60 & 1.90 &\ding{55} & \checkmark & \\
   & & 75 & 1.76 &\checkmark & n/a & \\
   & & 90 & 1.55 &\checkmark &n/a & blended with peak 3\\
   \hline
2 & 3 & 45 & 1.75 &\ding{55} & \checkmark & \\
 & & 60 & 2.34 &\ding{55} & \checkmark & \\
 & & 75 & 2.31 &\ding{55} & \checkmark & \\
 & & 90 & 2.06 &\checkmark & n/a & \\
 & 5 & 45 & 1.40 &\ding{55} & \ding{55} & 2 sub-clumps, centroid not coincident with any clump \\
 & & 60 & 2.09 &\ding{55} & \checkmark & \\
 & & 75 & 2.34 &\ding{55} & \checkmark & \\
 & & 90 & 2.26 &\ding{55} & \checkmark & \\
 & 7 & 45 & 1.29 &\ding{55} & \ding{55} & 3 sub-clumps, centroid not coincident with any clump \\
 & & 60 & 1.82 &\ding{55} & \checkmark & \\
 & & 75 & 2.24 &\ding{55} & \checkmark & \\
 & & 90 & 2.32 &\ding{55} & \checkmark & \\
 & 10 & 45 & 1.16 &\ding{55} & \ding{55} & 3 sub-clumps, centroid not coincident with any clump \\
 & & 60 & 1.50 &\ding{55} & \ding{55} & 2 sub-clumps, centroid not coincident with any clump \\
 & & 75 & 2.02 &\ding{55} & \checkmark & \\
 & & 90 & 2.27 &\ding{55} & \checkmark & \\
 \hline 
3 & 3 & 45 & 2.04 &\ding{55} & \checkmark & \\
  & 5 & 45 & 1.99 &\ding{55} & \checkmark & \\
 & & 60 & 1.95 &\ding{55} & \checkmark & \\
 & 7 & 45 & 1.89 &\ding{55} & \checkmark & \\
 & & 60 & 2.01 &\ding{55} & \checkmark & \\
 & & 75 & 1.86 &\ding{55} & \checkmark & \\
 & 10 & 45 & 1.76 &\ding{55} & \checkmark & \\
 & & 60 & 2.00 &\ding{55} & \checkmark & \\
 & & 75 & 1.98 &\ding{55} & \checkmark & \\
 \hline 
6 & 3 & 45 & 1.57 &\ding{55} & \checkmark & \\
 & & 60 & 1.43 &\ding{55} & \checkmark & \\
 & & 75 & 1.10 &\ding{55} & \ding{55} & $R_0$ overestimated by model; ${\cal S}$ very low\\
  & 5 & 45 & 1.62 &\ding{55} & \checkmark & \\
 & & 60 & 1.52 &\ding{55} & \checkmark & \\
 & & 75 & 1.34 &\ding{55} & \checkmark & \\
 & & 90 & 1.01 &\ding{55} & \ding{55} & $R_0$ overestimated by model; ${\cal S}$ very low\\
 & 7 & 45 & 1.62 &\ding{55} & \checkmark & \\
 & & 60 & 1.57 &\ding{55} & \checkmark & \\
 & & 75 & 1.46 &\ding{55} & \checkmark & \\
 & & 90 & 1.24 &\ding{55} & \checkmark & \\
 & 10 & 45 & 1.62 &\ding{55} & \checkmark & \\
 & & 60 & 1.61 &\ding{55} & \checkmark & \\
 & & 75 & 1.54 &\ding{55} & \checkmark & \\
 & & 90 & 1.41 &\ding{55} & \checkmark & \\
 \hline 
\end{tabular}
\caption{This table summarises the properties of each peak in each reconstruction in which it was detected with a signal to noise ${\cal S}_{\rm peak}>1$. The table lists the signal to noise value at the centroid used to compute the radial profile for each peak reconstruction, highlights those reconstructions exhibiting significant B-modes, and indicates whether the best-fit model exhibits values of $m_{\rm peak}$ and $R_0$ compatible with the data. The right-most column lists possible reasons that a peak with no noticeable B-mode signal might not have been included in the final $N_{\rm det}$ for the best-fit model. \label{tab:props}}
\end{table*}

\section{Discussion}
\label{sec:discuss}

It is important at this point to compare our results with other studies of this cluster, both to compare the performance of the flexion aperture mass statistic on small scales with other lensing techniques, and to ensure that prominent structures identified in previous works are correctly identified by our method. 

Whilst there is a wealth of published work on Abell 1689, we restrict our comparison to four works: 
\begin{itemize}
\item Leonard et al. (2007), included to ensure consistency in results obtained using two distinct flexion techniques on the same data.
\item Limousin et al. (2007) and Coe et al. (2010), both of which aim to quantify the substructure content of the cluster using strong and weak lensing measurements, and both of which make use of the same ACS dataset considered here, supplemented with wide field imaging for their weak lensing measurements.
\item Riemer-S{\o}renson et al. (2009), who consider the substructures seen in strong and weak lensing maps of the cluster from Limousin et al. (2007), and compare these to high-resolution X-ray temperature maps from Chandra. 
\end{itemize}

As in Leonard et al. (2007), we find that the central mass density concentration is undetected in all of our flexion reconstructions. As discussed in the earlier work, this results from the fact that our masking scheme strongly favours detection of background sources in the periphery of the image over those in the centre. The lack of data in this region significantly hinders any attempts to constrain the central density profile.

However, the peaks 2 and 6 shown in this paper are indeed detected in Leonard et al. (2007) -- although slightly offset between the two maps -- and the elongation in the Leonard et al. (2007) convergence map in the direction of the peak $[1,3]$ suggests a positive signal may have been seen in the earlier work. However, the parametric reconstruction carried out in that paper assumed that the flexion signal originated from known cluster members, and did not allow for the possibility of a significant dark clump, which the peak $[1,3]$ system appears to be. Similarly, the small offset in the detected positions of peaks 2 and 6 are unsurprising. Therefore, our results are entirely compatible with the earlier work and, moreover, offer the advantage that no assumptions were made regarding the locations of substructure peaks within the cluster.

Peak 2 also appears as a prominent feature in the parametric strong lensing reconstruction carried out by Limousin et al. (2007). Modelling this clump as an isothermal sphere, they find a best fit mass of $M_{200}=(1.25\pm 0.3) times10^{14}h^{-1}M_\odot$ within a radius of $1^\prime.56$, in good agreement with our constraints. As their model only constrains the central $\sim 2.6^\prime$ of the cluster, their mass model does not include our peak $[1,3]$ system or peak 6. However, given that both these systems lie well outside the critical region of the cluster, we do not expect a strong-lensing model to be particularly sensitive to these structures.

In Riemer-S{\o}renson et al., the lensing reconstruction performed by Limousin et al. (2007) is compared to a high-resolution X-ray map obtained using Chandra. The temperature profile of the cluster exhibits a substructure to the northeast of the centre of the cluster, in the same direction as,  but rather offset from, our peak 2. They argue that the X-ray and lensing substructures seen are coincident, with the gas mass amounting $\sim (9.7\pm 0.3)\times 10^{12}M_\odot$. 

No other significant substructure is seen in the X-ray map of the region covered by the ACS imaging. However, this is not entirely surprising given the coarser angular resolution of the Chandra images and the relative insensitivity of the X-ray temperature map to physically small, low-mass structures.

Coe et al. (2010) present an excellent, high resolution, non-parametric reconstruction of the cluster using their \textit{LensPerfect} software. Their reconstruction (c.f. their Figure\,5) shows a clear detection of our Peak 2, as well as smaller sub-clumps in the direction of (but offset from) our peak 6. They also appear to show a marginal detection of our peak 3, but no detection of peak 1. It is likely that this arises due to a lack of strong lensing data in this region of the image, which is substantially offset from the critical, strong lensing regime of the cluster.

\section{Summary}
\label{sec:conclusions}
In this paper, we have investigated in detail the complex internal structure of the galaxy cluster Abell 1689, using measurements of the flexion aperture mass statistic. While the shear aperture mass statistic is seen to be strongly contaminated by systematic errors, the flexion aperture mass statistic appears to hold a great deal of promise for substructure studies within clusters of galaxies. In the central 3.5$^{\prime}$ of the cluster, we identify three significant structures, two of which break down into smaller substructures as the resolution of the $\fmap$ reconstruction is made finer. These peaks are persistent across the $\fmap$ realisations, and largely show minimal B-mode contamination.

The identified structures were localised by comparing their detected positions across the 16 aperture mass reconstructions, and the $2\sigma$ dispersion in these measurements was consistently less than $\sim 10^{\prime\prime}$ in all cases, which highlights the power of the flexion aperture mass statistic to precisely locate substructures in a cluster environment. 

By comparison with template models, a pseudo-probability distribution was computed for each peak in order to determine which mass models best fit the data. In each case, the distribution of allowed masses was fairly narrow, and all peaks showed consistency between the masses obtained using NFW and SIS template models. This implies that our technique offers a method to determine the masses of substructures using only the aperture mass reconstructions and requiring no complicated parametric modelling. 

The results presented here show that Abell 1689 exhibits a rather complex internal structure. The presence of two dominant mass peaks (peaks [1,3] and 2) favour the idea that Abell 1689 has undergone a merger in its recent history (Andersson \& Madejski, 2004); these two mass concentrations may be the remnants of this interaction. Our results also highlight that standard weak lensing methods might not provide a complete description of the structure of a cluster of galaxies; the shear aperture mass reconstructions appear to favour a single lensing halo, whilst the $\fmap$ reconstructions clearly show in much more detail the sub-arcminute structure of the cluster. 

Our method is seen to show excellent agreement with strong lensing mass models of the cluster near the central, critical region of the cluster. Moreover, $\fmap$ techniques offer the possibility to detect significant substructure far removed from the critical region of the cluster, and where strong lensing measurements are not available.

We have demonstrated that even in the absence of high signal to noise $\fmap$ measurements, mass concentrations can be detected, and the allowed parameter space well constrained, by using a moderate number of $\fmap$ reconstructions, and incorporating a requirement of persistence of the structures across reconstructions and no B-mode signal in order to avoid contamination from peaks arising out of noise. This technique offers a simple and straightforward method for mapping and quantifying the substructure content of clusters of galaxies without the need for dynamical information or parametric modelling. It will prove very valuable in the analysis of data such as the HST Multi-Cycle Treasury Program ``CLASH" (P.I.: M. Postman and H. Ford), which will target 25 massive clusters. 
In combination with traditional weak lensing and strong lensing, flexion provides us with a multi-scale view of the distribution of dark matter on cluster scales.

\section{acknowledgments}
The authors would like to thank Marceau Limousin, Thomson Nguyen, Antony Lewis and Neal Jackson for useful discussions. We also acknowledge the anonymous referee, whose comments substantially improved the quality of this paper. 

In this work, AL has been supported in part by a BP/STFC Dorothy Hodgkin Postgraduate Award, and the European Research Council grant SparseAstro (ERC-228261). LJK is supported by a Royal Society University Research Fellowship. DMG is supported by NSF Award 0908307. AL and DMG also acknowledge support from HST Archival grant 10658.01-A.

\appendix
\section{Convergence, Shear and Flexion Profiles}
\label{app:profiles}

This appendix presents the expected shear and flexion profiles arising from two models commonly used in gravitational lensing: the singular isothermal sphere (SIS) model and the Navarro-Frenk-White (NFW) profile. Lasky \& Fluke (2009) provide a derivation of the expected shear and flexion signals from both of these profiles, as well as from a S\'ersic profile, and the results presented here are based in part on their work. 

\subsection{Singular Isothermal Sphere Model}
We begin by considering the SIS model, in which the density profile is defined by:
\begin{equation}
\rho(r)=\frac{\sigma_v^2}{2\pi Gr^2}\ ,
\end{equation}
where $\sigma_v$ is the three dimensional velocity dispersion of the lens. The convergence for this lens is given by:
\begin{equation}
\kappa(\bt)=\frac{\theta_E}{2|\bt|}\ ,
\end{equation}
where we have now made the transformation to angular coordinates ($\bxi=D_{\rm d}\bt$), and the Einstein radius of the lens, $\theta_E$, is given by:
\begin{equation}
\theta_E=4\pi\left(\frac{\sigma_v}{c}\right)^2\frac{D_{\rm ds}}{D_{\rm s}}\ ,
\end{equation}
The deflection angle, shear, and flexions are
\begin{eqnarray}
\alpha(\bt)&=&\theta_E\hat{\bt}\ ,\\
\gamma(\bt)&=&-\frac{\theta_E}{2|\bt|}e^{2i\phi}\ ,\\
{\cal F}(\bt)&=&-\frac{\theta_E}{2|\bt|^2}\hat{\bt}\ ,\\
{\cal G}(\bt) &=& \frac{3\theta_E}{2|\bt|^2}e^{3i\phi}\ ,
\end{eqnarray}
where $\phi$ is the position angle of the source with respect to the lens; i.e. $\phi=\arctan(\theta_2/\theta_1)$. 

\subsection{Navarro-Frenk-White Profile}

This profile has a density defined as
\begin{equation}
\rho(r)=\frac{\delta_c\rho_c}{(r/r_s)(1+r/r_s)^2}\ ,
\end{equation}
where $\rho_c$ is the critical density of the universe, and 
\begin{equation}
\delta_c=\frac{200}{3}\frac{c^3}{\ln(1+c)-c/(1+c)}\ .
\end{equation}
$c$ is the concentration parameter, defined as $R_{200}/r_s$. Before defining the convergence, it is useful to introduce some shorthand notation. First, we define $x\equiv |\bxi|/r_s=|\bt|/\theta_s$, where $|\bt|=|\bxi|/D_{\rm d}$. Second, we define a normalisation factor:
\begin{equation}
\kappa_c=\frac{2\rho_c\delta_cr_s}{\Sigma_c}\ .
\end{equation}
In this notation, the convergence is given by
\begin{equation}
\kappa(x)=\frac{\kappa_c}{\left(x^2-1\right)}\left[1-\Xi(x)\right]\ ,
\end{equation}
where
\begin{equation}
\Xi(x)=\begin{cases}
\frac{2}{\sqrt{1-x^2}} \arctan $h$\left(\sqrt{\frac{1-x}{1+x}}\right) & x<1\\ \\
\frac{2}{\sqrt{x^2-1}}\arctan\left(\sqrt{\frac{x-1}{x+1}}\right) & x>1
\end{cases}\ .
\end{equation} 

The lensing operators are:
\begin{eqnarray}
\alpha(x)&=&\frac{2\kappa_c\theta_s}{x}\left(\ln\left(\frac{x}{2}\right)+\Xi(x)\right)\hat{\bt}\ ,\\
\gamma(x)&=&\frac{\kappa_c}{(x^2-1)}\left[1-\Xi(x)-2\left(1-\frac{1}{x^2}\right)\left(\ln\left(\frac{x}{2}\right)+\Xi(x)\right)\right]e^{2i\phi}\ ,\\
{\cal F}(x)&=&-\frac{\kappa_c}{\theta_sx(x^2-1)^2}\left(2x^2+1-3x^2\Xi(x)\right)\hat{\bt}\ ,\\
{\cal G}(x)&=&\frac{\kappa_c}{\theta_s x(x^2-1)^2}\left[8\left(x-\frac{1}{x}\right)^2\ln\left(\frac{x}{2}\right)+3(1-2x^2)+\left(15x^2-20+\frac{8}{x^2}\right)\Xi(x)\right]e^{3i\phi},\nonumber\\
\end{eqnarray}
where, again, $\phi$ is the position angle of the source with respect to the centre of the lens. Note that in the limit of $x\rightarrow 1$, the lensing operators become:
\begin{eqnarray}
\kappa(1)&=&\frac{\kappa_c}{3}\ , \\
|\alpha|(1)&=&2\kappa_c\theta_s\left(1-\ln(2)\right)\ ,\\ 
|\gamma|(1)&=&\kappa_c\left(2\ln(2)-\frac{5}{3}\right)\ ,\\ 
|\f|(1)&=&\frac{2\kappa_c}{5\theta_s}\ , \\
|\g|(1)&=&\frac{2\kappa_c}{15\theta_s}\left(60\ln(2)-47\right)\ .
\end{eqnarray}

\label{lastpage}

\end{document}